\documentclass[11pt]{article}
\usepackage{amsmath}
\usepackage{amsfonts}
\usepackage{amssymb}
\usepackage{graphicx}
\usepackage[english]{babel}
\usepackage{amsmath}
\usepackage{amsfonts}
\usepackage{amssymb}
\usepackage{graphicx}   
\usepackage{times}
\def\psl {/\ \hskip -9pt p}       
\def\hbar{/\ \hskip -9pt $h$}     
\def\qsl {/\ \hskip -9pt q}       

\begin{document}
%
June 15th 2010   \hfill 

\null\vskip 2.5cm
{\baselineskip 20pt
\begin{center}
{\bf MIXING  AT 1-LOOP 
IN A $\boldsymbol{SU(2)_L}$ GAUGE THEORY OF WEAK INTERACTIONS}
\end{center}
}
\centerline{ B.~Machet
     \footnote[1]{LPTHE
(Laboratoire de Physique Th\'eorique et Hautes \'Energies),
tour 13-14, 4\raise 3pt \hbox{\tiny \`eme} \'etage,
          UPMC Univ Paris 06, BP 126, 4 place Jussieu,
          F-75252 Paris Cedex 05 (France),\\
         Unit\'e Mixte de Recherche UMR 7589 (CNRS / UPMC Univ Paris 06)}
    \footnote[2]{machet@lpthe.jussieu.fr}
     }
\vskip .7cm

\title{Mixing at 1-loop  in a
$\boldsymbol{SU(2)_L}$ gauge theory of weak interactions}

{\bf Abstract:}
Flavor mixing is scrutinized at 1-loop in a $SU(2)_L$ gauge theory of massive
fermions.
The main issue is to cope with kinetic-like,  momentum ($p^2$) dependent
effective interactions that arise at this order.
They spoil the unitarity of the connection between flavor and mass
states, which potentially alters the standard Cabibbo-Kobayashi-Maskawa
(CKM) phenomenology by giving rise, in
particular, to extra flavor changing neutral currents (FCNC).
We explore the conservative requirement
that these should be suppressed, which yields
relations between the CKM angles, the fermion and $W$ masses, and
a renormalization scale $\mu$.
For two generations, two solutions arise: either the mixing angle
of the fermion pair the closer to degeneracy is close to maximal while,
inversely, the mass and flavor states of the other pair are
quasi-aligned, or mixing angles in both sectors are very small.
For three generations, all mixing angles of neutrinos are predicted to be
large  ($|\theta_{23}| \approx$  maximal is the largest) and
the smallness of their mass differences induces mass-flavor quasi-alignment
for all charged leptons.
The hadronic sector differs in that the top quark is twice as heavy as the
$W$.  The situation  is, there, bleaker, as all angles come out too large,
but, nevertheless, encouraging,  because $\theta_{12}$ decreases as
the top mass increases.  Whether  other super-heavy
fermions could drag it down to realistic values stays an open issue,
together with the role of higher order corrections.
The same type of counterterms that turned off the 4th order
static corrections to the quark electric dipole moment are, here too, needed,
in particular to stabilize quantum corrections to mixing angles.

\smallskip

{\bf PACS:} 12.15.Ff \quad 12.15.Lk \quad 14.60.Pq  \hfill

\section{Introduction}

The origin of large mixing angles observed in leptonic charged currents
is still largely unknown \cite{Smirnov}.
A widespread belief is that it is linked to a
quasi-degeneracy of neutrinos, but this connection was never firmly
established. And it cannot be on simple grounds.
Indeed, the mixing angles that are ``observed'' in
neutrino oscillations are the ones
occurring in charged currents, which combine the individual mixing
matrices of fermions  with different electric charges 
\footnote{The electronic $(\nu_e)$, muonic $(\nu_\mu)$,
and tau $(\nu_\tau)$ neutrinos are
defined as the neutrinos that couple, inside charged currents,
 to the mass eigenstates of charged leptons. They are accordingly related
to the neutrino mass eigenstates by
 $ (\nu_e\ \nu_\mu\  \nu_\tau)^T = K_\ell^\dagger K_\nu
 (\nu_{em}\ \nu_{\mu m}\ \nu_{\tau m})^T$ 
 where $K_\ell$ and $K_\nu$ are the mixing matrices
 respectively of charged leptons and neutrinos.
This connection is seen to involve
the hermitian conjugate $K_\ell^\dagger K_\nu$  of the PMNS matrix.
\label{foot:nu}
}
; the
path that goes from the quasi-degeneracy of one of the two doublets to
large mixing in the PMNS matrix \cite{PMNS}
cannot thus be completely straightforward. 
 Furthermore, homographic
transformations on a (mass) matrix, while changing its eigenvalues,
do not change its eigenvectors, neither, accordingly, mixing angles;
an infinity of different mass spectra can thus be associated with a given
mixing angle.

We shall first focus  on two pairs of fermions, making up two generations.
For the sake of convenience (mainly for the simplicity of notations)
 we shall often call them generically
$(d,s)$ and $(u,c)$. The first will be supposed to be close to degeneracy
and the second largely split.
Results are transposed to the leptonic sector:
the Cabibbo angle $\theta_c$ \cite{Cabibbo} is then, in particular,
replaced by the corresponding entry $\theta_{PMNS}$ of the ($2 \times 2$)
PMNS matrix. Results which are specific to neutrinos will of course be
written with the adequate notations. 

This study, which finally supports  a relation  between
quasi-degeneracy and large mixing, rests on the following
argumentation.

 The  physical states are the eigenstates of the propagator at
its poles; in case of a coupled system of $n$ particles, like massive fermions
in the standard model of electroweak interactions \cite{GSW} which
are coupled through the scalar sector, the propagator, which is also the
inverse of the quadratic Lagrangian, is a $n \times n$
matrix;\newline 
 The determination of an orthogonal set of physical states accordingly 
requires the diagonalization of the sum
of the kinetic terms and of the mass terms in the Lagrangian;\newline
At the classical level, this procedure yields the standard
Cabibbo-Kobayashi-Maskawa (CKM) \cite{Cabibbo} \cite{CKM} phenomenology.
The classical Lagrangian
is written from the start devoid {\em a priori}, in bare flavor space,
 of FCNC.  In direct connection with the unitarity of mixing matrices,
in particular the Cabibbo matrix, the $SU(2)$ gauge algebra closes
on a diagonal ${\mathbb T}^3$ generator, which eliminates FCNC
at this order, in bare mass space as well as in bare flavor space
\footnote{The terminology FCNC is certainly not very good when dealing
with (bare) mass states. The reader should understand it as ``non-diagonal
currents in mass space''.}.
FCNC are generated at 1-loop among bare flavor or mass states (see Fig.~1),
but they are damped by the so-called ``Cabibbo suppression''.
This phenomenology is, up to now, in very good
agreement with experiment, and we choose to preserve it;\newline
 Subtle issues arise when considering the quadratic effective
 Lagrangian at 1-loop since, in particular, non-diagonal kinetic-like
transitions are generated (Fig,~2).
Then, the mandatory re-diagonalization of kinetic
terms, which is generally overlooked, exhibits two main features. 
First, due to the presence of mass-splittings, it unavoidably involves
slightly non-unitary transformations, which introduces in bare flavor space
at 1-loop, a new set of, mass and mixing (and $p^2$) dependent, FCNC.
Secondly, the 1-loop corrections to the mixing angles are non-perturbative
and present a high instability in the vicinity of degeneracy. This strongly
motivates the introduction of counterterms ``\`a la Shabalin''
 \cite{Shabalin} that cancel
1-loop non-diagonal transitions ``on mass-shell''.\newline
They restore a quasi-standard Cabibbo phenomenology, but for
the persistence of extra, mass and mixing dependent, FCNC in bare flavor space.
Their occurring is rooted in the non-degeneracy of fermions, which
counterterms  cannot turn off.
They are built to cancel non-diagonal 1-loop transitions when one of
the two concerned external fermions is on mass-shell, but the second
can, then, only be off mass-shell.
So, while 1-loop mass eigenstates, which result from the diagonalization of
the effective 1-loop Lagrangian, are, by definition, orthogonal 
and, as we show, do not exhibit FCNC
\footnote{with a subtlety, due to the dependence of $p^2$, that is
evoked in appendix \ref{subsec:nonorphys}.}, this is not exactly so
for bare mass states: orthogonality only
truly occurs among one on mass-shell and one off-mass shell
fermion.\newline  
We investigate at which condition these extra FCNC can get suppressed.
Such a requirement establishes a
connection between mass splittings and the Cabibbo angle $\theta_c$, which, for 
two generations and $m_u^2, m_d^2, m_s^2, m_c^2, p^2 \ll M_W^2$, writes
$\cos 2\theta_c \approx -\frac12 \frac{m_s^2 -
m_d^2}{m_c^2 - m_u^2}$. $\theta_c$ is seen to be quasi-maximal as soon as
$|m_s-m_d| \ll |m_c-m_u|$, that is, when one of the two fermion pair is
much closer to degeneracy than the second. A similar condition is  realized 
in the 2-generation leptonic sector, pushing to large values 
the similar angle of the PMNS matrix.  Thus, the conservative
requirement that the standard classical Cabibbo phenomenology
should be preserved at 1-loop provides, through FCNC, a connection between
large mixing and the quasi-degeneracy of two same-charge fermions.

Nature is however more complex: -- first, there are three and not only two
generations; secondly, in the quark sector, all mixing angles are small;
 -- last, while, in the lepton sector, the ``atmospheric''
angle $\theta_{23}$ seems actually close to maximal, this is not the case
for the ``solar'' angle $\theta_{12}$ which, though large, looks closer to
$35^o$, nor for $\theta_{13}$, which could be much smaller \cite{GiuntiKim}.
This is why the last part of this work
is dedicated to the 3-generations case, making in particular the
distinction between the leptonic case, where all known fermions
stand well below the electroweak scale $M_W$, and
the quark case where the top quark weights roughly $2M_W$.

This work is structured as follows. Sections \ref{section:fermions} to
\ref{section:FCNC} deal with two generations of fermions, first, from
section \ref{section:fermions} to \ref{section:rencab}, without introducing
Shabalin's counterterms, then, in sections \ref{section:stability} and
\ref{section:FCNC}, in their presence. Section
\ref{section:3gen} analyzes in detail the case of three generations.

In section \ref{section:fermions}, we explain the procedure to
re-diagonalize, at 1-loop, the quadratic Lagrangian (kinetic + mass terms)
of an $SU(2)_L$ gauge theory for several generations of massive fermions.
In subsection \ref{subsec:principle} we first briefly recall
the standard procedure to diagonalize, by a bi-unitary
transformation, the classical quadratic Lagrangian.
We then outline, taking the example of two
generations, how it is modified when 1-loop
transitions introduce non-diagonal, $p^2$-dependent, kinetic-like
interactions.
In  subsection \ref{subsec:1lt} we give the analytical formul{\ae} in the limit
$p^2 \ll m_W^2$, which then largely simplify when the four fermions masses
are  much smaller than the $W$ mass, too. Subsections \ref{subsec:diakin}
and \ref{subsec:diamass} are
respectively devoted to the re-diagonalization of kinetic terms, and of
mass terms. The first are shown to unavoidably introduce, because of mass
splittings, non-unitary transformations After these operations are done,
the whole effective quadratic
Lagrangian at 1-loop is back to diagonal, with its kinetic terms
proportional; to the unit matrix $\mathbb I$.

In section \ref{section:indivmix}, we focus on the (realistic)
case $|m_s-m_d| \ll |m_c-m_u|$. We  study  individual mixing matrices
({\em i.e.} the ones in the $(u,c)$ and $(d,s)$ sectors) and the two
corresponding mixing angles.

Section \ref{section:rencab} is devoted to the 1-loop Cabibbo matrix. First we
show how gauge invariance dictates the form of the 1-loop effective
Lagrangian, by, in particular, relating through the covariant derivative,
kinetic terms to gauge currents. We then demonstrate that, unlike
individual mixing matrices, the Cabibbo matrix stays unitary at 1-loop.

In section \ref{section:stability}, we first show that, in the absence of
counterterms, the 1-loop renormalization of the mixing angle for
degenerate $(d,s)$ is pathological. We then show how the introduction
of Shabalin's counterterms restore the stability and reliability of
1-loop corrections to mixing angles, in particular in the vicinity of
 degeneracy.  The 1-loop Cabibbo matrix still keeps unitary in their
presence.

In section \ref{section:FCNC}, still for two generations,
we show how extra FCNC arise, and we we solve the constraints controlling
their suppression, first in the absence of counterterms, then in their
presence.

Section \ref{section:3gen} is an extensive study of the 3-generation case,
in the presence of Shabalin's counterterms.
In subsection \ref{subsec:nofcnc}, we write the three equations which
guarantee that no extra FCNC is present in the bare flavor (or mass) space.
We then explicitly list all possible solutions. In subsection
\ref{subsec:anahij} we give analytical expressions concerning 1-loop
transitions between fermions when one among the six fermions making up
three generations (the top quark) is heavier than the $W$. In subsection
{\ref{subsec:6quarks} we solve the constraints for quarks. In subsection
\ref{subsec:3nu} we solve them for neutrinos.

The conclusions and outlook are given in section \ref{section:end}. We also
give, there, a comparison between this work and previous approaches
concerning the renormalization of mixing angles.

In appendix \ref{section:cterms}, we briefly comment on the dependence on
$p^2$ and some of its consequences, that we neglected in the core of the paper
where we considered the limit $p^2 \ll m_W^2$.

For the sake of simplicity (like in \cite{Shabalin}),
 we  work in a pure $SU(2)_L$ theory
of weak interactions instead of the standard $SU(2)_L \times U(1)$
electroweak model \cite{GSW}.
Since the theory is renormalizable, we  use the unitary gauge,
devoid of the intricacies due to scalar fields and which,
consistently working at order $g^2$, yields finite results for the
quantities of concern to us. While we cannot, accordingly, verify the
gauge independence of the results (independence on the $\xi$ parameter in
an $R_\xi$ gauge), gauge invariance is of primordial importance.

\section{1-loop transitions between non-degenerate fermions\,;
re-diagonalizing the quadratic Lagrangian}
\label{section:fermions}

\subsection{Principle of the method}
\label{subsec:principle}

At the classical level,  a bi-unitary transformation is used, in flavor
space, to diagonalize the sum of kinetic + mass terms
$\left(\begin{array}{cc}\bar d^0_f & \bar s^0_f\end{array}\right)
\left[ \psl\; {\mathbb I} - M^0_f\right]
\left(\begin{array}{c} d^0_f\cr s^0_f\end{array}\right)$  into
 $\left(\begin{array}{cc}\bar d^0_m & \bar s^0_m\end{array}\right)
\left[ \psl\; {\mathbb I} - \left(\begin{array}{cc}m_d & \cr & m_s \end{array}\right)\right]
\left(\begin{array}{c} d^0_m\cr s^0_m\end{array}\right)$. The two unitary
transformations, acting respectively on right- and left-handed fermions,
preserve the canonical form of both kinetic terms, which stay proportional 
to the unit matrix $\mathbb I$. This defines the
classical masses $m_d$ and $m_s$. The corresponding classical mass eigenstates
 $d^0_m$ and $s^0_m$ are orthogonal with respect to the classical
Lagrangian, which is akin to the property that no transition between them occurs
at the classical level.
In particular, the classical Lagrangian in flavor space is written devoid
{\em a priori} of FCNC; this is directly
related to the property that kinetic terms are proportional to the unit matrix,
since gauge currents are simply deduced by introducing the covariant derivative
with respect to the gauge group. The above
diagonalization leads to the standard Cabibbo (or CKM) phenomenology, in
which, in particular,  non-diagonal neutral gauge currents only get generated
at 1-loop (see Fig.~1), and are  damped, when expressed in bare
mass space, by the so-called ``Cabibbo suppression''.
This phenomenology is, up to now, in  agreement with experiment.

\vbox{\vskip 1cm
\begin{center}
\includegraphics[height=3.5truecm,width=5 truecm]{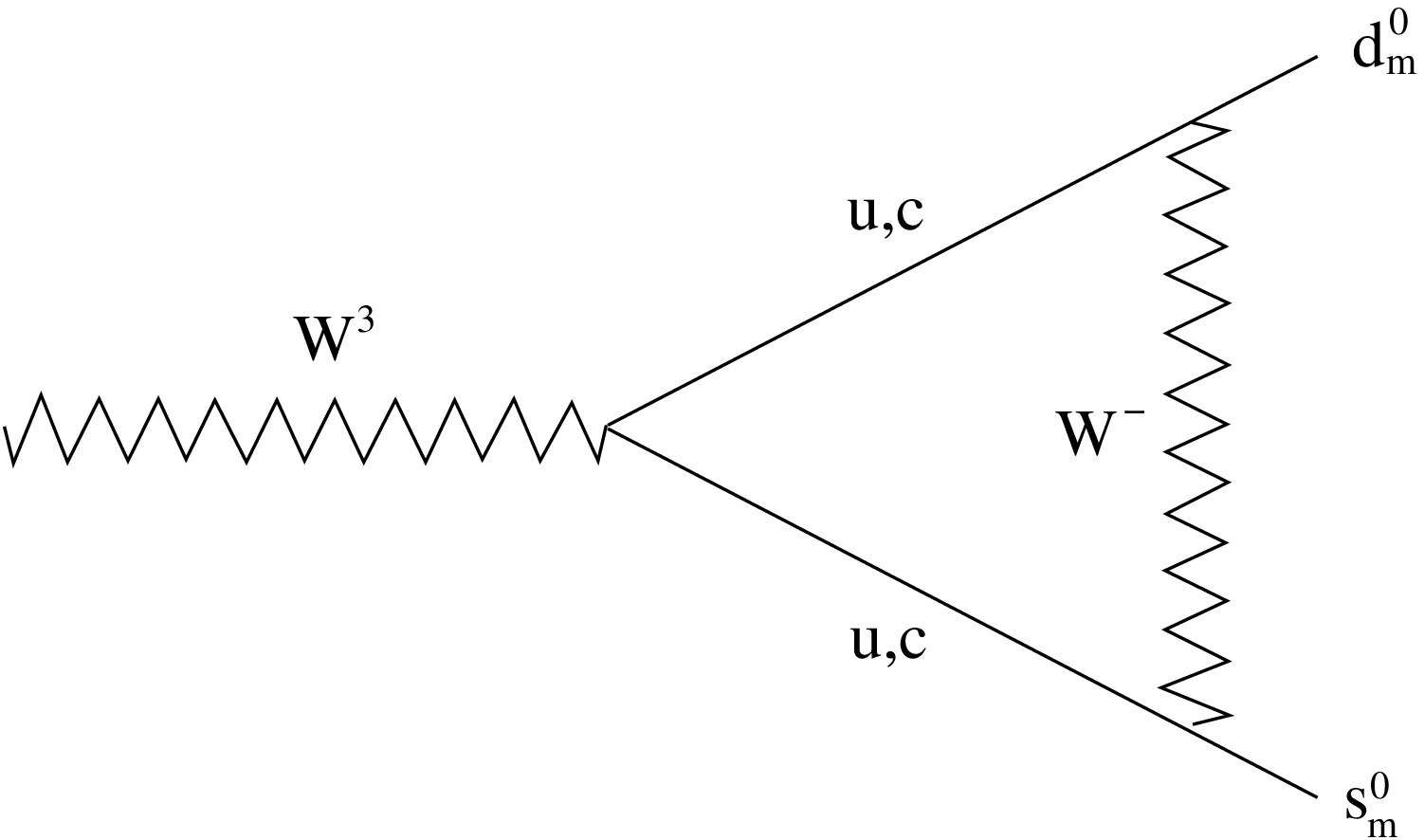}
\hskip 1cm
\includegraphics[height=3.5truecm,width=8 truecm]{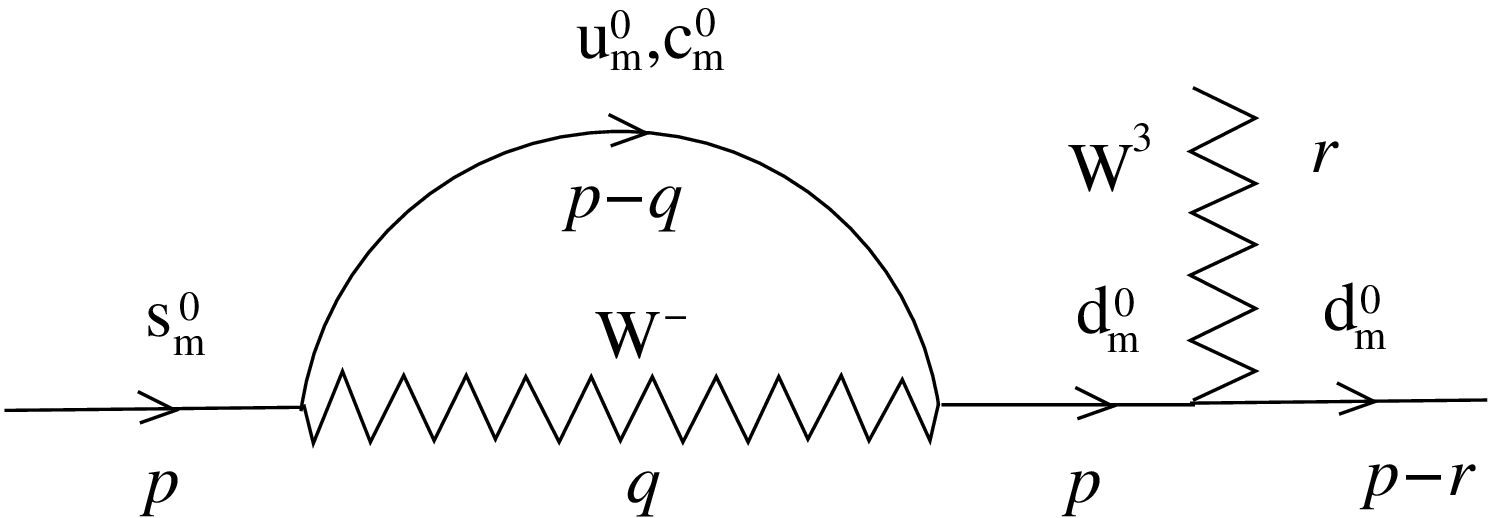}
\vskip 0pt
{\em Fig.~1: ``Standard'' flavor changing neutral currents at 1-loop}
\end{center}
}

However,  1-loop non-diagonal transitions, like $s^0_m \to d^0_m$
depicted in Fig.~2, trigger new phenomena which have not yet been
fully considered and which, in particular, also generate FCNC.
By the effect of the corresponding renormalization, the kinetic terms of
left-handed fermions stay indeed no
longer proportional to the unit matrix $\mathbb I$ but to some non-diagonal
 $K_d = {\mathbb I} + H_d, H_d={\cal O}(g^2)$, which depends on the classical
masses (fermions and gauge fields), on the classical Cabibbo mixing angle
$\theta_c$, and on $p^2$.

The pure kinetic terms $K_d$  for $(d^0_m, s^0_m)$ written
in (\ref{eq:Kd3})
\footnote{For the sake of convenience, we  work in the bare mass basis.}
 can be cast back to their canonical form by
a $p^2$-dependent non-unitary transformation  ${\cal V}_d(p^2, \ldots)$
according to
\begin{equation}
{\cal V}_{d}^\dagger\; K_d\; {\cal V}_{d} = {\mathbb I}.
\label{eq:kincon1}
\end{equation}
By (\ref{eq:kincon1}), which entails $K_d = ({\cal V}_d^{-1})^\dagger {\cal
V}_d^{-1}$,  the kinetic terms
\footnote{The subscript ``${_L}$'' refers  to left-handed fermions and
``${_R}$'' to right-handed ones.}
 $(\overline{d^0_{mL}},
\overline{s^0_{mL}})K_d \   /\ \hskip -10pt p
\left(\begin{array}{c} d^0_{mL} \cr s^0_{mL}
\end{array}\right)$ at 1-loop for left-handed $d$ and
$s$ in the bare mass basis rewrite $(\overline{d^0_{mL}},
\overline{s^0_{mL}})({\cal V}_d^{-1})^\dagger {\cal
V}_d^{-1}\  /\ \hskip -10pt p \left(\begin{array}{c} d^0_{mL} \cr s^0_{mL}
\end{array}\right)$, which leads to defining $d^1_{mL}$ and
$s^1_{mL}$ such that
%
$\left(\begin{array}{c} d^1_{mL} \cr s^1_{mL} \end{array}\right)
= {\cal V}_d^{-1}
\left(\begin{array}{c} d^0_{mL} \cr s^0_{mL} \end{array}\right)$.
%
The mass matrix, which had been made diagonal in the classical basis
$(d^0_m, s^0_m)$, is no longer so in the basis $(d^1_{mL}, s^1_{mL})$.
The second step of the procedure is accordingly to re-diagonalize it by a
second bi-unitary transformation. It leaves unchanged the canonical
form of the kinetic terms that has  been rebuilt in the first step of
the procedure. After the two steps have been completed,
the sum of kinetic + mass terms at 1-loop is diagonal.
The resulting basis of 1-loop mass eigenstates $(d_{mL}(p^2,
\ldots),s_{mL}(p^2, \ldots))$ is such that, at this order and at any
given $p^2$,
there exists no transition between $d_{mL}$ and $s_{mL}$. They
 are thus, by definition, orthogonal at 1-loop.

\subsection{1-loop transitions: explicit calculations}
\label{subsec:1lt}

We now explicitly calculate 1-loop transitions.
Gauge interactions induce  diagonal and non-diagonal
transitions between bare mass states. For example, Fig.~2 describes
 non-diagonal $s^0_m \to d^0_m$ transitions, mediated by the $W^\pm$
gauge bosons. Diagonal transitions are mediated either
by $W_\mu^\pm$ or by  $W_\mu^3$.

\vbox{
\begin{center}
\includegraphics[height=4truecm,width=8truecm]{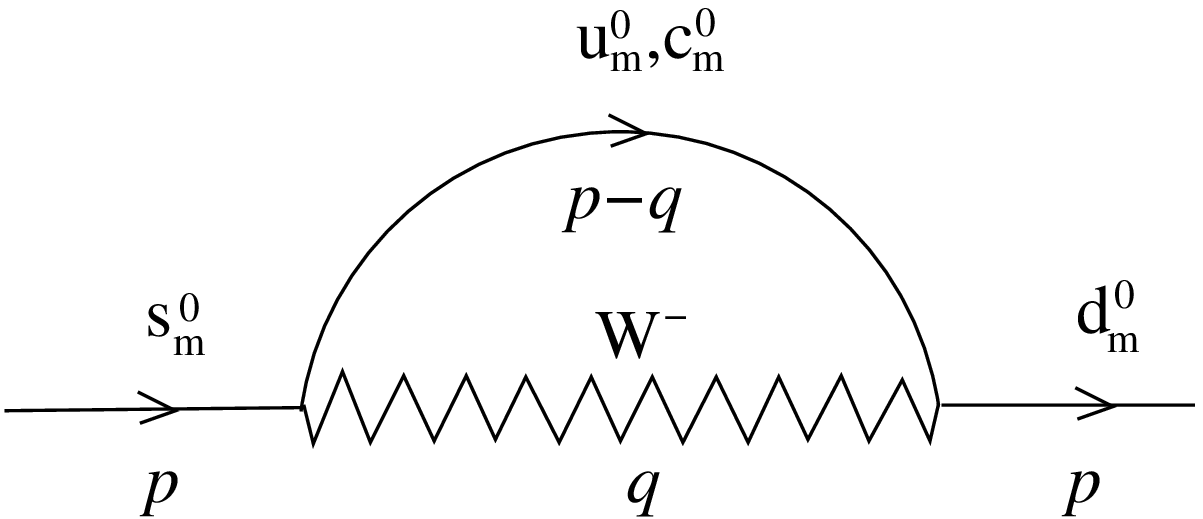}
\vskip 0pt
{\em Fig.~2: $s^0_m \to d^0_m$ transition at 1-loop}
\end{center}
}

The one depicted in Fig.~2 contributes as a left-handed, kinetic-like,
$p^2$-dependent interaction
\begin{equation}
{\cal A}_{sd}\, \bar d^0_m\, \psl (1-\gamma_5)\, s^0_m, \quad
{\cal A}_{sd} = \sin\theta_c \cos\theta_c \big(h(p^2, m_u, m_W)-h(p^2, m_c,
m_W)\big),
\label{eq:1loop}
\end{equation}
that we abbreviate, with shortened notations $\sin\theta_c \equiv s_c,
\cos\theta_c \equiv c_c$, into
\begin{equation}
{\cal A}_{sd} = s_c c_c (h_u-h_c).
\label{eq:1loopbis}
\end{equation}
It depends in particular on the classical Cabibbo angle
$\theta_c = \theta_d - \theta_u$.  The function $h$ is dimensionless.

 It is straightforward to deduce that  all
(diagonal and non-diagonal) 1-loop transitions  between $s^0_m$ and
$d^0_m$ mediated by $W^\pm$ gauge bosons  transform
their kinetic terms into
\begin{eqnarray}
&&\left(\begin{array}{cc} \bar d^0_m & \bar s^0_m\end{array}\right)
\left[ {\mathbb I}\; \psl +
\left(\begin{array}{cc}
c_c^2 h_u + s_c^2 h_c & s_c c_c(h_u -h_c) \cr
s_c c_c (h_u - h_c) & s_c^2 h_u + c_c^2 h_c
\end{array}\right)
\psl (1-\gamma_5)\right]
\left(\begin{array}{c} d^0_m \cr s^0_m
\end{array}\right)\cr
= && \left(\begin{array}{cc} \bar d^0_m & \bar s^0_m\end{array}\right)
\left[ {\mathbb I}\; \psl +
\left(\frac{h_u+h_c}{2} + (h_u-h_c)\; {\cal
T}_x(2\theta_c)\right)
\psl (1-\gamma_5)\right]
\left(\begin{array}{c} d^0_m \cr s^0_m
\end{array}\right),
\label{eq:Wcont}
\end{eqnarray}
where we noted
\begin{equation}
{\cal T}_x(\varphi) = \frac12\left(\begin{array}{rr}
\cos \varphi & \sin\varphi \cr \sin\varphi & -\cos\varphi \end{array}\right).
\label{eq:Tx}
\end{equation}
To the contributions (\ref{eq:Wcont}) we must add the diagonal
transitions mediated by the  $W^3_\mu$ gauge boson.
The kinetic terms for left-handed $d^0_m$ and $s^0_m$ quarks
then become
(omitting the fermionic fields and the dependence on
$p^2,\ldots$)
\footnote{From now onwards, to lighten the notations,
we shall frequently omit the dependence on $p^2$ and on the masses.}

\vbox{
\begin{eqnarray}
K_d &=& {\mathbb I} + H_d\; ;\cr
H_d &=& \left(\begin{array}{cc} {\cal A}_{dd} & {\cal A}_{ds}
\cr {\cal A}_{sd} & {\cal A}_{ss} \end{array}\right)
= \frac{h_u+h_c}{2} + (h_u-h_c)\; {\cal T}_x(2\theta_c)
+ \frac12\left(\begin{array}{cc}h_d & \cr & h_s\end{array}\right),
\label{eq:Kd3}
\end{eqnarray}
}

where $h_d = h(p^2, m_d, m_W)$ and $h_s = h(p^2, m_s, m_W)$.
Likewise, in the $(u,c)$ sector, one has

\vbox{
\begin{eqnarray}
K_u &=& {\mathbb I} + H_u\; ;\cr
H_u &=& \left(\begin{array}{cc}{\cal A}_{uu} & {\cal A}_{uc}
\cr {\cal A}_{cu} & {\cal A}_{cc}\end{array}\right) =
 \frac{h_d+h_s}{2} + (h_d-h_s)\; {\cal T}_x(-2\theta_c)
+ \frac12\left(\begin{array}{cc}h_u & \cr & h_c\end{array}\right).
\label{eq:Ku3}
\end{eqnarray}
}

Explicitly, one has

\vbox{
\begin{eqnarray}
{\cal A}_{sd} &=& \frac{g^2}{4}\int \frac{d^4q}{(2\pi)^4}\frac{1}{q^2-m_W^2}
\Big[(2-\epsilon)(\psl - \qsl) + \frac{2q.(p-q)\psl -q^2(\psl-\qsl)}{m_W^2}\Big]
(1-\gamma^5) \cr
&& \hskip 6cm \Big[\frac{V_{us} V_{ud}^\ast}{(p-q)^2 - m_u^2}+
\frac{V_{cs}V_{cd}^\ast}{(p-q)^2 - m_c^2}  \Big]\cr
&&\cr && \cr
&\stackrel{unitarity\ of\ V}{=}&
 \frac{g^2}{4}\int \frac{d^4q}{(2\pi)^4}\frac{1}{q^2-m_W^2}
\Big[(2-\epsilon)(\psl - \qsl) + \frac{2q.(p-q)\psl -q^2(\psl-\qsl)}{m_W^2}\Big]
(1-\gamma^5) \cr
&& \hskip 6cm V_{us}V_{ud}^\ast \frac{m_u^2-m_c^2}{\big[(p-q)^2 -
m_u^2\big]\big[(p-q)^2 - m_c^2\big]}.\cr
&&
\label{eq:Asd}
\end{eqnarray}
}

The factor $V_{us}V_{ud}^\ast$ in (\ref{eq:Asd}) is the $s_cc_c$ of
(\ref{eq:1loop}), which finally defines $(h_u - h_c)$ of (\ref{eq:1loopbis}).

All our forthcoming results depend on differences like $(h_i-h_j)$.
In the unitary gauge, after introducing 2 Feynman parameters $x$ and $y$,
the dimensionally (for $n = 4-\epsilon$ dimension) regularized expression
for $(h_i - h_j)$ writes ($\gamma \approx 0.572$ is the Euler constant)

\vbox{
\begin{eqnarray}
h_i-h_j &=& \frac{g^2}{4}\frac{i}{16\pi^2}(m_i^2-m_j^2)\int_0^1 dx \int_0^1
dy\; 2y \cr
&& \hskip -1.5cm\Bigg[-(1-y)\left( 1 + \frac{y^2 p^2}{2m_W^2}\right) \frac{1}{R^2}
+\frac{1}{m_W^2}\Bigg(-\left(-\frac{1+3y}{2}\right)\left(\frac{2}{\epsilon}+\ln
4\pi -\gamma \right) + \frac{1+y}{2} +
\frac{1+3y}{2}\ln\frac{R^2}{\mu^2}\Bigg) \Bigg],\cr
&& \cr
R^2 &=& -y(1-y)p^2 + y(1-x) m_j^2 + xy m_i^2 + (1-y)m_W^2.
\label{eq:hij1}
\end{eqnarray}
}

To obtain (\ref{eq:hij1}), the  relation $\gamma_\nu \gamma_\alpha
\gamma_\nu = -(2-\epsilon) \gamma_\alpha$ between the Dirac
matrices has been used. The scale $\mu$ originates from the necessity,
in $4-\epsilon$ dimensions, to replace $g^2$ by $g^2 \mu^\epsilon$.
The exact analytical expression for all values of $p^2, m_i^2, m_j^2$ cannot
be easily obtained, but, when $p^2\ll m_W^2$, $y(1-y)p^2$ can be safely
neglected with respect to $(1-y)m_W^2$ in $R^2$, such that
 (\ref{eq:hij1}) simplifies 
into (we write this time its expression once renormalized in the
$\overline{MS}$ scheme
which amounts to eliminating from (\ref{eq:hij1}) the pole in $1/\epsilon$
together with the terms proportional to $\ln 4\pi -\gamma$)
\begin{eqnarray}
h_i-h_j &\overset{p^2\ll m_W^2}{\underset{\overline{MS}}{\approx}}&
 \frac{g^2}{4}\frac{i}{16\pi^2}(m_i^2-m_j^2)
\int_0^1 dx \int_0^1 dy\; 2y 
\left[-\frac{1-y}{r^2}
+\frac{1}{2m_W^2}\left( (1+y) +
(1+3y)\ln\frac{r^2}{\mu^2}\right) \right],\cr
&& \cr
r^2 &=&  y(1-x) m_j^2 + xy m_i^2 + (1-y)m_W^2.
\label{eq:hij2}
\end{eqnarray}
The integration over $x$ can be done explicitly. This leads to the
 expression

\vbox{
\begin{eqnarray}
h_i-h_j &\overset{p^2\ll m_W^2}{\underset{\overline{MS}}{\approx}}&
\frac{g^2}{4}\frac{i}{16\pi^2}\int_0^1 dy
\left[-2(1-y) \ln\frac{ym_i^2+(1-y)m_W^2}{ym_j^2+(1-y)m_W^2}
-2y^2\frac{m_i^2 - m_j^2}{m_W^2}
\right.\cr
&& \hskip -1.5cm\left. y(1+3y)
\left(\frac{m_i^2}{m_W^2} \ln\frac{ym_i^2+(1-y)m_W^2}{\mu^2}
-\frac{m_j^2}{m_W^2} \ln\frac{ym_j^2+(1-y)m_W^2}{\mu^2}\right)
\right],
\label{eq:hij3}
\end{eqnarray}
}
which, like (\ref{eq:hij1}) and (\ref{eq:hij2}), vanishes when $m_i=m_j$.
After explicitly doing the $\int dy$ integration, one gets

\vbox{
\begin{eqnarray}
h_i-h_j &\overset{p^2\ll m_W^2}{\underset{\overline{MS}}{\approx}}&
\frac{g^2}{4}\frac{i}{16\pi^2}
\left[
-\frac23\frac{m_i^2-m_j^2}{m_W^2}
-2\left(\frac{m_W^2 \ln\frac{m_W^2}{\mu^2}-m_i^2
\ln\frac{m_i^2}{\mu^2}}{m_W^2-m_i^2} - (i\leftrightarrow j) \right)
\right.
\cr
&&
\left.
+\left(\left(2+\frac{m_i^2}{m_W^2}
\right)\left(-\frac{m_W^2}{m_W^2-m_i^2}
+ \frac{m_W^2\left( m_W^2\ln\frac{m_W^2}{\mu^2}-m_i^2\ln\frac{m_i^2}{\mu^2} \right)}
{(m_W^2-m_i^2)^2} \right. \right. \right. \cr
&& \hskip 2cm \left. \left.\left.
+\frac14\frac{m_W^2+m_i^2}{m_W^2-m_i^2}
-\frac12\frac{m_W^4\ln\frac{m_W^2}{\mu^2}-m_i^4\ln\frac{m_i^2}{\mu^2}}{(m_W^2-m_i^2)^2}
\right)- (i\leftrightarrow j)\right)
\right.
\cr 
&&
\left.
+ \left(\frac{m_i^2}{m_W^2}\frac{1}{(m_W^2-m_i^2)^2}
\left( -\frac{11m_W^4-7m_W^2m_i^2+2m_i^4}{6} \right. \right. \right. \cr
&& \hskip 2cm \left. \left. \left.
+\frac{m_W^6\ln\frac{m_W^2}{\mu^2}+\left(-3m_W^4m_i^2
+3m_W^2m_i^4-m_i^6\right)\ln\frac{m_i^2}{\mu^2}}{m_W^2-m_i^2} \right)
-(i \leftrightarrow j) \right)
\right].\cr && {}
\label{eq:hij4}
\end{eqnarray}
}

Eq.~(\ref{eq:hij4}) is only valid for $p^2 \ll m_W^2$ but its
dependence on the fermion masses $m_i$ and $m_j$ is then exact.
In the limit, always valid for 2 generations, when $m_i^2, m_j^2 \ll m_W^2$,
it drastically simplifies to
\begin{equation}
h_i-h_j \overset{p^2, m_i^2,m_j^2\ll m_W^2}{\underset{\overline{MS}}{\approx}}
\frac{g^2}{4}\frac{i}{16\pi^2} \frac{m_i^2-m_j^2}{m_W^2}
\left( -\frac{17}{4} + \frac32\ln\frac{m_W^2}{\mu^2} \right).
\label{eq:hij5}
\end{equation}
In the case of 3 generations of quarks, the top quark enters the game and
one is in the situation when $p^2, m_i^2 \ll m_W^2$ but
$m_j^2 \equiv m_t^2 \geq m_W^2$. The corresponding formul{\ae} will be given
 in  subsection \ref{subsec:anahij}.
Note that, in the approximation $p^2 \ll m_W^2$ that we are using, the
final result (\ref{eq:hij5}) no longer depends on $p^2$.

\subsection{First step: re-diagonalizing  kinetic terms
back to the unit matrix}
\label{subsec:diakin}

We shall now diagonalize the quadratic part of the
 effective 1-loop Lagrangian, which
means putting the pure kinetic terms back to the unit matrix and,
at the same time, re-diagonalizing the mass matrix.
This is accordingly a two-steps procedure.

Since the kinetic terms of right-handed fermions are not modified,
we shall only be concerned  with the left-handed ones.

The pure kinetic terms $K_d$  for $(d^0_m, s^0_m)$ written
in (\ref{eq:Kd3}) can be cast back to their canonical form by
a $p^2$-dependent non-unitary transformations  ${\cal V}_d(p^2, \ldots)$
according to (\ref{eq:kincon1}).

The procedure to find ${\cal V}_d$ is the following.
Let $(1+t_+^d)$ and $(1+t_-^d)$, $t_+^d, t_-^d = {\cal O}(g^2)$,
 be the eigenvalues of the symmetric matrix
 $K_d$; explicitly
\begin{equation}
t_\pm^d = \frac{h_u + h_c + \left[\frac{h_d + h_s}{2}\right]}{2}
\pm \frac12 \sqrt{\left(h_u - h_c\right)^2 + \left[\frac{h_d -
h_s}{2}\right]^2
+ 2\;(h_u - h_c)\;\left[\frac{h_d - h_s}{2}\right]\cos 2\theta_c}.
\label{eq:t+-}
\end{equation}
$K_d$ can be diagonalized by a rotation ${\cal
R}(\omega_d) \equiv\left(\begin{array}{rr} \cos\omega_d & \sin\omega_d \cr
-\sin\omega_d & \cos\omega_d
\end{array}\right)$ according to
\begin{equation}
{\cal R}(\omega_d)^\dagger\, K_d\, {\cal R}(\omega_d) = \left(\begin{array}{cc}
1+t_+^d & \cr &  1+t_-^d\end{array}\right),
\label{eq:diag1}
\end{equation}
with
\begin{equation}
\tan 2\omega_d = \displaystyle\frac{-(h_u -h_c)\sin
2\theta_c}{(h_u-h_c)\cos 2\theta_c + \left[\frac{h_d - h_s}{2}\right]},
\label{eq:tanomega}
\end{equation}
or, equivalently, 
\begin{equation}
\cos 2\omega_d = \frac{(h_u-h_c)\cos 2\theta_c + \left[\frac{h_d -
h_s}{2}\right]}{t_+^d-t_-^d},
\quad \sin 2\omega_d = -\frac{(h_u-h_c)\sin 2\theta_c}{t_+^d-t_-^d},
\label{eq:omegabis}
\end{equation}
in which $(t_+^d - t_-^d)$ can be immediately obtained from (\ref{eq:t+-})
\footnote{Eq.~(\ref{eq:tanomega}) also rewrites
$\frac{\sin 2(\omega_d +
\theta_c)}{\sin 2\omega_d}= -\frac{h_d - h_s}{h_u -h_c}$, which shows that
$\omega_d \to -\theta_c$ when $|m_s-m_d| \ll |m_u-m_c|$.
\label{foot:tanomega}}
.

Eq.~(\ref{eq:tanomega}) defines $\omega_d$ in particular as a function of
 $\theta_c$, $\omega_d = \omega_d(\theta_c,\ldots)$. Since both numerator
and denominator of (\ref{eq:tanomega}) are ${\cal O}(g^2)$, $\omega_d$
does not depend on the coupling constant $g$.

The diagonal matrix obtained in (\ref{eq:diag1}) is not yet the required
unit matrix, but one simply gets to it by  renormalizing the columns
of ${\cal R}(\omega_d)$ respectively by $\frac{1}{\sqrt{1+t_+^d}}$ and
$\frac{1}{\sqrt{1+t_-^d}}$.
The looked-for non-unitary matrix ${\cal V}_d$  writes finally
\begin{equation}
{\cal V}_d = \left(\begin{array}{rr}
\displaystyle\frac{c_{\omega_d}}{\sqrt{1+t_+^d}} &
\displaystyle\frac{s_{\omega_d}}{\sqrt{1+t_-^d}} \cr
-\displaystyle\frac{s_{\omega_d}}{\sqrt{1+t_+^d}} &
\displaystyle\frac{c_{\omega_d}}{\sqrt{1+t_-^d}}
\end{array}\right).
\label{eq:calVd}
\end{equation}
It differs from the rotation ${\cal R}(\omega_d)$ only at ${\cal O}(g^2)$
 and satisfies
\begin{equation}
{\cal V}_{d}\,{\cal V}_{d}^\dagger =
\frac{1}{(1+t_+^d)(1+t_-^d)}\left( {\mathbb I} + \frac{t_+^d+t_-^d}{2}
-(t_+^d-t_-^d)
\;{\cal T}_x(-2\omega_d)\right), \quad
{\cal V}_d^\dagger\,{\cal V}_d = \left(\begin{array}{cc}
\displaystyle\frac{1}{1+t_+^d} & \cr & \displaystyle\frac{1}{1+t_-^d}
\end{array}\right).
\label{eq:VV}
\end{equation}
%

For $|m_d^2-m_s^2| \ll |m_u^2-m_c^2|$,  $|h_d -h_s| \ll |h_u -h_c|$,
$(t_+-t_-) \approx (h_u-h_c)$ and the expression
 for $\sin 2\omega_d$ in (\ref{eq:omegabis}) shows
that $\omega_d(\theta_c) \approx -\theta_c$. So,
when the pair $(d,s)$ is close to degeneracy and $(u,c)$ far from it,
${\cal V}_d$  becomes  close to a rotation ${\cal R}(-\theta_c)$.
We shall come back  on this case in subsection \ref{subsec:degen}.

Eq.~(\ref{eq:VV}) shows that mass splittings $(t_+ \not= t_-)$ are responsible for the
non-unitarity of $\cal V$, and, so, for the non-unitary relation
between 1-loop and bare mass states (the same occurs in flavor space).
Note that this non-unitarity persists when $\omega_d \to 0$, which will be
the case when counterterms are introduced (see subsection
\ref{subsec:withct}).
 Unitarity can only be achieved for $t_+ = t_-$;
according to (\ref{eq:t+-}), this requires $ (h_u-h_c)^2 +
\left[\frac{h_d-h_s}{2}\right]^2
+ 2(h_u-h_c)\left[\frac{h_d-h_s}{2}\right]\cos 2\theta_c = 0$, which, since $\cos 2\theta_c \in
[-1, +1]$, can only eventually occur: -- either
 for $(h_u-h_c) = \frac{h_d-h_s}{2}$, that is,
for $(m_u-m_c) = \frac{m_d-m_s}{\sqrt{2}}$, in which case $\cos 2\theta_c =-1 \Leftrightarrow
\theta_c=\pi$; -- or for $h_u=h_c, h_d=h_s \Leftrightarrow
m_u=m_c, m_d=m_s$ (twice degenerate system).

\subsection{Second step: re-diagonalizing the mass matrix}
\label{subsec:diamass}

\subsubsection{1-loop mass eigenstates}
\label{subsub:m-states}

As mentioned in subsection \ref{subsec:principle}, the re-diagonalization
of kinetic terms leads to defining the basis $(d^1_{mL}, s^1_{mL})$, which
is related to the bare mass basis by the non-unitary relation ${\cal V}_d$.
In this basis, the mass terms
 $(\overline{d^0_{mL}}, \overline{s^0_{mL}}) M_d
\left(\begin{array}{c} d^0_{mR} \cr s^0_{mR}\end{array}\right) + h.c.$,
with $M_d = diag(m_d,m_s)$, rewrite
$(\overline{d^1_{mL}}, \overline{s^1_{mL}}) {\cal V}_d^\dagger M_d 
\left(\begin{array}{c} d^0_{mR} \cr s^0_{mR}\end{array}\right) + h.c.$.
Hence, the mass matrix that needs to be re-diagonalized is
${\cal V}_d^\dagger M_d$. It is done through two unitary transformations
${\cal R}(\xi_d)$ and $S(\xi_d)$ such that ${\cal R}(\xi_d)^\dagger
  ({\cal V}_d^\dagger M_d) S(\xi_d) =
diag (\mu_d, \mu_s)$.
Since ${\cal V}_d^\dagger M_d M_d^\dagger {\cal V}_d$ is a real symmetric matrix
\begin{eqnarray}
{\cal V}_d^\dagger\, M_d M_d^\dagger\, {\cal V}_d =
{\cal V}_{d}^\dagger
\left(\begin{array}{cc} m_d^2 & \cr & m_s^2\end{array}\right)
{\cal V}_{d}
= \left(\begin{array}{cc}
\displaystyle\frac{m_d^2\, c_{\omega_d}^2 + m_s^2 s_{\omega_d}^2}{1+t_+^d} &
-\displaystyle\frac{s_{\omega_d}
c_{\omega_d}(m_s^2-m_d^2)}{\sqrt{(1+t_+^d)(1+t_-^d)}} \cr
-\displaystyle\frac{s_{\omega_d}
c_{\omega_d}(m_s^2-m_d^2)}{\sqrt{(1+t_+^d)(1+t_-^d)}} &
\displaystyle\frac{m_d^2 s_{\omega_d}^2 + m_s^2 c_{\omega_d}^2}{1+t_-^d}
\end{array}\right),\cr &&
\label{eq:calM}
\end{eqnarray}
 ${\cal R}(\xi_d)$ can be taken as a rotation, according to
\begin{equation}
{\cal R}(\xi_d)^\dagger\,\left(
{\cal V}_d^\dagger\, M_d M_d^\dagger\, {\cal V}_d\right)
 \, {\cal R}(\xi_d) =
\left(\begin{array}{cc} \mu_d^2 & \cr & \mu_s^2 \end{array}\right).
\label{eq:diagM}
 \end{equation}
Being unitary, it  preserves the canonical form of the
kinetic terms that had been rebuilt in subsection \ref{subsec:diakin}.
It satisfies
\begin{equation}
\tan 2\xi_d = \displaystyle\frac
{-(m_d^2-m_s^2)\sqrt{(1+t_+^d)(1+t_-^d)}\sin 2\omega_d}
{(m_d^2-m_s^2)\left(1+\displaystyle\frac{t_+^d +t_-^d}{2} \right)\cos 2\omega_d
- (m_d^2+m_s^2)\displaystyle\frac{t_+^d-t_-^d}{2}}.
\label{eq:xid}
\end{equation}

Through $\omega_d(\theta_c,\ldots)$, (\ref{eq:xid}) defines $\xi_d$
in particular
as a function of  $\theta_c$, $\xi_d = \xi_d(\theta_c,\ldots)$.

Since the mass terms  rewrite
$(\overline{d^1_{mL}}, \overline{s^1_{mL}}) {\cal
R}(\xi_d)\;diag(\mu_d,\mu_s)\; S(\xi_d)^\dagger
\left(\begin{array}{c} d^0_{mR} \cr s^0_{mR}\end{array}\right) + h.c.$, 
the 1-loop left-handed mass  eigenstates $(d_{mL}, s_{mL})$ are defined by
$(\overline{d_{mL}}, \overline{s_{mL}}) =
(\overline{d^1_{mL}}, \overline{s^1_{mL}}) {\cal R}(\xi_d)$, which
leads to
\begin{equation}
\left(\begin{array}{c} d^0_{mL} \cr s^0_{mL}\end{array}\right) =
{\cal V}_d {\cal R}(\xi_d)
\left(\begin{array}{c} d_{mL} \cr s_{mL}\end{array}\right).
\label{eq:trans}
\end{equation}
By construction, at this order,
there exists no transition between $d_{mL}$ and $s_{mL}$, which are thus,
by definition, orthogonal.

\subsubsection{1-loop masses}
\label{subsub:renmass}

The re-diagonalization of kinetic terms indirectly contributes to
a renormalization of the masses: $m_d \to \mu_d, m_s \to
\mu_s$.  For   $\frac{t_+^d-t_-^d}{2}\frac{m_s^2 - m_d^2}{m_s^2 +
m_d^2}\cos 2\omega_d\ll1$ and $\frac{t_+^d-t_-^d}{2}\frac{m_s^2 + m_d^2}{m_s^2 -
m_d^2}\cos 2\omega_d\ll1$
\footnote{The first condition is immediately seen to be always satisfied.
The second too, unless $(d,s)$ are extremely close to degeneracy or
degenerate, which does not occur for any known fermions.
\label{foot:cond}},
 one gets, when $m_d \not= m_s$, from (\ref{eq:calM})
\begin{eqnarray}
\mu_s^2 &\approx& m_s^2
\left(1 - \frac{t_+^d+t_-^d}{2}\right) -m_d^2\, \frac{t_+^d-t_-^d}{2}\cos
2\omega_d,\cr
\mu_d^2 &\approx& m_d^2
\left(1 - \frac{t_+^d+t_-^d}{2}\right) + m_s^2\,
\frac{t_+^d-t_-^d}{2}\cos 2\omega_d.
\label{eq:muds}
\end{eqnarray}

This yields in particular, still when the two conditions mentioned at the
beginning of this subsection are satisfied,
\begin{equation}
\frac{\mu_s^2-\mu_d^2}{\mu_s^2+\mu_d^2}\approx
\frac{m_s^2-m_d^2}{m_s^2+m_d^2} - (t_+^d-t_-^d)\frac{m_s^4 +
m_d^4}{(m_s^2+m_d^2)^2}\cos 2\omega_d,
\label{eq:massratio}
\end{equation}
which becomes, for $m_s \approx m_d$ ($m_s \not= m_d$)

\vbox{
\begin{eqnarray}
\frac{\mu_s^2-\mu_d^2}{\mu_s^2+\mu_d^2}&\stackrel{m_s\approx m_d}{\approx}&
\frac{m_s^2-m_d^2}{m_s^2+m_d^2} - \frac{t_+^d-t_-^d}{2}\cos 2\omega_d\cr
&\stackrel{(\ref{eq:omegabis})}{\approx}&
\frac{m_s^2-m_d^2}{m_s^2+m_d^2}-\frac12 (h_u-h_c)\cos 2\theta_c
= \frac{m_s^2-m_d^2}{m_s^2+m_d^2} +
\frac{g^2}{16\pi^2}\frac{m_c^2-m_u^2}{m_W^2}\cos 2\theta_c.\cr
&&
\label{eq:massratio2}
\end{eqnarray}
}

Supposing $\cos 2\theta_c >0$ and $m_c > m_u$, 
$\frac{\mu_s^2-\mu_d^2}{\mu_s^2+\mu_d^2}$ goes to a minimum, identical to
its classical value, when $\theta_c$ becomes maximal.
A similar property is  satisfied in
the case of the MSW resonance (see for example \cite{GiuntiKim}). 

The classically degenerate case $m_d=m_s$ is most easily studied directly
from (\ref{eq:calM}). Degeneracy gets lifted at 1-loop since the
renormalized masses become, then,
$\mu_d^2=\frac{m_{d,s}^2}{1+t_+^d}, \mu_s^2=\frac{m_{d,s}^2}{1+t_-^d}$,
such that
$\frac{\mu_s^2 - \mu_d^2}{\mu_s^2 + \mu_d^2}\approx \frac{h_c-h_u}{2}
\approx \frac{g^2}{16 \pi^2}\frac{m_c^2 - m_u^2}{m_W^2}$. It turns out to
be the limit of (\ref{eq:massratio2}) for $m_d=m_s$ and vanishing $\theta_c$.

\section{Individual mixing matrices and mixing angles at 1-loop}
\label{section:indivmix}

\subsection{1-loop and classical mass eigenstates are non-unitarily
related} \label{subsec:non-unit}

According to (\ref{eq:trans}), the left-handed
1-loop mass eigenstates $(d_{mL}, s_{mL})$ are  related to
the bare ones via the product of a non-unitary transformation
${\cal V}_{d}$ by a  unitary one ${\cal R}(\xi_d)$.
The two bases are accordingly non-unitarily related \cite{mixing}.

We recall (see subsection \ref{subsec:diakin} after (\ref{eq:VV}))
that mass splittings are at the origin of the
non-unitarity of ${\cal V}_d$.
\cite{DuretMachet} \cite{DuMaVy} \cite{Benes}.

Since bare mass states are  related to bare flavor states by the classical
mixing matrix ${\cal C}_{d0} \equiv {\cal R}(\theta_d)$ of the $(d,s)$
pair, which is unitary, the physical mass eigenstates
 are also non-unitarily related to the latter. The relation is
\begin{eqnarray}
\left(\begin{array}{c} d^0_{fL} \cr s^0_{fL}\end{array}\right)
= {\cal C}_{d0} \left(\begin{array}{c}
d^0_{mL} \cr s^0_{mL} \end{array}\right)
\stackrel{(\ref{eq:trans})}{=}
 {\cal C}_{d0}\,{\cal V}_d\, {\cal R}(\xi_d) \left(\begin{array}{c}
d_{mL} \cr s_{mL} \end{array}\right),
\label{eq:mass2bare}
\end{eqnarray}

\subsection{Individual mixing matrices and mixing angles at 1-loop}
\label{subsec:indiv}

\subsubsection{The ${(d,s)}$ mixing angle}
\label{subsub:ds}

According to (\ref{eq:mass2bare}), the individual mixing  matrix at 1-loop is
 given by
\begin{equation}
{\cal C}_{d} = {\cal C}_{d0}\,{\cal V}_{d}\,{\cal R}(\xi_d)
= {\cal R}(\theta_d)\, {\cal V}_d\, {\cal R}(\xi_d).
\label{eq:Cd}
\end{equation}
Since ${\cal V}_d \approx {\cal R}(\omega_d) + {\cal O}(g^2)$ (see
(\ref{eq:calVd})), ${\cal C}_d$, though slightly non-unitary,
 stays nevertheless close to a rotation
\begin{equation}
{\cal C}_d
\approx {\cal R}(\theta_d +\omega_d +\xi_d) +{\cal O}(g^2).
\label{eq:indiv}
\end{equation}
The quantity  $(\omega_d + \xi_d)$ is seen to renormalize
the classical mixing angle $\theta_d$; it satisfies,
from (\ref{eq:xid}),  the relation (neglecting
the terms proportional to $\frac{t_+ + t_-}{2}$ which are of order
$g^{>2}$)
\begin{equation}
\tan 2(\omega_d +\xi_d) \approx\frac
{-\tan 2\omega_d\left[\frac{t_+^d-t_-^d}{2}\;\frac{m_d^2+m_s^2}{m_d^2-m_s^2}
\;\frac{1}{\cos2\omega_d}\right]}
{1 + \tan^2 2\omega_d
-\left[\frac{t_+^d-t_-^d}{2}\;\frac{m_d^2+m_s^2}{m_d^2-m_s^2}
\;\frac{1}{\cos2\omega_d}\right]}.
\label{eq:xiomega}
\end{equation}

In practice, $\tan 2(\omega_d + \xi_d)$ stays  small, and so does, accordingly,
$(\omega_d + \xi_d)$. Renormalization effects could become large
only close to the pole of (\ref{eq:xiomega}). It occurs for
\begin{equation}
\frac{1}{\cos 2\omega_d}=
\frac{t_+^d-t_-^d}{2}\frac{m_d^2 + m_s^2}{m_d^2 - m_s^2},
\label{eq:xiomegapole}
\end{equation}
that is, for $\frac{1}{\cos 2\omega_d}={\cal O}(g^2) \times
\frac{m_d^2 + m_s^2}{m_d^2 - m_s^2}$, which is usually unphysical because it
corresponds to $|\cos 2\omega_d| > 1$.
$|\cos 2\omega_d|$ could become smaller than $1$ only if, generically, 
$\left|\frac{m_d^2 - m_s^2}{m_d^2 + m_s^2}\right| <
\frac{t_+^d-t_-^d}{2}\approx
\frac{g^2}{16\pi^2}\frac{m_c^2 - m_u^2}{m_W^2}$, which is never satisfied
for known fermions, quarks or leptons
\footnote{For example, in the $(\nu_\mu, \nu_\tau, \nu,\tau)$ sector,
the condition writes $\left|\frac{m_{\nu_\tau}^2 - m_{\nu_\mu}^2}
{m_{\nu_\tau}^2 + m_{\nu_\mu}^2}\right|
< \frac{g^2}{16\pi^2}\frac{m_\tau^2 - m_\mu^2}{m_W^2}$, the r.h.s. of which
$\approx 1.9\,10^{-7}$, while the l.h.s. is experimentally known to be
 ${\cal O}(10^{-3})$ if one considers that the neutrino mass scale is ${\cal O}
(eV)$. The mismatch is similar in the $(\nu_e, \nu_\tau, e,\tau)$ sector and
worse in the $(\nu_e, \nu_\mu, e,\mu)$ sector.} .

From (\ref{eq:xiomega}), (\ref{eq:tanomega}) and (\ref{eq:omegabis})
one also  gets  $\tan 2(\omega_d + \xi_d)$ as a function
of $\theta_c$ and the classical masses
\begin{equation}
\tan 2(\omega_d + \xi_d) \approx \frac{\frac12 \frac{m_d^2 + m_s^2}{m_d^2 -
m_s^2}(h_u - h_c)\sin 2\theta_c}
{1 - \frac12 \frac{m_d^2 + m_s^2}{m_d^2 - m_s^2}\big(
(h_u - h_c)\cos 2\theta_c + \left[\frac{h_d-h_s}{2}\right]\big)}.
\label{eq:xiom2}
\end{equation}

\subsubsection{The ${(u,c)}$ mixing angle}
\label{subsub:uc}

In the same configuration $|m_d-m_s| \ll |m_u-m_c|$,
from the expression equivalent to (\ref{eq:tanomega}) in the $(u,c)$
sector, $\tan 2\omega_u = \frac{(h_d -h_s)\sin
2\theta_c}{(h_d-h_s)\cos 2\theta_c + \left[\frac{h_u - h_c}{2}\right]}$,
one deduces that, since $|h_u - h_c| \gg |h_d-h_s|$,
 $\omega_u \to 0$.
Then, from the equivalent of (\ref{eq:xiom2}), one gets $\tan 2(\omega_u +
\xi_u) \approx \frac12 (h_d-h_s) \sin 2\theta_c$, which is very small (see
(\ref{eq:hij5})).

\section{The 1-loop Cabibbo  matrix ${\mathfrak C}(p^2,\ldots)$}
\label{section:rencab}

\subsection{The effective Lagrangian at 1-loop (in the bare mass basis)}
\label{subsec:Leff}

$SU(2)_L$ gauge invariance demands the replacement, in the Lagrangian,
of the partial derivative $\partial$ by the covariant derivative $D$.
This is how, at the classical level and in the bare mass basis,
 calling $\Psi_m^{0\ T} = (u^0_{mL},
c^0_{mL},d^0_{mL},s^0_{mL})$, the kinetic + gauge terms write in their
standard form \hfill\break
\hbox{$i\,\overline\Psi^0_m\, \overleftrightarrow{D_\mu}
 \gamma^\mu  \Psi^0_m\break \equiv
\frac{i}{2} \Big(\overline\Psi^0_m \gamma^\mu (D_\mu \Psi^0_m) -
\overline{(D_\mu \Psi^0_m)}\gamma^\mu\,\Psi^0_m \Big)$},
 such that
\begin{equation}
{\cal L}_{class} =
\overline\Psi^0_m \big({\mathbb I}\; (i\partial_\mu) + g \vec T. \vec W_\mu\big)
\gamma^\mu\Psi^0_m
+ \ldots
\label{eq:Lclass}
\end{equation}
The $T$'s are the (Cabibbo rotated) $SU(2)$ generators
\begin{equation}
T^3 = \frac12
\left(\begin{array}{ccc}
1   & \vline &   \cr
\hline
& \vline & -1    \end{array}\right),
T^+ = \left(\begin{array}{ccc}
 &   \vline &  {\cal C}_{0} \cr
\hline
& \vline & \end{array}\right),
T^- = \left(\begin{array}{ccc}
 &   \vline   &  \cr
\hline
{\cal C}^\dagger_{0} & \vline &  \end{array}\right),
\label{eq:TTT}
\end{equation}
where ${\cal C}_0$ is the classical Cabibbo matrix
\begin{equation}
{\cal C}_0 = {\cal R}(\theta_c)
= \left(\begin{array}{rr}
\cos\theta_c & \sin\theta_c \cr -\sin\theta_c & \cos\theta_c
\end{array}\right)
 = {\cal C}_{u0}^\dagger\, {\cal C}_{d0} =
{\cal R}(\theta_u)^\dagger\, {\cal R}(\theta_d).
\label{eq:bareCab}
\end{equation}
Gauge currents and their $SU(2)_L$ algebra
 are thus directly related to kinetic terms by gauge
invariance and the resulting Lagrangian is both gauge invariant
and hermitian.

We shall use the same procedure to determine the 
Lagrangian after 1-loop transitions have been accounted for.
Still in the bare mass basis $\Psi^0_m$, we have seen in subsection
\ref{subsec:1lt} that  the kinetic terms, which are classically
proportional, in momentum space, to ${\mathbb I}\,
\psl$ get renormalized at 1-loop into $A(p^2, m_i, m_W)\, \psl$, with
\begin{equation}
A(p^2, \ldots) =
 \left(\begin{array}{ccc}
K_u(p^2,\ldots) & \vline &   \cr \hline  & \vline & K_d(p^2,\ldots) \end{array}\right)
={\mathbb I} + \left(\begin{array}{ccc}
H_u(p^2,\ldots) & \vline &   \cr \hline  & \vline & H_d(p^2,\ldots) \end{array}\right);
\label{eq:kcterm}
\end{equation}
$p_\mu$ stands, there, for the common momentum of the ingoing and outgoing
fermions, as depicted in Fig.~1.

The 1-loop kinetic + gauge Lagrangian  that we will
hereafter consider is accordingly
$i\,\overline \Psi^0_m\,  \overleftrightarrow {AD_\mu}\,\gamma^\mu \Psi^0_m \equiv
\frac{i}{2} \Big(\overline \Psi^0_m \gamma^\mu (AD_\mu\, \Psi^0_m) -
\overline{(AD_\mu\Psi^0_m)}\,\gamma^\mu \Psi^0_m \Big)$,
which yields
\begin{equation}
{\cal L}_{1-loop} = \overline\Psi^0_m
\left( A\, (i\partial_\mu) +\frac{g}{2} (A\, \vec T + \vec T A). \vec W_\mu
\right)\gamma^\mu\Psi^0_m + \ldots
\label{eq:Lcharged}
\end{equation}
It has the required properties of gauge invariance and, thanks to the
presence of the symmetric expression $A\vec T + \vec T A$, of hermiticity
(hermiticity is, instead,  not achieved if one considers
a kinetic Lagrangian of the form $i\,\overline \Psi^0_m\,
\overrightarrow {AD_\mu}\,\gamma^\mu \Psi^0_m$ (with ``$\rightarrow$'' instead
of ``$\leftrightarrow$'' on top of $AD_\mu$)).
Gauge invariance has in particular dictated the 1-loop expression of the
gauge currents, from which we shall now deduce that of the 1-loop
Cabibbo matrix.

\subsection{The Cabibbo matrix ${\mathfrak C}(p^2, \ldots)$
 stays unitary}
\label{subsec:cabmat}

The 1-loop  Cabibbo matrix in the bare mass basis
can be read directly from the expression $ \frac{g}{2} \overline\Psi^0_m
(A\, \vec T + \vec T A) \gamma^\mu\Psi^0_m$ of the gauge currents
that results from (\ref{eq:Lcharged}). This yields  
\begin{equation}
{\cal C}^{bm}(p^2,\ldots) = \frac12\big[
 \underbrace{({\mathbb I} + H_u)}_{K_u(p^2,\ldots)}\, {\cal C}_{0}
+{\cal C}_{0}\,\underbrace{({\mathbb I} + H_d)}_{K_d(p^2,\ldots)}\big].
\label{eq:Cab2}
\end{equation}
A naive calculation
could erroneously lead to the conclusion that ${\cal C}^{bm}$
is non-unitary.
Indeed, using ${\cal C}_0 = {\cal R}(\theta_d-\theta_u)$ and
the expressions (\ref{eq:Kd3}) (\ref{eq:Ku3}) for $K_d$ and $K_u$,
one finds ${\cal C}^{bm}({\cal C}^{bm})^\dagger \not = {\mathbb
I}$. However,
these expressions are written in a  basis which is 
non-orthogonal at 1-loop. Consider indeed, for example, the relation
${\cal C}^\ast_{11}{\cal C}_{12}+{\cal C}^\ast_{21}{\cal C}_{22}\not=0$.
It traduces the non-orthogonality of the two
vectors ${\cal C}\left(\begin{array}{c} 0 \cr 1 \end{array}\right)
\equiv\left(\begin{array}{c} {\cal C}_{12} \cr {\cal
C}_{22}\end{array}\right)$ 
and ${\cal C}\left(\begin{array}{c} 1 \cr 0 \end{array}\right)
\equiv\left(\begin{array}{c} {\cal C}_{11} \cr {\cal
C}_{21}\end{array}\right)$ when their scalar product is evaluated with the
metric $(1,1)$. However, this is the appropriate metric  only at the
classical level, where
$\left(\begin{array}{c} 0 \cr 1 \end{array}\right)$ and
$\left(\begin{array}{c} 1 \cr 0 \end{array}\right)$, which represent
fermions in bare mass space,  are orthogonal since no
transition occurs between the two of them; but it is no longer so at
1-loop (see Fig.~1)
\footnote{Likewise, for any matrix
$U$, the relation $UU^\dagger =1$ traduces unitarity only if $U$ is
expressed in an orthogonal basis of states ({\em i.e.} no transition
exists between them at the order that is considered).}. 
The pure kinetic terms in (\ref{eq:Lcharged}) are, in particular, not
normalized to $\mathbb I$ but to the non-diagonal matrix $A$. 
It is thus necessary, before drawing any conclusion, to go  to the 
orthogonal basis of
1-loop mass eigenstates by using the relation (\ref{eq:trans}).
Because of the unitarity of the ${\cal R}(\xi)$ rotations, one has
$]{\cal V}_{u,d}{\cal R}(\xi_{u,d})]^\dagger K_{u,d} [{\cal V}_{u,d}{\cal
R}(\xi_{u,d})] \equiv
{\cal R}(\xi_{u,d)}^\dagger [{\cal V}_{u,d}^\dagger K_{u,d} {\cal V}_{u,d}]
{\cal R}(\xi_{u,d})
\stackrel{(\ref{eq:kincon1})}{=} {\cal R}(\xi_{u,d})^\dagger {\cal
R}(\xi_{u,d}) = {\mathbb I}$,
 such that the pure kinetic terms get now normalized to $\mathbb I$.
And, as we show next,
the 1-loop Cabibbo matrix ${\mathfrak C}(p^2,\ldots)$ rewrites, then,
as a rotation.  It becomes indeed in this basis
\begin{eqnarray}
 {\mathfrak C}(p^2,\ldots) = [{\cal V}_u\, {\cal R}(\xi_u)]^\dagger\, {\cal
C}^{bm}(p^2,\ldots)\,
[{\cal V}_d\, {\cal R}(\xi_d)].
\label{eq:cabrenor}
\end{eqnarray}
Transforming the general expressions (\ref{eq:cabrenor}) and
(\ref{eq:Cab2}) with the help of (\ref{eq:kincon1}) which
entails $K_d = ({\cal V}_d^{-1})^\dagger {\cal V}_d^{-1}$
($K_u=({\cal V}_u^{-1})^\dagger {\cal V}_u^{-1}$),
yields
\begin{equation}
{\mathfrak C} = \frac12 {\cal R}(\xi_u)^\dagger \left[
{\cal V}_u^{-1} {\cal C}_0 {\cal V}_d + {\cal V}_u^\dagger {\cal C}_0
({\cal V}_d^{-1})^\dagger\right]{\cal R}(\xi_d)
= \frac12 {\cal R}(\xi_u)^\dagger \left[
{\cal V}_u^{-1} {\cal C}_0 {\cal V}_d + \big(({\cal V}_u^{-1} {\cal C}_0 {\cal
V}_d)^{-1}\big)^\dagger\right]{\cal R}(\xi_d).
\label{eq:calC}
\end{equation}
Using the expression (\ref{eq:calVd}) for the ${\cal V}$'s, one gets\hfill\break
${\cal V}_u^{-1} {\cal C}_0 {\cal V}_d =
\left(\begin{array}{rr}
\cos(\theta_c-\omega_u+\omega_d)\sqrt{\frac{1+t^u_{+}}{1+t^d_{+}}}   &
\sin(\theta_c-\omega_u+\omega_d) \sqrt{\frac{1+t^u_{+}}{1+t^d_{-}}} \cr
-\sin(\theta_c-\omega_u+\omega_d)\sqrt{\frac{1+t^u_{-}}{1+t^d_{+}}}  &
\cos(\theta_c-\omega_u+\omega_d)\sqrt{\frac{1+t^u_{-}}{1+t^d_{-}}}
\end{array}\right)$ and
 $\left[\left({\cal V}_u^{-1} {\cal C}_0 {\cal V}_d\right)^{-1}\right]^\dagger =
\left(\begin{array}{rr}
\cos(\theta_c-\omega_u+\omega_d)\sqrt{\frac{1+t^d_{+}}{1+t^u_{+}}}   &
\sin(\theta_c-\omega_u+\omega_d) \sqrt{\frac{1+t^d_{+}}{1+t^u_{-}}} \cr
-\sin(\theta_c-\omega_u+\omega_d)\sqrt{\frac{1+t^d_{-}}{1+t^u_{+}}}  &
\cos(\theta_c-\omega_u+\omega_d)\sqrt{\frac{1+t^d_{-}}{1+t^u_{-}}}
\end{array}\right)$ 
which leads finally to
\begin{equation}
{\mathfrak C}(p^2,\ldots) = {\cal R}\Big(\big(\theta_d +
\omega_d + \xi_d\big) -
\big(\theta_u + \omega_u + \xi_u\big)\Big)
+{\cal O}(g^{(>2)}).\  q.e.d.
\label{eq:renC}
\end{equation}
${\mathfrak C}(p^2)$ stays thus unitary for any common value of $p^2$ at
which its entries are evaluated
\footnote{This may not be in
contradiction with the non-unitarity claimed in \cite{DuretMachet} and
\cite{Benes} when the two external fermions legs are on different
mass-shell, since, then, two different $p^2$ are involved. See also appendix
\ref{subsec:nonorphys}}.
(\ref{eq:renC}) shows that the Cabibbo angle $\theta_c = \theta_d-\theta_u$
 gets renormalized by
$(\omega_d + \xi_d) -(\omega_u + \xi_u)$.

In the basis of 1-loop
mass eigenstates, the Lagrangian $\cal L$ rewrites
\begin{equation}
{\cal L} =
\left(\begin{array}{cccc} \overline{u_{mL}} & \overline{c_{mL}} & \overline
{d_{mL}}  &\overline{s_{mL}}\end{array}\right)(p^2, \ldots)
\left( \psl +g\;  \vec {\mathfrak T}(p^2,\ldots) . \vec W_\mu
\, \gamma^\mu + \ldots \right)
\left(\begin{array}{c} u_{mL} \cr c_{mL} \cr d_{mL} \cr s_{mL}
\end{array}\right)(p^2, \ldots) + \ldots,
\label{eq:Lren}
\end{equation}
with ``1-loop'' $SU(2)_L$ generators $\vec{\mathfrak T}(p^2, \ldots)$
depending now on $p^2$ and on the masses
\begin{equation}
{\mathfrak T}^3(p^2,\ldots) = \frac12
\left(\begin{array}{ccc}
1   & \vline &   \cr
\hline
& \vline & -1    \end{array}\right),
{\mathfrak T}^+(p^2,\ldots) = \left(\begin{array}{ccc}
 &   \vline &  {\mathfrak C}(p^2,\ldots) \cr
\hline
& \vline & \end{array}\right),
{\mathfrak T}^-(p^2,\ldots) = \left(\begin{array}{ccc}
 &   \vline   &  \cr
\hline
{\mathfrak C}^\dagger(p^2,\ldots) & \vline &  \end{array}\right).
\label{eq:TTTren}
\end{equation}
Our procedure has accordingly preserved the
$SU(2)_L$ structure of  gauge currents  at 1-loop, which  guarantees in
particular that the corresponding Ward identities are satisfied.

We keep mentioning the dependence on $p^2$, reminding that it only goes
away (becoming sub-leading in powers of $\frac{p^2}{m_W^2}$ when $p^2 \ll
m_W^2$. Since we are not able to get the exact dependence on this variable,
we shall keep on working in this approximation, which is only justified at
energies well below the electroweak scale. Some remarks concerning the
$p^2$ dependence are given in appendix \ref{section:cterms}.

{\em Note:} One can easily demonstrate that
${\mathfrak C}(p^2,\ldots) = {\cal C}_u^\dagger\, {\cal C}_d + {\cal
O}(g^2)$, reminiscent of the classical relation ${\cal C}_0 =
{\cal C}^\dagger_{u0}\,{\cal C}_{d0}$, as follows.
Since $H_{u}$ and $H_d$ in (\ref{eq:Cab2}) are ${\cal O}(g^2)$, the
terms proportional to them in (\ref{eq:cabrenor}) can be calculated with
the expressions of ${\cal R}(\xi_d)$ and ${\cal V}_d$ at ${\cal O}(g^0)$,
that is, for $t_+ = 0 =t_-$; one can accordingly take in there
${\cal R}(\xi_d) \stackrel{(\ref{eq:xid})}{\to} {\cal R}(-\omega_d)$
and ${\cal V}_d \stackrel{(\ref{eq:calVd})}{\to} {\cal R}(\omega_d)$, such
that ${\cal V}_d {\cal R}(\xi_d) \to {\mathbb I}$. The
same approximation can be done in the $(u,c)$ sector.
The resulting expression for $\mathfrak C$ is
\begin{equation}
{\mathfrak C}(p^2,\ldots) \stackrel{{\cal O}(g^2)}{\approx}
{\cal R}(\xi_u)^\dagger\, {\cal V}_u^\dagger\;\,
{\cal C}_0\; {\cal V}_d\, {\cal R}(\xi_d)
+ \frac12 \big( \underbrace{H_u\, {\cal C}_0 + {\cal C}_0\,
H_d}_{{\cal O}(g^2)} \big),
\label{eq:cabrenor2}
\end{equation}
which leads to the announced formula after using
(\ref{eq:bareCab}), and
(\ref{eq:Cd}) and its equivalent for ${\cal C}_u$. Since ${\mathfrak
C}(p^2)$ is unitary,  the non-unitarity
of ${\cal C}_u^\dagger {\cal C}_d$ gets compensated by that of $\frac12(H_u
{\cal C}_0 + {\cal C}_0 H_d)$.

\section{Restoring ``perturbative stability'': canceling non-diagonal
 transitions at 1-loop with counterterms}
\label{section:stability}

\subsection{Instability close to degeneracy}
\label{subsec:degen}

Quasi-degenerate systems are known to be unstable with respect to small
perturbations. This property is easily verified here, through the amount by
which classical mixing angles are renormalized when 1-loop transitions are
accounted for. It undergoes indeed large variations when the classical
masses span a very small interval in the neighborhood of degeneracy:
we first consider the case of exact classical degeneracy ($m_d=m_s$),
secondly the pole of (\ref{eq:xiomega}), which corresponds to a situation
where $d$ and $s$ are extremely close to degeneracy (see subsection
\ref{subsec:indiv}), and, last, the pole of
$\tan 2\xi$, which also corresponds to quasi-degenerate fermions, but not
as close as previously.

$\bullet$\quad For exact classical degeneracy  
$h_d=h_s$ such that, by the expression of $\sin 2\omega_d$
in (\ref{eq:omegabis}), $\omega_d = -\theta_c$.
(\ref{eq:calM}) shows then that ${\cal V}_d^\dagger M_d M_d^\dagger
{\cal V}_d$ stays diagonal, and, so,  $\xi_d = 0$
\footnote{This is in agreement with (\ref{eq:xid}) 
which shows that $\tan 2\xi_d$ has no pole when $m_d=m_s$.}
.
The classical $(d,s)$ mixing angle $\theta_d$ is  renormalized
(see (\ref{eq:indiv})) by $(\omega_d + \xi_d) = -\theta_c$  and becomes
$\theta_d - \theta_c = \theta_u$, the classical mixing angle of the $(u,c)$
pair.

According to (\ref{eq:renC}), the Cabibbo mixing angle gets renormalized
from its classical value
$\theta_c$ to $\theta_c +(\omega_d +\xi_d) - (\omega_u + \xi_u)
= -(\omega_u + \xi_u)$. This is vanishing by the equivalent of
(\ref{eq:tanomega}) which yields $\omega_u=0$ for $h_d=h_s$, and then
by that of (\ref{eq:xid}) which entails $\xi_u=0$ for $\omega_u=0$.
 To such a system is accordingly associated
a vanishing 1-loop Cabibbo angle.  Renormalization effects can thus be large.

$\bullet$\quad At the pole of (\ref{eq:xiomega}), by definition,
the renormalization of $\theta_d$ becomes maximal $(\pm \frac{\pi}{4})$. 

$\bullet$\quad At the pole of $\tan 2\xi_d$,
it becomes instead minimally small (see subsection \ref{subsub:ds}).

So, in a close neighborhood of degeneracy, the renormalization $(\omega_d +
\xi_d)$ of $\theta_d$ undergoes large variations. So does the one of the
Cabibbo angle.

\subsection{The counterterms of Shabalin}
\label{subsec:shabal}

Let us now add to the classical Lagrangian in bare mass
space the counterterms which were first
proposed by Shabalin in his study \cite{Shabalin} of the electric dipole
moment of quarks. They are devised to cancel the
($p^2$-dependent) $s^0_m \leftrightarrow d^0_m$ transitions when either
$p^2= m_d^2$ or $p^2 = m_s^2$ ($d$ or $s$  on mass-shell). So, an on
mass-shell $s^0_m$ cannot anymore transmute into a $d^0_m$ with the same
virtuality, and {\em vice versa}.
They were also introduced in \cite{DuMaVy} and
\cite{shortDuMaVy}.
In the short letter \cite{shortDuMaVy}, the inclusion of these counterterms
was proposed as a solution to rescue the standard CKM phenomenology.
In \cite{DuMaVy}, only the classical Lagrangian + the
counterterms were re-diagonalized, but the effective 1-loop transitions were
not included. This completion is the goal of the lines below.
We shall go through the same steps as previously, 
re-diagonalizing simultaneously the effective kinetic and mass terms up to
${\cal O}(g^2)$, including Shabalin's counterterms.

Following \cite{DuMaVy}, let us accordingly add to the bare Lagrangian
the kinetic and mass-like counterterms which concern both chiralities
of fermions
\begin{equation}
-A_d\, \overline{d^{0}_{m}}\, \psl \,(1-\gamma^5) \,s^{0}_{m}
- B_d\, \overline{d^{0}_{m}}\, (1-\gamma^5)\, s^{0}_m
- E_d\, \overline{d^{0}_{m}}\, \psl \,(1+\gamma^5)\, s^{0}_{m}
- D_d\, \overline{d^{0}_{m}}\, (1+\gamma^5)\, s^{0}_m.
\label{eq:cterms}
\end{equation}
Requesting that
$s^0_m \to d^0_m$ transitions vanish when either $s^0_m$ or $d^0_m$ is on
mass-shell yields (see Appendix A of \cite{DuMaVy})
\begin{eqnarray}
A_d &=& s_cc_c \frac{m_d^2\, (h_u-h_c)_{p^2 = m_d^2}- m_s^2\, (h_u-h_c)_{p^2
= m_s^2}}{m_d^2 - m_s^2}
\approx s_cc_c\left( (h_u-h_c)_{p^2=m_d^2} + m_s^2
\frac{\partial(h_u-h_c)}{\partial p^2}\Big|_{p^2=m_d^2}\right),\cr
E_d &=&  s_cc_c\frac{m_s m_d \left((h_u-h_c)_{p^2 = m_d^2} - (h_u-h_c)_{p^2
= m_s^2}\right)}{m_d^2 - m_s^2}
\approx s_cc_c m_sm_d \frac{\partial(h_u-h_c)}{\partial p^2}\Big|_{p^2=m_d^2},\cr
&& \cr
&& B_d = -m_s\, E_d,\quad D_d= -m_d\, E_d,
\label{eq:BD}
\end{eqnarray}
The re-diagonalization of the left-handed kinetic terms at 1-loop
is operated via a non-unitary transformation ${\cal V}_d$ of the same form as
(\ref{eq:calVd}). Counterterms only induce the replacement of
$s_cc_c\,(h_u-h_c)(p^2,\ldots)$ with $s_cc_c\,(h_u-h_c)(p^2,\ldots) - A_d$,
such that the angle $\omega_d$ changes from (\ref{eq:tanomega}) to
\begin{equation}
\tan 2\omega_{dL}(p^2,\ldots) = \frac{-2 \big(s_cc_c\,(h_u-h_c)(p^2,\ldots) - A_d\big)}{(h_u-h_c)(p^2,\ldots)\cos
2\theta_c + \left[\frac{(h_d-h_s)(p^2,\ldots)}{2}\right]},
\label{eq:tanomega2}
\end{equation}
in which we have added a subscript ``$_L$'' to $\omega_d$ to distinguish it from
its counterpart $\omega_{dR}$  associated with right-handed fermions.

The quantity $\big(s_cc_c\,(h_u-h_c)(p^2, \ldots)-A_d\big)$, which will be often
encountered, writes
\begin{eqnarray}
s_cc_c\,(h_u-h_c)(p^2, \ldots) - A_d &\approx& s_cc_c\,
\left( (h_u-h_c)(p^2, \ldots) - (h_u-h_c)_{p^2=m_d^2} -m_s^2
\frac{\partial(h_u-h_c)}{\partial p^2}\Big|_{p^2=m_d^2}\right)\cr
&\approx& s_cc_c\left(p^2 - (m_d^2+m_s^2) \right)
\frac{\partial(h_u-h_c)}{\partial p^2}\Big|_{p^2=m_d^2},
\label{eq:f-A}
\end{eqnarray}
in which we have taken $p^2 \sim m_d^2 \sim m_s^2$.

By differentiating (\ref{eq:hij3}) with respect to $p^2$, one gets, still in
the limit $p^2, m_i^2, m_j^2 \ll m_W^2$ and in the $\overline{MS}$ scheme
\begin{equation}
\frac{\partial(h_i-h_j)}{\partial p^2} 
 \overset{p^2, m_i^2, m_j^2\ll m_W^2}{\underset{\overline{MS}}{\approx}}
3\,\frac{g^2}{4} \frac{i}{16\pi^2}  \frac{m_i^2 -m_j^2}{m_W^4}.
\label{eq:hder}
\end{equation}
One has now (we added a superscript ``$d$'' to $t_+$
and $t_-$ because $A_d \not = A_u$, such that $t^d_+ \not= t^u_+, t^u_-
\not= t^d_-$, and also a subscript ``$_L$'' to recall that they concern
left-handed fields)

\vbox{
\begin{eqnarray}
 t^d_{\pm L}(p^2,\ldots) &=& \frac{h_u + h_c + \left[\frac{h_d +
h_s}{2}\right]}{2}(p^2,\ldots)\cr
&&\hskip -2cm \pm \frac12 \sqrt{\left((h_u - h_c)(p^2,\ldots)\cos 2\theta_c +
\left[\frac{(h_d-h_s)(p^2,\ldots)}{2}\right]\right)^2 + 
4\big(s_cc_c\,(h_u-h_c)(p^2,\ldots) -A_d\big)^2},\cr
&&
\label{eq:t+-bis}
\end{eqnarray}
}

which gives back (\ref{eq:t+-}) when $A_d$ is set to zero.

As far as the right-handed kinetic terms are concerned, they are controlled
by the matrix $\left(\begin{array}{cc} 1 & -E_d \cr -E_d & 1
\end{array}\right)$ and are accordingly re-diagonalized into the unit
matrix by a non-unitary transformation ${\cal U}_d$
\begin{equation}
{\cal U}_d^\dagger \left(\begin{array}{cc} 1 & -E_d \cr -E_d & 1
\end{array}\right) {\cal U}_d = {\mathbb I}, \quad
 {\cal U}_d = \frac{1}{\sqrt{2}}\left(
\begin{array}{cc} \frac{1}{\sqrt{1+E_d}}  & \frac{1}{\sqrt{1-E_d}} \cr
\frac{-1}{\sqrt{1+E_d}}  &    \frac{1}{\sqrt{1-E_d}}  \end{array}\right)
\Rightarrow {\cal U}_d{\cal U}_d^\dagger = \frac{1}{1-E_d^2}
\left(\begin{array}{cc} 1 & E_d \cr E_d & 1 \end{array}\right).
\end{equation}
It corresponds to $\omega_{dR} = \frac{\pi}{4}, t^d_{+R} = E_d, t^d_{-R} =
-E_d$.

The mass matrix to diagonalize is now ${\cal V}_d^\dagger M_d \,
{\cal U}_d$, where, including the counterterms, $M_d$ is now given by
\begin{equation}
M_d = \left(\begin{array}{cc} m_d & D_d \equiv-m_dE_d \cr
B_d \equiv-m_s E_d & m_s\end{array}\right).
\end{equation}
The rotation ${\cal R}(\xi_{dL})$ will accordingly diagonalize the matrix
$({\cal V}_d^\dagger M_d {\cal U}_d)( {\cal U}_d^\dagger M_d^\dagger {\cal
V}_d)$.

Neglecting irrelevant terms proportional to $E^{\geq 2}$ and to $g^{>2}$, one gets
\begin{eqnarray}
{\cal V}_d^\dagger M_d {\cal U}_d {\cal U}_d^\dagger M_d^\dagger {\cal V}_d
&=& \left(\begin{array}{cc}
\frac{m_d^2 c_{\omega_{dL}}^2 + m_s^2 s_{\omega_{dL}}^2 + 4 m_dm_s E_d
s_{\omega_{dL}} c_{\omega_{dL}}}{1+t^d_{+L}} & 
\frac{(m_d^2-m_s^2) s_{\omega_{dL}} c_{\omega_{dL}} -2m_dm_s
E_d(c_{\omega_{dL}}^2-s_{\omega_{dL}}^2)}{\sqrt{(1+t^d_{+L})(1-t^d_{-L})}} \cr
  \frac{(m_d^2-m_s^2) s_{\omega_{dL}} c_{\omega_{dL}} -2m_dm_s
E_d(c_{\omega_{dL}}^2-s_{\omega_{dL}}^2)}{\sqrt{(1+t^d_{+L})(1-t^d_{-L})}} & 
\frac{m_d^2 s_{\omega_{dL}}^2 + m_s^2 c_{\omega_{dL}}^2 -4m_dm_sE_d
s_{\omega_{dL}} c_{\omega_{dL}} }{1+t^d_{-L}}
\end{array}\right) \cr
&+& m_dm_s E_d \left( \begin{array}{cc}
-\sin 2\omega_{dL}   & \cos 2\omega_{dL} \cr  \cos 2\omega_{dL}  & \sin
2\omega_{dL} 
\end{array}\right).
\label{eq:VMUUMV}
\end{eqnarray}

The expression (\ref{eq:xid}) for $\tan 2\xi_d$ gets replaced by
\begin{equation}
\tan 2\xi_{dL}(p^2,\ldots)
= \frac{-(m_d^2-m_s^2)\sin 2\omega_{dL} +2m_dm_sE_d \cos
2\omega_{dL}}
{(m_d^2-m_s^2) \cos 2\omega_{dL} + 2 m_dm_s E_d\sin2\omega_{dL} 
-(m_d^2+m_s^2)\frac{t^d_{+L}-t^d_{-L}}{2}},
\label{eq:xid2}
\end{equation}
in which we have neglected factors $(1 +\alpha t^d_{+L} +\beta t^d_{-L}),
\alpha,\beta ={\cal O}(1)$, which
yield contributions of unwanted higher order in $g$.

Unless $\cos 2\theta_c \approx -\frac12\frac{h_d -h_s}{h_u-h_c}
\stackrel{(\ref{eq:hij5})}{\approx}
-\frac{m_d^2-m_s^2}{m_u^2-m_c^2}$,
(\ref{eq:tanomega2}), (\ref{eq:f-A}) and (\ref{eq:hder}),
show that, when $p^2 \ll m_W^2$ and since $m_u^2, m_c^2 \ll
m_W^2$, $\omega_{dL} \sim m_s^2/m_W^2$ is very small.
Then,  using
$\sin 2\omega_{dL} \approx \tan 2\omega_{dL}$, the expression for $E_d$
in (\ref{eq:BD}) and the one for $t^d_{+L}-t^d_{-L}$ coming from
(\ref{eq:t+-bis}) (in which we neglect the term $4(s_cc_c(h_u-h_d)-A_d)$),
 (\ref{eq:xid2}) rewrites (the term $2m_dm_s E_d \sin
2\omega_{dL}$ in its denominator  can always be
neglected)
\begin{equation}
\tan 2\xi_{dL} \approx 2s_cc_c
\frac{\partial(h_u-h_c)}{\partial p^2}\left(
\frac{(m_d^2-m_s^2)\left(p^2-(m_d^2+m_s^2)\right) +m_d^2 m_s^2}
{(m_d^2-m_s^2) - \frac{m_d^2+m_s^2}{2}\big((h_u-h_c)\cos
2\theta_c + \left[\frac{h_d-h_s}{2}\right] \big)}\right),
\label{eq:xid3}
\end{equation}
showing, with (\ref{eq:hder}), that $\xi_{dL} \sim (p^2,m^2)/m_W^2$
is also very small.

When $\cos 2\theta_c \approx -\frac12\frac{h_d -h_s}{h_u-h_c}
\stackrel{(\ref{eq:hij5})}{\approx}
-\frac{m_d^2-m_s^2}{m_u^2-m_c^2}$, $\tan 2\omega_{dL} \to \infty$, which
corresponds to $\omega_{dL}$ maximal.
Then, (\ref{eq:xid2}) and (\ref{eq:t+-bis}) yield
$\tan 2\xi_{dL} \to -\frac{m_d^2 - m_s^2}{2m_dm_s E_d -
(m_d^2+m_s^2)\big(s_cc_c(h_u-h_d)-A_d\big)}$, which, using (\ref{eq:BD}) and
(\ref{eq:f-A}), is finally equivalent to
$\tan 2\xi_{dL} = -\frac{m_d^2-m_s^2}{s_cc_c \frac{\partial
(h_u-h_c)}{\partial p^2}}\frac{1}{2m_d^2m_s^2 -
(m_d^2+m_s^2)\big(p^2-(m_d^2+m_s^2)\big)}$. Unless $d$ and $s$ are exactly
degenerate (in which case $\xi_{dL}$ shrinks to $0$),
 this yields a quasi-maximal $\xi_{dL}$, because of the very
small value of $\frac{\partial (h_u-h_c)}{\partial p^2}$, given in
(\ref{eq:hder}).

 This is however not true when the
numerator of (\ref{eq:tanomega2}) vanishes, which occurs for $s_cc_c (h_u-h_c)
-A_d=0$, or, likewise, by (\ref{eq:t+-bis}), for $t^+_{dL}= t^-_{dL}$. In
this case, $\omega_{dL}$ is undetermined and can be taken to vanish, since
the matrix of kinetic terms is proportional to the unit matrix. One then
finds a very small $\tan 2\xi_{dL} = \frac{2m_dm_s E_d}{m_d^2 - m_s^2}$
(see (\ref{eq:BD}) and (\ref{eq:hder})).

The expressions  obtained in the $(u,c)$ channel are very similar. One
gets:
\begin{eqnarray}
A_u &=& -s_cc_c \frac{m_u^2\, (h_d-h_s)_{p^2 = m_u^2}- m_c^2\, (h_d-h_s)_{p^2
= m_c^2}}{m_u^2 - m_c^2}
\approx -s_cc_c\left( (h_d-h_s)_{p^2=m_u^2} + m_c^2
\frac{\partial(h_d-h_s)}{\partial p^2}\Big|_{p^2=m_u^2}\right);\cr
E_u &=&  -s_cc_c\frac{m_u m_c \left((h_d-h_s)_{p^2 = m_u^2} - (h_d-h_s)_{p^2
= m_c^2}\right)}{m_u^2 - m_c^2}
\approx s_cc_c m_sm_d \frac{\partial(h_d-h_s)}{\partial p^2}\Big|_{p^2=m_u^2};\cr
&& \cr
&& B_u = -m_c\, E_u,\quad D_u= -m_u\, E_u;
\label{eq:BDu}
\end{eqnarray}
\begin{equation}
\tan 2\omega_{uL}(p^2,\ldots) =
\frac{-2 \big(-s_cc_c\,(h_d-h_s)(p^2,\ldots) - A_u\big)}
{(h_d-h_s)(p^2,\ldots)\cos 2\theta_c +
\left[\frac{(h_u-h_c)(p^2,\ldots)}{2}\right]};
\label{eq:tanomegau}
\end{equation}
\begin{eqnarray}
-s_cc_c\,(h_d-h_s)(p^2, \ldots) - A_u &\approx& -s_cc_c\,
\left( (h_d-h_s)(p^2, \ldots) - (h_d-h_s)_{p^2=m_u^2} -m_c^2
\frac{\partial(h_d-h_s)}{\partial p^2}\Big|_{p^2=m_u^2}\right)\cr
&\approx& -s_cc_c\left(p^2 - (m_u^2+m_c^2) \right)
\frac{\partial(h_d-h_s)}{\partial p^2}\Big|_{p^2=m_u^2};
\label{eq:f-Au}
\end{eqnarray}
\begin{eqnarray}
 t^u_{\pm L}(p^2,\ldots) &=& \frac{\left[\frac{h_u + h_c}{2}\right] + h_d + h_s}{2}(p^2,\ldots)\cr
&&\hskip -1.5cm \pm \frac12 \sqrt{\left((h_d - h_s)(p^2,\ldots)\cos 2\theta_c +
\left[\frac{(h_u-h_c)(p^2,\ldots)}{2}\right]\right)^2 + 
4\big(-s_cc_c\,(h_d-h_s)(p^2,\ldots) -A_u\big)^2};\cr
&&
\label{eq:t+-bisu}
\end{eqnarray}
\begin{eqnarray}
\tan 2\xi_{uL}(p^2,\ldots)
&=& \frac{-(m_u^2-m_c^2)\sin 2\omega_{uL} +2m_um_cE_u \cos
2\omega_{uL}}
{(m_u^2-m_c^2) \cos 2\omega_{uL} + 2 m_um_c E_u\sin2\omega_{uL} 
-(m_u^2+m_c^2)\frac{t^u_{+L}-t^u_{-L}}{2}}\cr
&& \cr
&\approx&
-2s_cc_c
\frac{\partial(h_d-h_s)}{\partial p^2}\left(
\frac{(m_u^2-m_c^2)\left(p^2-(m_u^2+m_c^2)\right) +m_u^2 m_c^2}
{(m_u^2-m_c^2) - \frac{m_u^2+m_c^2}{2}\left((h_d-h_s)\cos
2\theta_c + \left[\frac{h_u-h_c}{2}\right] \right)}\right).\cr
&&
\label{eq:xiu2}
\end{eqnarray}
Unlike in the $(d,s)$ sector, because $|m_d-m_s| < |m_u-m_c|$, $\tan
2\omega_{ul}$ given by (\ref{eq:tanomegau}) cannot have any pole. This
makes $\omega_{uL}$ always very small and, likewise, $\xi_{uL}$.
Furthermore, the equality $t^+_{uL}= t^-_{uL}$ can never be achieved (see
also section \ref{section:FCNC}). These results stay true when
$m_d=m_s$, in which case $h_d=h_s$, which entails that
 $A_u, E_u, B_u, D_u$,  $\omega_{uL}$ and $\xi_{uL}$ vanish.

\subsection{Stability is restored}
\label{subsec:stabil}

We now check that  Shabalin's counterterms stabilize 1-loop mixing angles
in the vicinity of $d-s$ degeneracy.

Still except when $\cos 2\theta_c = -\frac12 \frac{h_d-h_s}{h_u - h_c}$, which
corresponds, when $m_d=m_s$, to $\theta_c$ maximal (see also subsection
\ref{subsec:withct}), $\omega_{dL}$ stays small when $m_d \approx m_s$.
From (\ref{eq:tanomega2}), (\ref{eq:f-A}), (\ref{eq:hder}), one gets 
\begin{equation}
\tan 2\omega_{dL} 
\overset{p^2, m_d^2\sim m_s^2, m_u^2, m_c^2\ll m_W^2}{\underset{\overline{MS}}{\approx}}
 -3\, \frac{p^2 - 2m_d^2}{m_W^2}\tan 2\theta_c,
\end{equation}
and so does $\xi_{dL}$, which, from (\ref{eq:xid3}), becomes 
\begin{equation}
\tan 2\xi_{dL}
\overset{p^2, m_d^2=m_s^2, m_u^2, m_c^2\ll m_W^2}{\underset{\overline{MS}}{\approx}}
-3\,\frac{m_d^2}{m_W^2}\frac{1}{\left(-\frac{17}{4}+ \frac32
\ln\frac{m_W^2}{\mu^2}\right)}\tan 2\theta_c,
\end{equation}
since, for $\mu^2 \in [m_K^2, m_D^2]$, $\left(-\frac{17}{4}+ \frac32
\ln\frac{m_W^2}{\mu^2}\right) \in [7,12]$.

So, when $m_d \simeq m_s$, the mixing angle $\theta_{dL}$ is accordingly
renormalized at 1-loop by the small quantity
$\omega_{dL} + \xi_{dL} \approx \frac12 (\tan 2\omega_{dL} +
\tan 2\xi_{dL}) \sim \frac{m_d^2}{m_W^2}\tan 2\theta_c$.

In the $(u,c)$ sector, $E_u=0=A+u$ when $m_d=m_s$ and one gets
\begin{eqnarray}
\tan 2\xi_u \approx -\tan 2\omega_{uL} = -\frac{4A_u}{h_u-h_c} =0,
\end{eqnarray}
such that $\theta_{uL}$ is not renormalized at all.

\subsection{The Cabibbo matrix
${\mathfrak C}(p^2, \ldots)$ still stays unitary}

 The expression for ${\mathfrak C}(p^2, \ldots)$ is still given by (\ref{eq:calC}),
but one must
now accounts for $t^u_{\pm L} \not = t^d_{\pm L}$ since $A_u \not = A_d$.
One gets now ${\cal V}_u^{-1} {\cal C}_0 {\cal V}_d =
\left(\begin{array}{rr}
\cos(\theta_c-\omega_u+\omega_d)\sqrt{\frac{1+t^u_{+L}}{1+t^d_{+L}}}   &
\sin(\theta_c-\omega_u+\omega_d) \sqrt{\frac{1+t^u_{+L}}{1+t^d_{-L}}} \cr
-\sin(\theta_c-\omega_u+\omega_d)\sqrt{\frac{1+t^u_{-L}}{1+t^d_{+L}}}  &
\cos(\theta_c-\omega_u+\omega_d)\sqrt{\frac{1+t^u_{-L}}{1+t^d_{-L}}}
\end{array}\right)$ and
 $\left[\left({\cal V}_u^{-1} {\cal C}_0 {\cal V}_d\right)^{-1}\right]^\dagger =
\left(\begin{array}{rr}
\cos(\theta_c-\omega_u+\omega_d)\sqrt{\frac{1+t^d_{+L}}{1+t^u_{+L}}}   &
\sin(\theta_c-\omega_u+\omega_d) \sqrt{\frac{1+t^d_{+L}}{1+t^u_{-L}}} \cr
-\sin(\theta_c-\omega_u+\omega_d)\sqrt{\frac{1+t^d_{-L}}{1+t^u_{+L}}}  &
\cos(\theta_c-\omega_u+\omega_d)\sqrt{\frac{1+t^d_{-L}}{1+t^u_{-L}}}
\end{array}\right)$, which leads to the same formula (\ref{eq:renC})
as before for  ${\mathfrak C}(p^2, \ldots)$, which is unitary. Accordingly,
like in the
absence of Shabalin's counterterms, the classical Cabibbo angle $\theta_c$
gets renormalized at 1-loop by \hfill\break
 $(\omega_{dL} + \xi_{dL})(p^2,m_d^2, m_s^2, m_u^2, m_c^2,
m_W^2) -(\omega_{uL} + \xi_{uL})(p^2, m_d^2, m_s^2, m_u^2, m_c^2, m_W^2)$.

For more remarks concerning the $p^2$ dependence, see appendix
\ref{section:cterms}.

\section{Suppressing extra flavor changing neutral currents} 
\label{section:FCNC}

The absence of flavor changing neutral currents 
is classically implemented {\em ab initio} in bare flavor space
by the canonical choice of the kinetic terms, proportional to the unit
matrix, and by that  of the $SU(2)_L$ generators which, in the $(u,c,d,s)$ basis,
 write  $T^3 = \frac12 \left(\begin{array}{ccc}
1   & \vline &   \cr \hline & \vline & -1    \end{array}\right),
T^+ = \left(\begin{array}{ccc} & \vline & 1 \cr \hline
& \vline & \end{array}\right), T^- = \left(\begin{array}{ccc}
 & \vline &  \cr \hline 1 & \vline & \end{array}\right)$. The diagonality
of the $T^3$ generator ensures  that the $W^3$ gauge boson only couples, in
both $(u,c)$ and $(d,s)$ sectors, to diagonal fermionic currents: no FCNC
occurs classically.  That this property  is
preserved in bare mass space is the essence of the GIM mechanism: the
closure of the $SU(2)_L$ algebra (\ref{eq:TTT}) on the same  $T^3$ as above
is ensured by the unitarity of  the classical Cabibbo matrix ${\cal C}_0$.
The situation is different at 1-loop since vertex corrections with an
internal charged gauge boson induce non-diagonal couplings of the $W^3$
gauge field (see Fig.~1 left) and also, for example,
the non-diagonal $s \to d$ transition of Fig.~2 inserted on one of the
two external fermion legs of a  $W^3 s\bar s$ vertex triggers:
-- 1-loop FCNC's if one considers $s^0_f \to d^0_f$ transitions,
-- their equivalent for mass states if one considers,
like we did, $s^0_m \to d^0_m$ transitions (see Fig.~1 right).

We have seen with (\ref{eq:TTTren}), and this stays valid in the
presence of Shabalin's counterterms, that, in the 1-loop mass basis, the
$SU(2)_L$ algebra closes on the ``canonical'' ${\mathfrak T}^3 \equiv T^3 =
\frac12 \left(\begin{array}{ccc} 1 & \vline & \cr \hline & \vline & -1
\end{array}\right)$. So, after 1-loop transitions of the type of Fig.~2,
have been accounted for, one is back to a situation similar to the
classical one. 1-loop non-diagonal neutral gauge currents are triggered
by vertex corrections. As for the second origin of FCNC, insertion of
Fig.~2 on one of the external leg of a $W^2 f \bar f$ vertex (Fig.~1 right),
 it is
important to recall, as was demonstrated in \cite{DuMaVy} (Appendix B),
that the introduction of Shabalin's counterterms do not modify
transitions of the type $s \to d W^3$:
 the counterterms do cancel
the non-diagonal transitions on external legs, but $s \to d W^3$ transitions are
re-created with the same amplitude through the covariant derivative that
has to be used inside them.

Is the situation strictly identical to the standard one? The
answer is ``not exactly'', and this is what we investigate now.
The issue is that of the existence of mass splittings, which are
responsible for two facts:\newline
* the slight non-unitarity of the connection between the orthogonal set of
1-loop mass eigenstates and bare mass (or flavor) states;\newline
* that the two fermions concerned by 1-loop non-diagonal transitions
(Fig.~2) cannot be both on mass-shell, such that Shabalin's counterterms
can only restore 1-loop orthogonality between  one on mass-shell fermion
and a second one which is off mass-shell.

Since, by construction, 1-loop mass eigenstates as we defined them, by the
diagonalization of the 1-loop quadratic effective Lagrangian (kinetic +
mass terms), are orthogonal, the non-unitarity of their connection to bare
mass states (and, thus, to bare flavor states, since the last two are
unitarily connected) makes FCNC still occur in bare flavor (or mass) space.
This trivially appears by transforming back the $W^3 f \bar f$ coupling in
the space of 1-loop mass states, that we emphasized to be ``canonical''
(proportional to $T^3$), to bare flavor space. So, we face a situation
where, because of (unavoidable) mass splittings, the standard situation in
bare flavor space is spoiled.

We adopt a conservative point of view,  require that the
phenomenology should not differ from the standard one, and 
therefore that these extra FCNC vanish or, at least, are strongly damped.

\subsection{When no counterterm is added}
\label{subsec:noct}

As soon as 1-loop transitions Fig.~2 are accounted for, the  bare flavor 
(or mass) states do not form anymore an orthogonal set, such that requesting
the absence of FCNC in this basis appears somewhat academic. In spite of
this, and since the principle of the method and formulae will keep valid when
counterterms are introduced, we proceed with this first case.

To that purpose, it is enough to use the relation
(\ref{eq:mass2bare}) between 1-loop mass eigenstates and bare flavor states
(and its equivalent in the $(u,c)$ sector), which leads to the expression
(\ref{eq:Cd}) for the 1-loop mixing matrix ${\cal C}_d$.
Neutral gauge currents in the space of 1-loop mass eigenstates being
proportional to $T^3$, their expression in bare
flavor space gets simply proportional to
$({\cal C}_d^{-1})^\dagger {\cal C}_d^{-1}
= ({\cal C}_d {\cal C}_d^\dagger)^{-1} \stackrel{(\ref{eq:Cd})}{=}
 ({\cal C}_{d0}{\cal V}_d {\cal
V}_d^\dagger {\cal C}_{d0}^\dagger)^{-1}$, and a similar expression in the
$(u,c)$ sector.
From the expression (\ref{eq:calVd}) of ${\cal V}_d$, it is easy matter to get
($T_x$ is defined in (\ref{eq:Tx}))
\begin{eqnarray}
{\cal C}_{d0}{\cal V}_d {\cal V}_d^\dagger {\cal C}_{d0}^\dagger
&=& \frac{1}{(1+t^d_{+L})(1+t^d_{-L})}\Big(
1+ \frac{t^d_{+L} + t^d_{-L}}{2} - (t^d_{+L} - t^d_{-L}){\cal
T}_x\big(-2(\theta_{dL}+\omega_{dL})\big)\Big)\cr
\Rightarrow ({\cal C}_{d0}{\cal V}_d {\cal V}_d^\dagger {\cal
C}_{d0}^\dagger)^{-1} &\approx &
(1+t^d_{+L})(1+t^d_{-L})\Big(
1- \frac{t^d_{+L} + t^d_{-L}}{2} + (t^d_{+L} - t^d_{-L}){\cal
T}_x\big(-2(\theta_{dL}+\omega_{dL})\big)\Big),\cr
&&
\label{eq:nofcnc}
\end{eqnarray}
which makes  FCNC's proportional to $-(t^d_{+L} - t^d_{-L}) \sin
2(\theta_{dL} + \omega_{dL})$ (the sine function corresponds to the
non-diagonal terms of ${\cal T}_x$, as it appears in (\ref{eq:Tx})),
and an equivalent expression in the $(u,c)$ sector.
According to
(\ref{eq:nofcnc}), in both the $(d,s)$ and $(u,c)$ sectors,
their suppression  requires that $(t^{u,d}_{+L} - t^{u,d}_{-L}) \sin
2(\theta_{uL,dL}+\omega_{uL,dL})$ vanishes or, at least, that it be
as small as possible.

$\bullet$\ According to (\ref{eq:t+-}), the equality of $t^d_{+L}$ and
$t^d_{-L}$ requires \hfill\break 
 $\cos 2\theta_c =
-\frac12\left(\frac{h_u-h_c}{h_d-h_s}+ \frac{h_d-h_s}{h_u-h_c} \right)
\approx -\frac12 \left(\frac{m_c^2-m_u^2}{m_s^2-m_d^2} +
\frac{m_s^2-m_d^2}{m_c^2-m_u^2}  \right)$. This corresponds to $|\cos
2\theta_c|>1$, which can never be satisfied.

$\bullet$\ FCNC's can accordingly only be suppressed if $(\omega_{dL} +
\theta_{dL})\approx 0$ and an equivalent condition in the $(u,c)$ sector. As
already mentioned in subsection \ref{subsec:diakin}, when $(d,s)$ are much closer
to degeneracy that $(u,c)$, $\omega_{dL} \approx -\theta_c$ such that the
condition for FCNC suppression rewrites $\theta_{dL} \approx \theta_c$.
One also finds that   $\theta_{uL} \approx -\omega_{uL}$ becomes small
(see subsection \ref{subsec:stabil}}).
So, FCNC's get suppressed when bare flavor and mass states for
the fermion pair which is the farthest from degeneracy get close to
alignment. No condition  on $\theta_c$ arises in this case.

\subsection{In the presence of Shabalin's counterterms}
\label{subsec:withct}

If bare flavor states were a set of truly orthogonal states at 1-loop,
they could
only be unitarily connected with 1-loop mass eigenstates since the latter
are constructed as being orthogonal. Then, the absence of FCNC would
naturally translate from one basis to the other.
That, instead, non-unitarity persists even in the presence of counterterms
can be traced out in the expression (\ref{eq:calVd}) for ${\cal V}_d$, to
the relations (\ref{eq:VV}), and
is due to $t^+_{dL} \not = t^-_{dL}$.

Relations (\ref{eq:nofcnc}) keep valid such that the discussion stays
formally the same as in subsection \ref{subsec:noct}). Results are  
different because the expression of
$\omega_{dL}$ has changed into (\ref{eq:tanomega2}); so has
the formula for $t_{\pm}$ which is now given by (\ref{eq:t+-bis}).
Unlike previously, maximal mixing turns out to be one of the two
types of solutions that arise.

$\bullet$\ While, in the absence of counterterms,  neither
$t^+_{dL}= t^-_{dL}$,
nor $t^+_{uL}= t^-_{uL}$ could be satisfied, in their presence
 the first relation now can be.
According to (\ref{eq:t+-bis}), the equality of $t^d_{+L}$ and $t^d_{-L}$
requires both $\cos 2\theta_c = -\frac12\frac{h_d-h_s}{h_u-h_c}\approx
-\frac12\frac{m_d^2-m_s^2}{m_u^2-m_c^2}$ and $(s_cc_c(h_u-h_c) -A_d)=0$. This
corresponds to a Cabibbo angle close to maximal and, according to
(\ref{eq:f-A}), to  $p^2= m_d^2+m_s^2$.
At these values of $\theta_c$ and $p^2$, the 1-loop kinetic terms for the
d-type fermions become
proportional to $\left(1 +
\frac{h_u+h_c+\left[\frac{h_d+h_s}{2}\right]}{2}\right){\mathbb I}$,
making $\omega_{dL}$ undetermined. It can be in particular taken to
vanish, such that, according to (\ref{eq:xid2}), $\xi_{dL}$ is then
 very small.

In the $(u,c)$ channel, since $(m_c-m_u)> (m_s-m_d)$, one can never have
$t^u_{+L} = t^u_{-L}$ because this would correspond to $|\cos
2\theta_c|>1$. So, FCNC's can only be suppressed, there, for
$\theta_{uL}=-\omega_{uL}(p^2, \ldots)$. Strictly speaking, since
$\theta_{uL}$ is a constant and $\omega_{uL}$ a function of $p^2$ and of
the masses, the equality can only take place at one value of $p^2$.
However, since all dependence's on $p^2$ are always very weak, 
$(\theta_{uL} + \omega_{uL})$ will only deviate very little from zero
when $p^2$ varies.
 Since  $(-s_cc_c(h_d-h_s) -A_u)$
is always very small,  the equivalent of (\ref{eq:tanomega2}) entails
that so is $\omega_{uL}(p^2, \ldots)$, and, by the equivalent of
(\ref{eq:xid2}), so is $\xi_{uL}(p^2, \ldots)$.

The set $(t^d_{+L} = t^d_{-L}, \theta_{uL}=-\omega_{uL})$ constitutes the
first possibility to suppress  FCNC's at 1-loop.  It corresponds to a
quasi-maximal Cabibbo angle, to small $\theta_{uL}$, small $\omega_{uL}$,
to $\omega_{dL}=0$ and to small $\xi_{dL}$. Accordingly, $\theta_{dL}$ is also
quasi-maximal, and all angles get renormalized at 1-loop by small
quantities, which makes this solution perturbatively safe. Note that, since
$\theta_{uL}$ is small and stays so at 1-loop, this corresponds to a
quasi-alignment of flavor and mass states in the channel with the largest
mass splitting.

For the same $\theta_c$ (close to maximal) but when
$p^2 \not = m_d^2+ m_s^2$,
$(s_cc_c(h_u-h_c) - A_d)$ stays very small (see (\ref{eq:f-A}),
(\ref{eq:hder})).
$\tan 2\omega_{dL}$ given by (\ref{eq:tanomega2}) becomes infinite, which
corresponds to $\omega_{dL}$ maximal.  The FCNC's can be 
taken to vanish (neglecting  a very weak dependence on $p^2$)
for $\theta_{dL}= -\omega_{dL}$, which is then maximal, too 
(like in the previous case). $\theta_{dL}$ gets renormalized at 1-loop into
$\theta_{dL} + \omega_{dL} +\xi_{dL} = \xi_{dL}$ such that
$\tan 2\xi_{dL}\stackrel{(\ref{eq:xid2})}{\approx}
-\frac{m_d^2-m_s^2}{2m_dm_s
E_d - (m_d^2+m_s^2)\frac{f_d(p^2,\ldots)-A_d}{2}}$, which is very large.
So, $\xi_{dL}$ becomes close to maximal, too. This makes the classical
 and maximal $\theta_{dL}$ renormalized by a small amount, which however
results from the cancellation between two large angles.
  In the $(u,c)$ channel, things
are like previously: small $\theta_{uL} = -\omega_{uL}$, and small
$\xi_{uL}$.

This case is thus similar to the previous one in the sense that $\theta_c$
has the same large value, $\theta_{dL}$ too, that $\theta_{uL}$ is small,
and that all of them are renormalized at 1-loop by small quantities.
However, that the renormalization of $\theta_{dL}$ results from the
cancellation between two large angles raises the question whether this
situation is perturbatively safe.
The answer is positive for two reasons:\newline
* a small variation in $p^2$ away from $(m_d^2+m_s^2)$,
that is outside any of the two
concerned mass-shells, is not expected to change the nature of the
perturbative series;\newline
* the 1-loop calculation that we performed in the bare mass basis can as
well be done in the bare flavor basis; since the two are related by a
unitary transformation ${\cal R}(\theta_{dL})$, such a transformation
cannot change either the character of the perturbative series. Going through the
same steps, one easily finds that $\omega_{dL}$ gets replaced by
$(\omega_{dL}+ \theta_{dL})$, which is now very small. In the bare flavor
basis, one finds that the maximal $\theta_{dL}$ still gets, of course,
renormalized by a small amount, but this now results from the sum
of two small quantities, which is a perturbatively safe situation. 

$\bullet$\ Like in the absence of counterterms, from (\ref{eq:nofcnc}),
FCNC's can also be canceled when the two  conditions, respectively
$\theta_{dL}= -\omega_{dL}(p^2, \ldots)$ in the $(d,s)$ channel,
 and $\theta_{uL}= - \omega_{uL}(p^2, \ldots)$ in the $(u,c)$ channel,
 are satisfied (or very close to this, because of
the very weak dependence on $p^2$), without, now,  any 
relation connecting $(t^{d}_{+L} - t^{d}_{-L})$ and $\theta_c$.
Then, since, for $p^2, m^2 < m_W^2$,
$(s_cc_c(h_u-h_d) - A_d)$ and $(s_cc_c(h_d-h_s) - A_u)$ are small,
so are $\omega_{uL,dL}(p^2,\ldots)$ and $\xi_{uL,dL}(p^2,\ldots)$.
Accordingly, $\theta_{uL}$ and
$\theta_{dL}$ are both small and  renormalized at 1-loop by small
quantities. This corresponds to a small $\theta_c$, which is also
renormalized by a small quantity. This configuration is perturbatively safe.

This discussion can be straightforwardly transposed to the leptonic case.

In addition to stabilizing the 1-loop renormalization of mixing angles in
the vicinity of degeneracy, the introduction of Shabalin's counterterms has
been seen to promote maximal mixing (in one channel, accompanied with
quasi-alignment in the other channel) as one of the two natural solutions
 to the suppression of extra FCNC in the bare flavor basis. Maximal mixing
cannot play this role in their absence.

A delicate issue is of course to discriminate between the two types of solutions,
and to determine why one or the other should be preferred.
Since $t_{\pm L}$ are the eigenvalues of the 1-loop kinetic terms,
the equality $t_{+L} = t_{-L}$ corresponds to the
case where, up to an overall  renormalization $\frac{1}{\sqrt{1+t_\pm}}$,
they can be re-diagonalized by a unitary $\cal V$ (see (\ref{eq:calVd}));
in the corresponding channel, which corresponds to the fermionic pair the
closest to degeneracy, the individual mixing matrix $[{\cal C}_{0d} {\cal
V}_{d} {\cal R}(\xi_{d})](p^2, \ldots)$  becomes unitary, too (such that,
in addition to the suppression of FCNC, neutral gauge currents also
 satisfy the property of universality).
A quasi-maximal
Cabibbo (or PMNS) angle corresponds to a minimization of FCNC's, to the
smallest possible deviation from unitarity of the individual mixing matrix
in the channel which is  the closest to degeneracy, to a quasi-maximal
individual mixing in this same channel, and to the quasi-alignment of
flavor and mass eigenstates in the other channel.
This situation corroborates a common argumentation that mass and flavor
eigenstates of charged leptons coincide \cite{Akhmedov}.

In the quark sector, reversely, the distinction
between the two types of fermions,  both charged,  and which,
furthermore, are not observed as particles, is less clear.
The second solution to the suppression of FCNC's, in which both mixing
angles are small, and
which treats the two channels on an equal footing, looks
then more adapted to the situation.

Note that the landscape that we obtain in this work is similar to the one
present in \cite{DuMaVy}. Two types of solutions to the unitarization
equation were uncovered there: the so-called ``Cabibbo-like'' solutions, in
which no constraint occurred for the Cabibbo angle, and maximal mixing. The
Cabibbo angle could then only be constrained by additional assumption; it
turned out, there, that a suitable one was that universality and the
absence of FCNC were violated with the same strength.

\section{The case of 3 generations}
\label{section:3gen}

Our goal is now to generalize the previous calculations to the case of 3
generations of fermions, asking in particular that no extra (with respect
to the ``standard'' phenomenology) FCNC is present
at 1-loop in the basis of bare flavor states in the presence of Shabalin's
counterterms.

A major difference with the case of two generations is, in the quark
sector, the presence of the heavy top quark $m_t \simeq 2 m_W$. This makes
in particular invalid the approximation $m \ll m_W$ for all fermion masses $m$,
that we used in this case.

\subsection{Conditions for suppressing extra FCNC (in the presence of
counterterms)}
\label{subsec:nofcnc}

Like in the case of two generations, extra FCNC will be absent in the
$(d,s,b)$ sector iff ${\cal
C}_{d0} {\cal V}_d {\cal V}_d^\dagger {\cal C}^\dagger_{d0}=
diag(\alpha^d,\beta^d,\gamma^d)$ diagonal.
(not necessarily proportional to the unit matrix),
where ${\cal C}_{d0}$ represents now the $3 \times 3$ classical mixing
matrix for $(d,s,b)$ quarks. Similar expressions occur in the $(u,c,t)$
sector and for the two leptonic ones.

$K_d$ being the kinetic terms of $(d,s,b)$ at 1-loop (eventually including
Shabalin's counterterms), (\ref{eq:kincon1}) $\Rightarrow {\cal V}_d {\cal
V}_d^\dagger = K_d^{-1}$, such that $K_d^{-1} = {\cal C}_{d0}^\dagger\,
diag(\alpha^d,\beta^d,\gamma^d)\,{\cal C}_{d0}$. Now, Shabalin's counterterms are
precisely devised so as to (nearly, that is, up to a very weak dependence
in $p^2$) cancel non-diagonal terms in $K_d$,
which originate from 1-loop transitions of the type depicted in Fig.~2.
Accordingly, in their presence, $K_d$, and thus $K_d^{-1}$, too, are
practically diagonal. The condition for suppressing extra FCNC rewrites
accordingly ${\cal C}_{d0}^\dagger\,
diag(\alpha^d,\beta^d,\gamma^d)\,{\cal C}_{d0}= diagonal$. and we insist that
it is only valid in the presence of counterterms.

Since ${\cal C}_{d0}$ is unitary, the
condition rewrites: $\alpha^d {\mathbb I} + {\cal C}_{d0}^\dagger\, diag(0,
u^d\equiv\beta^d-\alpha^d, v^d\equiv\gamma^d-\alpha^d)\, {\cal C}_{d0}$ diagonal. The first term,
proportional to $\alpha^d$, being already diagonal, the condition applies to
the second contribution. Forgetting, as we always did, about $CP$
violating phases, it is convenient to parametrize ${\cal C}_{d0} = {\cal
R}_{23} {\cal R}_{13} {\cal R}_{12}$, with ${\cal R}_{23} =
\left(\begin{array}{rrr} 1 & 0 & 0 \cr 0& c^d_{23} & s^d_{23} \cr 0 &
-s^d_{23} & c^d_{23}
\end{array}\right), {\cal R}_{13} = \left(\begin{array}{rrr} c^d_{13} & 0 &
s^d_{13} \cr 0 & 1 & 0 \cr -s^d_{13} & 0 & c^d_{13} \end{array}\right), {\cal
R}_{12} \ \left(\begin{array}{rrr} c^d_{12} & s^d_{12} & 0 \cr -s^d_{12} &
c^d_{12}
& 0 \cr 0 & 0 & 1 \end{array}\right)$, to search for eventual solutions
different from $\alpha^d = \beta^d = \gamma^d$ ($u^d=0=v^d$). Equating to zero the 3
non-diagonal entries of the symmetric matrix  ${\cal C}_{d0}^\dagger\,
diag(\alpha^d,\beta^d,\gamma^d)\,{\cal C}_{d0}$ yields the 3 equations:
\begin{subequations}
\begin{equation}\label{subeq:c1}
(u^d+v^d)\, s^d_{12}c^d_{12}(c^d_{13})^2 = (u^d-v^d)\left[-s^d_{13}\sin 2\theta^d_{23} \cos
2\theta^d_{12} - s^d_{12}c^d_{12} \cos 2\theta^d_{23}(1+(s^d_{13})^2) \right];
\end{equation}
\begin{equation}\label{subeq:c2}
(u^d+v^d)\, c^d_{12}s^d_{13}c^d_{13} = (u^d-v^d)\, c^d_{13} \left[c^d_{12}s^d_{13} \cos
2\theta^d_{23} -s^d_{12}\sin 2\theta^d_{23}   \right];
\end{equation}
\begin{equation}\label{subeq:c3}
(u^d+v^d)\, s^d_{12}s^d_{13}c^d_{13} = (u^d-v^d)\, c^d_{13}\left[s^d_{12}s^d_{13}
\cos 2\theta^d_{23}
+ c^d_{12} \sin 2\theta^d_{23}  \right],
\end{equation}
\label{eq:c123}
\end{subequations}
that we now solve.

First make the ratio of (\ref{subeq:c2}) and (\ref{subeq:c3}). For
$s^d_{13}c^d_{13} \not=0$ and $c^d_{13}\not=0$, it yields
$\frac{s^d_{12}}{c^d_{12}} \stackrel{s^d_{13}c^d_{13}\not=0, c^d_{13}\not=0}{=} 
\frac{c^d_{12} s^d_{13} \cos 2\theta^d_{23} - s^d_{12}\sin
2\theta^d_{23}}{s^d_{12}s^d_{13}\cos 2\theta^d_{23} + c^d_{12} \sin
2\theta^d_{23}}
\Rightarrow \sin 2\theta^d_{23}=0 \Rightarrow \theta^d_{23}=0\ or\ \frac{\pi}{2}$.

For $\theta^d_{23}=0$ (\ref{eq:c123}) become
\begin{subequations}
\begin{equation}\label{subeq:c1b}
(u^d+v^d)\, s^d_{12}c^d_{12}(c^d_{13})^2= -(u^d-v^d)\,s^d_{12}c^d_{12}(1+(s^d_{13})^2);
\end{equation}
\begin{equation}\label{subeq:c2b}
(u^d+v^d)\, c^d_{12}s^d_{13}c^d_{13} = (u^d-v^d)\, c^d_{12}s^d_{13}c^d_{13};
\end{equation}
\begin{equation}\label{subeq:c3b}
(u^d+v^d)\, s^d_{12}s^d_{13}c^d_{13} = (u^d-v^d)\, s^d_{12}s^d_{13}c^d_{13}.
\end{equation}
\label{eq:c123b}
\end{subequations}
Since $s^d_{13}c^d_{13} \not= 0$, (\ref{subeq:c2b}) and (\ref{subeq:c3b}) demand
$v^d=0$ which, plugged into (\ref{subeq:c1b}), yields $2u^d s^d_{12}c^d_{12}=0$,
requiring either $u^d=0$ or $[s^d_{12}c^d_{12}=0 \Rightarrow \theta^d_{12}=0\ or\
\theta^d_{12}= \frac{\pi}{2}]$.

For $\theta^d_{23}=\frac{\pi}{2}$ (\ref{eq:c123}) become
\begin{subequations}
\begin{equation}\label{subeq:c1c}
(u^d+v^d)\, s^d_{12}c^d_{12}(c^d_{13})^2= (u^d-v^d)\,s^d_{12}c^d_{12}(1+(s^d_{13})^2);
\end{equation}
\begin{equation}\label{subeq:c2c}
(u^d+v^d)\, c^d_{12}s^d_{13}c^d_{13} = -(u^d-v^d)\, c^d_{12}s^d_{13}c^d_{13};
\end{equation}
\begin{equation}\label{subeq:c3c}
(u^d+v^d)\, s^d_{12}s^d_{13}c^d_{13} = -(u^d-v^d)\, s^d_{12}s^d_{13}c^d_{13}.
\end{equation}
\label{eq:c123c}
\end{subequations}
Since $s^d_{13}c^d_{13} \not= 0$, (\ref{subeq:c2c}) and (\ref{subeq:c3c}) demand
$u^d=0$ which, plugged into (\ref{subeq:c1c}), yields $2v^d s^d_{12}c^d_{12}=0$,
requiring either $v^d=0$ or $[s^d_{12}c_{12}=0 \Rightarrow \theta^d_{12}=0\ or\
\theta^d_{12}= \frac{\pi}{2}]$.

$c^d_{13}=0$ is a trivial solution of
(\ref{subeq:c2}) and (\ref{subeq:c3}); (\ref{subeq:c1}) becomes,
then,\newline
$(u^d-v^d)\left[ \sin 2\theta^d_{23} \cos 2\theta^d_{12} + \sin 2\theta^d_{12}\cos
2\theta^d_{23}\right]=0 \Rightarrow \theta^d_{12}=-\theta^d_{23}+\frac{n\pi}{2}\
or \ u=v$.

 For $s^d_{13}=0$, (\ref{subeq:c2}) and (\ref{subeq:c3}) entail again $[\sin
2\theta^d_{23}=0 \Rightarrow \theta^d_{23}=0 \ or\ \theta^d_{23}= \frac{\pi}{2}]$,
or $u^d=v^d$, while (\ref{subeq:c1}) becomes
$(u^d+v^d)s^d_{12}c^d_{12} = -(u^d-v^d)s^d_{12}c^d_{12}\cos
2\theta^d_{23}$. For $u^d=v^d$ this
requires $\theta^d_{12}=0\ or\ \frac{\pi}{2}$,
for $\theta_{23}=0$, this requires either
$u^d=0$ or $[\theta^d_{12}=0\ or\ \frac{\pi}{2}]$ and,
 for $\theta^d_{23}=\frac{\pi}{2}$, this requires either $v^d=0$ or
$[\theta^d_{12}=0\ or\ \frac{\pi}{2}]$. 

To summarize, the solutions to the suppression of FCNC at 1-loop in bare
flavor space are the following:\newline\newline
\begin{eqnarray}
(a)&:&\quad u^d=0=v^d\ (\Leftrightarrow \alpha^d = \beta^d=\gamma^d);\cr
(b)&:&\quad \theta^d_{12} = 0 =\theta^d_{23}= \theta^d_{13}:\ 
\text{general mass-flavor alignment (trivial solution)};\cr
(c)&:&\quad \theta^d_{13}=0=\theta^d_{12}, \theta^d_{23}=\frac{\pi}{2};\cr
(d)&:&\quad \theta^d_{13}=0, \theta^d_{23}=\frac{\pi}{2}=
\theta^d_{12};\cr
(e)&:&\quad \theta^d_{13}=0=\theta^d_{23},
\theta^d_{12}=\frac{\pi}{2};\cr
(f)&:&\quad \theta^d_{13}=\frac{\pi}{2},
\theta^d_{23}=-\theta^d_{12}+\frac{n\pi}{2};\cr
(g)&:&\quad \theta^d_{13}=\frac{\pi}{2}, u^d=v^d\ (\Leftrightarrow
\beta^d=\gamma^d);\cr
(h)&:&\quad \theta^d_{12}=0 = \theta^d_{23}, v^d=0\ (\Leftrightarrow
\alpha^d=\gamma^d);\cr
(i)&:&\quad \theta^d_{12}=0, \theta^d_{23}=\frac{\pi}{2}, u^d=0\ (\Leftrightarrow
\alpha^d=\beta^d);\cr
(j)&:&\quad \theta^d_{12}=0=\theta^d_{13}, u^d=v^d\ (\Leftrightarrow
\beta^d=\gamma^d);\cr
(k)&:&\quad  \theta^d_{12}= \frac{\pi}{2}, \theta^d_{23}=0, v^d=0\ (\Leftrightarrow
\alpha^d=\gamma^d);\cr
(l)&:&\quad \theta^d_{12}=\frac{\pi}{2}= \theta^d_{23}, u^d=0\ (\Leftrightarrow
\alpha^d=\beta^d);\cr
(m)&:&\quad \theta^d_{12}=\frac{\pi}{2}, \theta^d_{13}=0, u^d=v^d\
(\Leftrightarrow \beta^d=\gamma^d);\cr
(n)&:&\quad \theta^d_{23}=0=\theta^d_{13},  u^d=0\ (\Leftrightarrow
\alpha^d=\beta^d);\cr
(o)&:&\quad \theta^d_{23}=\frac{\pi}{2}, \theta^d_{13}=0, v^d=0\ (\Leftrightarrow
\alpha^d=\gamma^d).
\label{eq:a2o}
\end{eqnarray}
Note that $\theta_{13}=0=\theta_{23}$ is a solution of (\ref{eq:c123})
included in (g).
These solutions correspond  to the following ${{\cal C}_{d0}}$'s:
\begin{eqnarray}
&& (a)\stackrel{\alpha^d=\beta^d=\gamma^d}{\to}
any;\ (b)\to {\mathbb I};\  (c)\to \left(\begin{array}{ccc}
1&0&0 \cr 0&0&1 \cr 0& -1& 0 \end{array}\right);\ 
(d)\to \left(\begin{array}{ccc} 0&1&0 \cr 0&0&1 \cr 1&0&0
\end{array}\right);\ 
(e)\to\left(\begin{array}{ccc}
0&1&0\cr -1 & 0 & 0 \cr 0 & 0 & 1
\end{array}\right);\cr 
&& \hskip -1cm (f)\stackrel{n=1}{\to} \left(\begin{array}{ccc}
0 & 0 & 1 \cr -1 & 0 & 0 \cr 0 & -1 & 0
\end{array}\right)\ \stackrel{n=2}{or}\ \left(\begin{array}{ccc} 0 & 0 & 1 \cr 0 & -1 & 0
\cr 1 & 0 & 0 \end{array}\right)\ \stackrel{n=3}{or}\ \left(\begin{array}{ccc}
0 & 0 & 1 \cr 1 & 0 & 0 \cr 0 & 1 & 0\end{array}\right);\ 
(g)\stackrel{\beta^d=\gamma^d}{\to}\left(\begin{array}{ccc}
0 & 0 & 1\cr -s^d_{12+23} & c^d_{12+23} & 0 \cr -c^d_{12+23} & -s^d_{12+23} & 0
\end{array}\right);\cr
&& (h)\stackrel{\alpha^d=\gamma^d}{\to} {\cal R}_{13};\ 
(i)\stackrel{\alpha^d=\beta^d}{\to} \left(\begin{array}{ccc}
c^d_{13} & 0 & s^d_{13}\cr -s^d_{13} & 0 & c^d_{13}\cr 0 & -1 & 0
\end{array}\right);\ 
(j)\stackrel{\beta^d=\gamma^d}{\to}{\cal R}_{23};\ 
(k)\stackrel{\alpha^d=\gamma^d}{\to} \left(\begin{array}{ccc}
0 & c^d_{13} & s^d_{13}\cr -1 & 0 & 0 \cr 0 & -s^d_{13} & c^d_{13}
\end{array}\right);\cr 
&&\hskip -1.5cm (l)\stackrel{\alpha^d=\beta^d}{\to}\left(\begin{array}{ccc}
0 & c^d_{13} & s^d_{13} \cr 0 & -s^d_{13} & c^d_{13} \cr 1 & 0 & 0
\end{array}\right);\ 
(m)\stackrel{\beta^d=\gamma^d}{\to}\left(\begin{array}{ccc}
0 & 1 & 0 \cr -c^d_{23} & 0 & s^d_{23} \cr s^d_{23} & 0 & c^d_{23}
\end{array}\right);\  
(n)\stackrel{\alpha^d=\beta^d}{\to} {\cal R}_{12};\
(o)\stackrel{\alpha^d=\gamma^d}{\to} \left(\begin{array}{ccc}
c^d_{12} & s^d_{12} & 0\cr 0 & 0 & 1\cr s^d_{12} & -c^d_{12} & 0
\end{array}\right).\cr
&&
\label{eq:aRo}
\end{eqnarray}
Similar formul{\ae} are obtained in the $(u,c,t)$ sector. The relevant
parameters will be then given a superscript ``$u$'' instead of ``$d$''.

We see that the configurations that suppress FCNC are described by two
possible sets of conditions: the ones which concern the $(d,s,d)$ mixing
angles $\theta^d_{ij}$, fixing the mass-flavor relations in this channel
(partial or total alignment {\em etc}),
and the ones concerning $\alpha^d,\beta^d, \gamma^d$ which establish connections
between the masses (fermions, $W$, $\mu$) and the CKM angles $\theta_{ij},
\delta$. Solution (a) is of the second type; (b), (c), (d), (e), (f)
are of the first type; all others are mixed.

The physical mixing  patterns that are observed exhibit, in addition to
approximate alignment as one goes up the generations, some peculiar
values of some of CKM angles. This is why we shall focus in the following
on the solutions that possibly constrain the latter, {\em i.e.} 
(a) and (g) to (o).

The conditions of the second type may not be possible to achieve.
The first task is accordingly to scrutinize the conditions $\alpha=\beta,
\beta=\gamma, \alpha=\gamma$ in both channels, $(d,s,b)$ and $(u,c,t)$,
 and to select the ones that
can be fulfilled. If, for example, in the $(d,s,b)$ channel, only
$\alpha^d = \beta^d$ can be achieved, one has to choose among the 7
solutions (b), (c), (d), (e), (f),  (i), (l), (n).
 The first four are very
constrained solutions. For (b), there is total mass-flavour alignment in
this sector. For (c), (d) and (e), the 3 angles in the $(d,s,b)$ sector are
either vanishing of equal to $\frac{\pi}{2}$. For (f),
$\theta^d_{13}=\frac{\pi}{2}$ while the sum of the 2 other angles is a
multiple of $\frac{\pi}{2}$. In (i) and (l),
$\theta^d_{12}$ and
$\theta^d_{23}$ are constrained, respectively to $0$ or $\frac{\pi}{2}$ and
to $\frac{\pi}{2}$, leaving $\theta^d_{13}$ free,
 while in (n), $\theta^d_{13}$ and $\theta^d_{23}$ are both constrained to
$0$, while $\theta^d_{12}$ is left free.

Still with the example of the $(d,s,b)$ channel, the conditions $\alpha^d =
\beta^d, \beta^d=\gamma^d, \alpha^d=\gamma^d$ write respectively
\begin{eqnarray}
{\cal A}^\pm_{dd} + {\cal A}^3_{dd} &=& {\cal A}^\pm_{ss} + {\cal A}^3_{ss},\cr
{\cal A}^\pm_{ss} + {\cal A}^3_{ss} &=& {\cal A}^\pm_{bb} + {\cal A}^3_{bb},\cr
{\cal A}^\pm_{dd} + {\cal A}^3_{dd} &=& {\cal A}^\pm_{bb} + {\cal A}^3_{bb},
\label{eq:condsb1}
\end{eqnarray}
in which, like in subsection \ref{subsec:1lt},
 ${\cal A}^\pm_{ii}$ and ${\cal A}^3_{ii}$ denote the
1-loop  amplitudes for the diagonal transition $i \to i$ mediated
respectively by $W^\pm$ and $W^3$.

It is simple matter,  using the unitarity of $V$, to get
\begin{equation}
{\cal A}^3_{ii} - {\cal A}^3_{jj} = \frac12 (h_i - h_j).
\label{eq:A1}
\end{equation}

\vbox{
\begin{eqnarray}
{\cal A}^\pm_{dd} &=& |V_{ud}|^2(h_u-h_t) + |V_{cd}|^2(h_c-h_t),\cr
{\cal A}^\pm_{ss} &=& |V_{us}|^2(h_u-h_t) + |V_{cs}|^2(h_c-h_t),\cr
{\cal A}^\pm_{bb} &=& |V_{ub}|^2(h_u-h_t) + |V_{cb}|^2(h_c-h_t),\cr
{\cal A}^\pm_{uu} &=& |V_{ud}|^2(h_d-h_b) + |V_{us}|^2(h_s-h_b),\cr
{\cal A}^\pm_{cc} &=& |V_{cd}|^2(h_d-h_b) + |V_{cs}|^2(h_s-h_b),\cr
{\cal A}^\pm_{tt} &=& |V_{td}|^2(h_d-h_b) + |V_{ts}|^2(h_s-h_b).
\label{eq:A2}
\end{eqnarray}
}

The 6 non-trivial conditions (3 in the $(d,s,b)$ sector and 3 in the
$(u,c,t)$ sector) that we need consider write accordingly

\vbox{
\begin{subequations}
\begin{equation}\label{subeq:6cond1}
\alpha^u=\beta^u:\frac12 (h_d-h_s) + (|V_{ud}|^2 - |V_{us}|^2)(h_u-h_t) 
    + (|V_{cd}|^2 - |V_{cs}|^2)(h_c-h_t) = 0,
\end{equation}
\begin{equation}\label{subeq:6cond2}
\beta^u=\gamma^u: \frac12 (h_s-h_b) + (|V_{us}|^2 - |V_{ub}|^2)(h_u-h_t) 
    + (|V_{cs}|^2 - |V_{cb}|^2)(h_c-h_t) = 0,
\end{equation}
\begin{equation}\label{subeq:6cond3}
\alpha^u=\gamma^u:\frac12 (h_d-h_b) + (|V_{ud}|^2 - |V_{ub}|^2)(h_u-h_t) 
    + (|V_{cd}|^2 - |V_{cb}|^2)(h_c-h_t) = 0,
\end{equation}
\begin{equation}\label{subeq:6cond4}
\alpha^d=\beta^d:\frac12 (h_u-h_c) + (|V_{ud}|^2 - |V_{cd}|^2)(h_d-h_b) 
    + (|V_{us}|^2 - |V_{cs}|^2)(h_s-h_b) = 0,
\end{equation}
\begin{equation}\label{subeq:6cond5}
\beta^d=\gamma^d: \frac12 (h_c-h_t) + (|V_{cd}|^2 - |V_{td}|^2)(h_d-h_b) 
    + (|V_{cs}|^2 - |V_{ts}|^2)(h_s-h_b) = 0,
\end{equation}
\begin{equation}\label{subeq:6cond6}
\alpha^d=\gamma^d: \frac12 (h_u-h_t) + (|V_{ud}|^2 - |V_{td}|^2)(h_d-h_b) 
    + (|V_{us}|^2 - |V_{ts}|^2)(h_s-h_b) = 0.
\end{equation}
\label{eq:6cond}
\end{subequations}
}
The 6 equations  (\ref{eq:6cond}) include only 2 pairs of independent
conditions ((\ref{subeq:6cond1}) +(\ref{subeq:6cond1})=(\ref{subeq:6cond3}),
 (\ref{subeq:6cond4})+(\ref{subeq:6cond5})=(\ref{subeq:6cond6})).

The particular case of 2 generations, that we studied before, is easily
recovered. One has, then, $|V_{ud}|^2 = c_c^2 = |V_{cs}|^2, |V_{us}|^2 =
s_c^2 = |V_{cd}|^2$.
(\ref{eq:6cond}) shrinks to
\begin{eqnarray}
\alpha^u=\beta^u:\frac12 (h_d-h_s) + (c_c^2-s_c^2) (h_u-h_c) &=& 0,\cr
\alpha^d=\beta^d:\frac12 (h_u-h_c) + (c_c^2 - s_c^2)(h_d - h_s) &=& 0,
\label{eq:2cond}
\end{eqnarray}
of which only the first can be realized, leading to a large (quasi-maximal)
 Cabibbo angle,
and leaving mass-flavor alignment as the only possibility in the $(u,c)$
sector.

\subsection{Coping with the top quark: analytic expressions
 for  $(h_i-h_t)$}
\label{subsec:anahij}

The approximate expression of $(h_i-h_j)$ for $m_i^2, m_j^2, p^2 \ll
m_W^2$ is given by (\ref{eq:hij5}). It is valid for $u,d,s,c,b$ quarks, all
leptons, but it is not valid when the top quark
is involved. In this case,
an approximate expression for $(h_i - h_t)$ can still  be obtained  from
(\ref{eq:hij4}), which  is valid for for $m_i^2, p^2 \ll m_W^2$, and keeps exact
in the top quark mass dependence $m_t$:

\vbox{
\begin{eqnarray}
h_i-h_t &\approx& \frac{g^2}{4} \frac{i}{16\pi^2}\left(
-\frac32 - \ln\frac{m_W^2}{\mu^2} + \frac{m_i^2}{m_W^2}\left(
-\frac{17}{4} + \frac32\ln\frac{m_W^2}{\mu^2}\right) + t_{terms}
\right),\cr
&& \cr
&& \cr
t_{terms} &\approx& 
\frac23 \frac{m_t^2}{m_W^2} + \frac72\frac{m_W^2}{m_W^2-m_t^2}
+\frac14 \left( 5 - \frac{m_t^2}{m_W^2}
\right)\frac{m_t^2}{m_W^2-m_t^2}\cr
&& + 2\frac{m_W^2
\ln\frac{m_W^2}{\mu^2}-m_t^2\ln\frac{m_t^2}{\mu^2}}{m_W^2-m_t^2}
 -\frac12 \left(2+ \frac{m_t^2}{m_W^2}  \right)
\frac{m_W^4 \ln\frac{m_W^2}{\mu^2} -
m_t^4\ln\frac{m_t^2}{\mu^2}}{(m_W^2-m_t^2)^2}\cr
&&\hskip -2cm -\frac{m_t^2}{m_W^2}\frac{1}{(m_W^2-m_t^2)^2}\left(
-\frac{11m_W^4 - 7 m_W^2m_t^2 + 2m_t^4}{6}
+\frac{m_W^6\ln\frac{m_W^2}{\mu^2} +(-3m_t^2m_W^4+
3m_t^4m_W^2-m_t^6)\ln\frac{m_t^2}{\mu^2}}{m_W^2 - m_t^2}\right).\cr
&&
\label{eq:hit}
\end{eqnarray}
}
When $m_t$ becomes larger and larger, $t_{terms}$ scale like
\begin{equation}
t_{terms} \stackrel{m_t \gg m_W}{\sim}
\frac{m_t^2}{m_W^2}\left(\frac{7}{12} -
\frac12\,\ln\frac{m_t^2}{\mu^2}\right).
\label{eq:bigt}
\end{equation}

In practice, according to (\ref{eq:6cond}), one needs $(h_u-h_t)$ and
$(h_c-h_t)$.

\subsection{Solving the constraints for 3 generations of quarks}
\label{subsec:6quarks}

The CKM matrix we parametrize  as
\begin{equation}
V=
\left(\begin{array}{ccc}
V_{ud} & V_{us} & V_{ub} \cr
V_{cd} & V_{cs} & V_{cb} \cr
V_{td} & V_{ts} & V_{tb} \cr
\end{array}\right)
= \left(\begin{array}{ccc}
c_{12}c_{13} & s_{12}c_{13} & s_{13}\,e^{-i\delta} \cr
-s_{12}c_{23}-c_{12}s_{23}s_{13}\,e^{i\delta} &
c_{12}c_{23}-s_{12}s_{23}s_{13}\,e^{i\delta} &
s_{23}c_{13} \cr
s_{12}s_{23} -c_{12}c_{23}s_{13}\,e^{i\delta} &
-c_{12}s_{23}-s_{12}c_{23}s_{13}\,e^{i\delta} &
c_{23}c_{13}
\end{array}\right),
\label{eq:V}
\end{equation}
such that
\begin{eqnarray}
|V_{ud}|^2 - |V_{us}|^2 &=& c_{13}^2\cos 2\theta_{12};\cr
|V_{cd}|^2 - |V_{cs}|^2 &=& \cos 2\theta_{12}(-c_{23}^2 + s_{13}^2
s_{23}^2) +\sin 2\theta_{12}\sin 2\theta_{23} s_{13}\cos\delta;\cr
|V_{us}|^2 - |V_{ub}|^2 &=& s_{12}^2 c_{13}^2 - s_{13}^2;\cr
|V_{ud}|^2 - |V_{ub}|^2 &=& c_{12}^2 c_{13}^2 - s_{13}^2;\cr
|V_{cs}|^2 - |V_{cb}|^2 &=& c_{12}^2 c_{23}^2 +s_{23}^2(-c_{13}^2 + s_{12}^2
s_{13}^2) -\frac12 \sin 2\theta_{12}\sin 2\theta_{23} s_{13}\cos\delta;\cr
|V_{ud}|^2 - |V_{cd}|^2 &=& c_{12}^2(c_{13}^2 -s_{23}^2s_{13}^2) -
s_{12}^2c_{23}^2
 -\frac12 \sin2\theta_{12}\sin 2\theta_{23} s_{13}\cos\delta;\cr
|V_{us}|^2 - |V_{cs}|^2 &=& s_{12}^2(c_{13}^2 -s_{23}^2s_{13}^2) -
c_{12}^2c_{23}^2 +\frac12 \sin2\theta_{12}\sin 2\theta_{23}
s_{13}\cos\delta;\cr
|V_{cd}|^2 - |V_{cb}|^2 &=& s_{12}^2 c_{23}^2 + s_{23}^2 ( c_{12}^2
s_{13}^2 -c_{13}^2)+\frac12 \sin 2\theta_{12}\sin 2\theta_{23} s_{13}\cos\delta;\cr
|V_{cd}|^2 - |V_{td}|^2 &=& \cos 2\theta_{23}(s_{12}^2 - c_{12}^2 s_{13}^2)
+  \sin2\theta_{12}\sin 2\theta_{23} s_{13}\cos\delta;\cr
|V_{cs}|^2 - |V_{ts}|^2 &=& \cos 2\theta_{23}(c_{12}^2 -s_{12}^2 s_{13}^2)
-  \sin2\theta_{12}\sin 2\theta_{23} s_{13}\cos\delta;\cr
|V_{ud}|^2 - |V_{td}|^2 &=& c_{12}^2 (c_{13}^2 -c_{23}^2 s_{13}^2)
-s_{12}^2 s_{23}^2 +\frac12 \sin2\theta_{12}\sin 2\theta_{23}
s_{13}\cos\delta;\cr
|V_{us}|^2 - |V_{ts}|^2 &=& s_{12}^2 (c_{13}^2 - c_{23}^2 s_{13}^2)
-c_{12}^2 s_{23}^2 -\frac12 \sin2\theta_{12}\sin 2\theta_{23}
s_{13}\cos\delta.
\end{eqnarray}
The constraints (\ref{eq:6cond})
become (we remind that $t_{terms}$ is given in (\ref{eq:hit}))
\begin{subequations}
\begin{eqnarray}\label{subeq:6const1}
\alpha^u = \beta^u &:& \frac12 \frac{m_d^2-m_s^2}{m_W^2}\left(-\frac{17}{4}+\frac32
\ln\frac{m_W^2}{\mu^2}\right) = \cr
&&  -c_{13}^2\cos 2\theta_{12}
\left[\left(
-\frac32 - \ln\frac{m_W^2}{\mu^2} + \frac{m_u^2}{m_W^2}\left(
-\frac{17}{4} + \frac32\ln\frac{m_W^2}{\mu^2}\right) + t_{terms}
\right) \right]\cr
&& -\left[\cos 2\theta_{12}(-c_{23}^2 + s_{13}^2
s_{23}^2) +\sin 2\theta_{12}\sin 2\theta_{23} s_{13}\cos\delta\right]\cr
&& \hskip 4cm
\left[ \left(
-\frac32 - \ln\frac{m_W^2}{\mu^2} + \frac{m_c^2}{m_W^2}\left(
-\frac{17}{4} + \frac32\ln\frac{m_W^2}{\mu^2}\right) + t_{terms}
\right)  \right];\cr
&&
\end{eqnarray}
\begin{eqnarray}\label{subeq:6const2}
\beta^u = \gamma^u &:& 
\frac12 \frac{m_s^2-m_b^2}{m_W^2}\left(-\frac{17}{4}+\frac32
\ln\frac{m_W^2}{\mu^2}\right) =\cr
&& - (s_{12}^2 c_{13}^2 - s_{13}^2)
\left[\left(
-\frac32 - \ln\frac{m_W^2}{\mu^2} + \frac{m_u^2}{m_W^2}\left(
-\frac{17}{4} + \frac32\ln\frac{m_W^2}{\mu^2}\right) + t_{terms}
\right) \right]\cr
&&- \left[ c_{12}^2 c_{23}^2 +s_{23}^2(-c_{13}^2 + s_{12}^2
s_{13}^2) -\frac12 \sin 2\theta_{12}\sin 2\theta_{23} s_{13}\cos\delta
\right]\cr
&& \hskip 4cm
\left[ \left(
-\frac32 - \ln\frac{m_W^2}{\mu^2} + \frac{m_c^2}{m_W^2}\left(
-\frac{17}{4} + \frac32\ln\frac{m_W^2}{\mu^2}\right) + t_{terms}
\right)  \right];\cr
&&
\end{eqnarray}
\begin{eqnarray}\label{subeq:6const3}
\alpha^u = \gamma^u &:& 
\frac12 \frac{m_d^2-m_b^2}{m_W^2}\left(-\frac{17}{4}+\frac32
\ln\frac{m_W^2}{\mu^2}\right) = \cr
&& - (c_{12}^2 c_{13}^2 - s_{13}^2)
\left[\left(
-\frac32 - \ln\frac{m_W^2}{\mu^2} + \frac{m_u^2}{m_W^2}\left(
-\frac{17}{4} + \frac32\ln\frac{m_W^2}{\mu^2}\right) + t_{terms}
\right) \right]\cr
&& - \left[ 
 s_{12}^2 c_{23}^2 + s_{23}^2 ( c_{12}^2
s_{13}^2 -c_{13}^2)+\frac12 \sin 2\theta_{12}\sin 2\theta_{23} s_{13}\cos\delta
\right]\cr
&& \hskip 4cm
\left[ \left(
-\frac32 - \ln\frac{m_W^2}{\mu^2} + \frac{m_c^2}{m_W^2}\left(
-\frac{17}{4} + \frac32\ln\frac{m_W^2}{\mu^2}\right) + t_{terms}
\right)  \right];\cr
&&
\end{eqnarray}

\vbox{
\begin{eqnarray}\label{subeq:6const4}
\alpha^d = \beta^d &:&   \frac12 (m_u^2-m_c^2) = -(m_d^2-m_s^2)
 \left[c_{12}^2(c_{13}^2 -s_{23}^2s_{13}^2) -
s_{12}^2c_{23}^2
 -\frac12 \sin2\theta_{12}\sin 2\theta_{23} s_{13}\cos\delta\right]\cr
&& \hskip 1cm- (m_s^2-m_b^2)\left[s_{12}^2(c_{13}^2 -s_{23}^2s_{13}^2) -
c_{12}^2c_{23}^2 +\frac12 \sin2\theta_{12}\sin 2\theta_{23}
s_{13}\cos\delta\right];
\end{eqnarray}
}

\begin{eqnarray}\label{subeq:6const5}
\beta^d = \gamma^d &:& \frac12\left[ \left(
-\frac32 - \ln\frac{m_W^2}{\mu^2} + \frac{m_c^2}{m_W^2}\left(
-\frac{17}{4} + \frac32\ln\frac{m_W^2}{\mu^2}\right) + t_{terms}
\right)  \right] = \cr
&&  -\frac{m_d^2-m_b^2}{m_W^2}\left(-\frac{17}{4}+\frac32
\ln\frac{m_W^2}{\mu^2}\right)\left[\cos 2\theta_{23}(s_{12}^2 - c_{12}^2 s_{13}^2)
+  \sin2\theta_{12}\sin 2\theta_{23} s_{13}\cos\delta\right]\cr
&&   -\frac{m_s^2-m_b^2}{m_W^2}\left(-\frac{17}{4}+\frac32
\ln\frac{m_W^2}{\mu^2}\right)
\left[  \cos 2\theta_{23}(c_{12}^2 -s_{12}^2 s_{13}^2)
-  \sin2\theta_{12}\sin 2\theta_{23} s_{13}\cos\delta  \right];\cr
&&
\end{eqnarray}
\begin{eqnarray}\label{subeq:6const6}
\alpha^d = \gamma^d &:& \frac12\left[ \left(
-\frac32 - \ln\frac{m_W^2}{\mu^2} + \frac{m_u^2}{m_W^2}\left(
-\frac{17}{4} + \frac32\ln\frac{m_W^2}{\mu^2}\right) + t_{terms}
\right)  \right] = \cr
&&  -\frac{m_d^2-m_b^2}{m_W^2}\left(-\frac{17}{4}+\frac32
\ln\frac{m_W^2}{\mu^2}\right)
\left[ c_{12}^2 (c_{13}^2 -c_{23}^2 s_{13}^2)
-s_{12}^2 s_{23}^2 +\frac12 \sin2\theta_{12}\sin 2\theta_{23}
s_{13}\cos\delta \right]\cr
&&   -\frac{m_s^2-m_b^2}{m_W^2}\left(-\frac{17}{4}+\frac32
\ln\frac{m_W^2}{\mu^2}\right)
\left[ s_{12}^2 (c_{13}^2 - c_{23}^2 s_{13}^2)
-c_{12}^2 s_{23}^2 -\frac12 \sin2\theta_{12}\sin 2\theta_{23}
s_{13}\cos\delta \right].\cr
&&
\end{eqnarray}
\label{eq:6const}
\end{subequations}
Notice that (\ref{subeq:6const4}) is the only equation which is not
influenced by the large mass of the top quark.

For $\theta_{23}=0=\theta_{13}$, (\ref{subeq:6const1}) reduces to
$\frac12 (m_d^2-m_s^2) = (m_c^2-m_u^2)\cos 2\theta_{12}$, which is the
constraint on the Cabibbo angle when 2 generations only are present (the
first of eqs.~(\ref{eq:2cond})).

Once the masses of the fermions, the one of the $W$ gauge boson, and the
renormalization scale $\mu$ are fixed, they constitute a system of 4
equations for the 4 CKM angles $\theta_{12}, \theta_{23}, \theta_{13}$ and
$\delta$.

Some simplifications can be performed. First, even in the large interval
$\mu \in [100\,MeV, m_W]$, the $t_{terms}$ largely
dominate over $\frac{m_{u,c}^2}{m_W^2}\left(
-\frac{17}{4} + \frac32\ln\frac{m_W^2}{\mu^2}\right)$, at least by a factor
$1000$. The latter can thus always be neglected. The same
$t_{terms}$ dominate over $\ln\frac{m_W^2}{\mu^2}$ by at least a factor 3,
and over $\frac32$ by at least a factor $6$. It is accordingly a reasonable
approximation to only consider their contribution inside the corresponding
$[\ ]$ brackets.  Secondly, it is also reasonable to neglect $m_d^2 \ll m_b^2,
m_s^2 \ll m_b^2, m_u^2 \ll m_c^2$ and, even, $m_d^2 \ll m_s^2$. The system
(\ref{eq:6const}) then simplifies to 
\begin{subequations}
\begin{eqnarray}\label{subeq:6cb1}
\alpha^u =\beta^u &:& \frac{m_s^2}{m_W^2}\left(-\frac{17}{4}+\frac32
\ln\frac{m_W^2}{\mu^2}\right)
\approx  2\,t_{terms}\,\Big(
(c_{13}^2 -c_{23}^2 + s_{23}^2 s_{13}^2)\cos 2\theta_{12}
+s_{13}\sin 2\theta_{12}\sin 2\theta_{23} \cos\delta\Big);\cr
&&
\end{eqnarray}

\vbox{
\begin{eqnarray}\label{subeq:6cb2}
\beta^u = \gamma^u &:& \frac{m_b^2}{m_W^2}\left(-\frac{17}{4}+\frac32
\ln\frac{m_W^2}{\mu^2}\right)
\approx  2\,t_{terms}\,\Big(s_{12}^2 c_{13}^2 - s_{13}^2
+ c_{12}^2 c_{23}^2 +s_{23}^2(-c_{13}^2 + s_{12}^2
s_{13}^2)\cr
&& \hskip 6cm -\frac12 s_{13}\sin 2\theta_{12}\sin 2\theta_{23} \cos\delta
\Big);
\end{eqnarray}
}

\begin{eqnarray}\label{subeq:6cb3}
\alpha^u = \gamma^u &:& \frac{m_b^2}{m_W^2}\left(-\frac{17}{4}+\frac32
\ln\frac{m_W^2}{\mu^2}\right) \approx
2\,t_{terms}\,\Big(
c_{12}^2 c_{13}^2 - s_{13}^2
+  s_{12}^2 c_{23}^2 + s_{23}^2 ( c_{12}^2
s_{13}^2 -c_{13}^2) \cr
&& \hskip 6cm +\frac12 \sin 2\theta_{12}\sin 2\theta_{23} s_{13}\cos\delta
\Big);
\end{eqnarray}
\begin{eqnarray}\label{subeq:6cb4}
\alpha^d = \beta^d &:& m_c^2 \approx 
 -2\Bigg(m_s^2\Big[ \cos 2\theta_{12}
\left(c_{23}^2 + c_{13}^2 -s_{23}^2 s_{13}^2\right)
- s_{13} \sin2\theta_{12}\sin 2\theta_{23} \cos\delta \Big] \cr
&& +m_b^2 \Big[s_{12}^2 (c_{13}^2 -s_{23}^2 s_{13}^2) -c_{12}^2 c_{23}^2
 +\frac12 \, s_{13} \sin2\theta_{12}\sin 2\theta_{23} \cos\delta\Big] 
\Bigg);
\end{eqnarray}
\begin{eqnarray}\label{subeq:6cb5}
\beta^d = \gamma^d &:& t_{terms} \approx 
2\frac{m_b^2}{m_W^2}\left(-\frac{17}{4}+\frac32
\ln\frac{m_W^2}{\mu^2}\right)
c_{13}^2\cos 2\theta_{23};
\end{eqnarray}
\begin{eqnarray}\label{subeq:6cb6}
\alpha^d = \gamma^d &:& t_{terms} \approx 
2\frac{m_b^2}{m_W^2}\left(-\frac{17}{4}+\frac32
\ln\frac{m_W^2}{\mu^2}\right)
\left( c_{13}^2 - s_{13}^2 -s_{23}^2c_{13}^2
\right).
\end{eqnarray}
\label{eq:6cb}
\end{subequations}
It is important to stress that the system (\ref{eq:6cb}) is only
approximate, while (\ref{eq:6const}) is exact; this why, in particular,
while the simultaneous fulfillment of (\ref{subeq:6const2}) and
(\ref{subeq:6const3}) (resp. (\ref{subeq:6const5}) and
(\ref{subeq:6const6}))  entails that of (\ref{subeq:6const1})
(resp. (\ref{subeq:6const4})), the same does
not occur for (\ref{subeq:6cb2}), (\ref{subeq:6cb3}) and
(\ref{subeq:6cb1}) (resp. (\ref{subeq:6cb5}), (\ref{subeq:6cb6}) and
(\ref{subeq:6cb4})) .

As a short numerical calculation shows,
(\ref{subeq:6cb5}) can never be satisfied,
because it would correspond to $|c_{13}^2 \cos
2\theta_{23}| > 300$ (still for $\mu \in [100\, MeV, m_W]$).
The same argumentation shows that (\ref{subeq:6cb6}) cannot be satisfied
either.
So, in the $(d,s,b)$ sector, only $\alpha^d=\beta^d$ 
can eventually be satisfied and solutions (b), (c), (d), (e), (f), (i), 
(l), (n) are the only ones that should be considered.

Summing (\ref{subeq:6cb2}) and (\ref{subeq:6cb3}) yields a constraint which
does not include $\theta_{12}$ nor $\delta$:
\begin{equation}
\frac{1}{t_{terms}} \frac{m_b^2}{m_W^2}\left(-\frac{17}{4}+\frac32
\ln\frac{m_W^2}{\mu^2}\right) = 3 c_{13}^2(1+s_{23}^2) -1,
\label{eq:c13}
\end{equation}
such that the quantity $3 c_{13}^2(1+s_{23}^2) -1$ must be a small
number, the modulus of which does not exceed $1.5\, 10^{-3}$. The
condition $0 \leq s_{23}^2 \leq 1$ entails
\begin{equation}
\frac16 \leq c_{13}^2 \leq \frac13 \stackrel
{\theta_{13}\in[0,\frac{\pi}{2}]}{\Rightarrow} 55^0 \leq \theta_{13} \leq
66^o. 
\end{equation}
which is not compatible with the observed value of $\theta_{13}$ in the CKM
matrix.
(\ref{subeq:6cb2}) and (\ref{subeq:6cb3}) are not either individually
compatible with the observed values of the
CKM angles. Indeed, plugging in these values, their r.h.s. come close
to $2 t_{terms}$, which is much larger than their l.h.s. 

 Let us now consider (\ref{subeq:6cb1}) and
(\ref{subeq:6cb4}).  Since, for $\mu \in [100\,MeV, m_W]$,
 $\frac{m_s^2}{m_W^2}\left(-\frac{17}{4}+\frac32
\ln\frac{m_W^2}{\mu^2}\right)\ll 2t_{terms}$, (\ref{subeq:6cb1}) rewrites
\begin{eqnarray}
\alpha^u=\beta^u: (s_{23}^2 -s_{13}^2 + s_{23}^2 s_{13}^2)\cos 2\theta_{12}
+s_{13}\sin 2\theta_{12}\sin 2\theta_{23} \cos\delta \approx 0,
\label{eq:acon}
\end{eqnarray}
which is presumably only trustable for $\delta=0$ since we did not
introduce any $CP$-violating phase in the partial rotations ${\cal R}_{12},
{\cal R}_{23}, {\cal R}_{13}$.

In case  (\ref{subeq:6cb1}) and
(\ref{subeq:6cb4})  are simultaneously satisfied, eliminating the
$CP$-violating phase $\delta$ between the two of them yields 
\begin{equation}
m_c^2 \approx 
 -2\left(2\,m_s^2 c_{13}^2 \cos 2\theta_{12} 
+m_b^2 \Big[s_{12}^2 (c_{13}^2 -s_{23}^2 s_{13}^2) -c_{12}^2 c_{23}^2 
 -\frac12 \,
(c_{13}^2 -c_{23}^2 + s_{23}^2 s_{13}^2)\cos 2\theta_{12}\Big] \right),
\label{eq:bcon}
\end{equation}
from which  one deduces that very small values of $\theta_{23}$
and $\theta_{13}$, like observed in the quark sector,
 are only compatible with $\theta_{12}$ quasi-maximal:
$\cos 2\theta_{12} \approx \frac{m_c^2}{2(m_b^2-2m_s^2)}\ (\theta_{12}
\approx 44^o)$, which is not the observed value ($\theta_{12}\approx 13^o$)
of the Cabibbo angle. Consequently, a rather small Cabibbo angle can only be
achieved if at least one among the two angles $\theta_{23}$ and
$\theta_{13}$ is not very small. As we saw  by summing (\ref{subeq:6cb2})
 and (\ref{subeq:6cb3}), this must be the case of $\theta_{13}$.
From (\ref{eq:c13}) and (\ref{eq:bcon}), one gets, after neglecting
$\frac{2m_s^2}{3(1+s_{23}^2)}\ll \frac{m_b^2}{3}$
\begin{equation}
s_{23}^2  \approx \frac43 c_{12}^2 + \frac {m_c^2}{2m_b^2}
\approx \frac43 c_{12}^2 + 4.5\,10^{-2},
\end{equation}
which entails in particular $s_{23}^2 \geq 4.5\,10^{-2} \Rightarrow
\theta_{23} \geq 12^o$ and $c_{12}^2 \leq \frac34 \Rightarrow
\theta_{12} \geq 30^o$.

To summarize, the only equations that can eventually be simultaneously
 satisfied are
(\ref{subeq:6cond1}) to (\ref{subeq:6cond4}). They lead to CKM angles which
are not the ones observed in the quark sector, and which are all fairly
large (except $\theta_{23}$ which can go as low as $12^o$).

There are of course other possibilities, which are to be looked for among
the solutions (a) to (o) in each of the two sectors $(d,s,b)$ and $(u,c,t)$.

It is appropriate to consider solution (b) which means global mass-flavor
alignment, in one of the two sectors, first, for example $(u,c,t)$. 
The only left over constraint from the demanded suppression of extra FCNC is
accordingly (\ref{subeq:6cb4}), which corresponds to $\alpha^d=\beta^d$
 (we recall that
(\ref{subeq:6cb5}) and (\ref{subeq:6cb6}) can never be satisfied). Only
solutions (b), (c), (d), (e), (f), (i), (l), (n) are thus to be considered.
They apply to mixing angles of the $(d,s,b)$ sector, but these can be
identified with CKM angle due to the alignment in the u-type sector.
(b) corresponds to global mass-flavor alignment in the $(d,s,b)$ sector,
too. (c), (d), (e), (f) correspond to the CKM matrices represented in
(\ref{eq:aRo}). They offer no special interest, mixing
angles being $0$ or $\frac{\pi}{2}$. (i), with $\theta_{12}=0,
\theta_{23}=\frac{\pi}{2}$, yields $\cos 2\theta_{13} \approx
-\frac{m_c^2}{2m_s^2}$ which is impossible because it is $>1$. (l), with
$\theta_{12}=\frac{\pi}{2}=\theta_{23}$, corresponds to
$\cos 2\theta_{23} = -\frac{m_c^2}{2(m_b^2-m_s^2)}$ very small, such that
$\theta_{13}$ is close to maximal. (n), with $\theta_{13}=0=\theta_{23}$,
corresponds to $\cos 2\theta_{12} \approx -\frac{m_c^2}{2(m_b^2-2m_s^2)}$,
such that $\theta_{12}$ is close to maximal.

Let us then choose global mass-flavor alignment in the $(d,s,b)$ sector.
Only (\ref{subeq:6cond1}), (\ref{subeq:6cond2}) and (\ref{subeq:6cond3})
can then be considered as eventual constraints to suppress extra FCNC, and
we shall consider them for $\delta=0$, neglecting $CP$-violation effects.
If the 3 of them are realized, we have already seen that $\theta_{13}$ will
be large $55^o \leq \theta_{13}\leq 66^o$. Since this is in contradiction
with observation, we have to relax at least one of the three constraints.
Since they are not independent, at least 2 of them must be relaxed,
otherwise the 3rd would be automatically satisfied. Keeping only
(\ref{subeq:6cond2}) or only (\ref{subeq:6cond3}) cannot accommodate for
very small $\theta_{13}$ and $\theta_{23}$ (see (\ref{eq:a2o})), such that,
if one looks for solutions close to reality,
it looks appropriate to relax both of them and only keep
(\ref{subeq:6cond1}), associated with the constraint $\alpha^u=\beta^u$.
Among the solutions associated with the latter,
(n)  (see (\ref{eq:a2o})) is specially  worth investigating because the exact
suppression of extra FCNC corresponds then to vanishing
$\theta_{23}$ and $\theta_{13}$. In this case, as we already mentioned,
(\ref{subeq:6const1}) reduces to the 2-generation constraint 
$\cos 2\theta_{12}= \frac12 \frac{m_d^2-m_s^2}{m_c^2-m_u^2}$,
which corresponds to a Cabibbo angle close to maximal.
A not fully complete suppression can be thought to possibly
accommodate for small values of $\theta_{23}$ and $\theta_{13}$.

Instead of working on the approximate system
(\ref{eq:6cb}), let us rather consider the exact one (\ref{eq:6const}) and,
more specifically,  (\ref{subeq:6const1}) in
a realistic situation when $\theta_{23}$ and $\theta_{13}$
are not strictly vanishing but only very small. Solution (n)
is not, then, exactly satisfied at 1-loop, but it could be
at higher orders.  More precisely, let us determine which values of
$\theta_{12}$ are compatible with 
(\ref{subeq:6const1}) and realistic values of $\theta_{23}$ and
$\theta_{13}$.  (\ref{subeq:6const1}) rewrites (for $\delta=0$)
\begin{eqnarray}
&& \frac12 \frac{m_d^2-m_s^2}{m_W^2} 
-\frac{m_c^2-m_u^2}{m_W^2} \cos 2\theta_{12}
\approx \cr
&&\hskip 1.5cm-\frac{m_c^2}{m_W^2}\sin 2\theta_{23}s_{13}\sin 2\theta_{12}
+ \Big(s_{13}^2 \frac{m_u^2}{m_W^2}
-s_{23}^2(1+s_{13}^2)\frac{m_c^2}{m_W^2} \Big)\cos 2\theta_{12}\cr
&&\hskip 1.5cm+ \Big( (s_{13}^2 - s_{23}^2 - s_{13}^2 s_{23}^2) \cos 2\theta_{12}
-\sin 2\theta_{23}s_{13} \sin 2\theta_{12} \Big)\;T(m_t, m_W, \mu),\cr
&& \cr
&& T(m_t,m_W,\mu) =
 \frac{-\frac32 -\ln\frac{m_W^2}{\mu^2} + t_{terms}}{-\frac{17}{4} +
\frac32\ln\frac{m_W^2}{\mu^2}} \stackrel{m_t\gg m_W}{\sim}
 \frac{-\frac32 -\ln\frac{m_W^2}{\mu^2} + \frac{m_t^2}{m_W^2}\left(\frac{7}{12} -
\frac12\,\ln\frac{m_t^2}{\mu^2}\right)}{-\frac{17}{4} +
\frac32\ln\frac{m_W^2}{\mu^2}}.\cr
&&
\label{eq:ss}
\end{eqnarray}
The expression for  $t_{terms}$ is given in (\ref{eq:hit}) and its
behaviour as $m_t$ grows, which we used in the r.h.s. of (\ref{eq:ss}),
has been given in (\ref{eq:bigt}).

The prediction for 2 generations is obtained by putting the r.h.s. of
(\ref{eq:ss}) to $0$, that is, for example, by setting  $s_{23}=0 =s_{13}$.

The modulus of $T$ is larger than $1.45$ as soon as $\mu \geq 10\,
MeV$, while $\frac{m_c^2}{m_W^2} \approx 3.5\,10^{-4}$. So, we can neglect
$2\frac{m_c^2}{m_W^2}s_{23}s_{13}\sin 2\theta_{12}$ with respect to 
$2T s_{23}s_{13}\sin 2\theta_{12}$ is the r.h.s. of (\ref{eq:ss}).
As for the terms proportional to $\cos 2\theta_{12}$, $s_{23}^2 s_{13}^2
\frac{m_c^2}{m_W^2} \ll s_{23}^2 s_{13}^2 T$, such that 
(\ref{eq:ss}) can be approximated by
\begin{eqnarray}
\frac12 \frac{m_d^2-m_s^2}{m_W^2} 
-\frac{m_c^2-m_u^2}{m_W^2} \cos 2\theta_{12}
&\approx& \frac{s_{13}^2 m_u^2 -s_{23}^2 m_c^2}{m_W^2} \cos 2\theta_{12}
\cr
&& \hskip -4cm+ 
\Big( (s_{13}^2 - s_{23}^2 - s_{13}^2 s_{23}^2) \cos 2\theta_{12}
-\sin 2\theta_{23}s_{13} \sin 2\theta_{12} \Big)T(m_t, m_W, \mu).
\label{eq:ss1}
\end{eqnarray}
The vanishing of the l.h.s. of (\ref{eq:ss1}) is the condition for no extra FCNC
for 2 generations only (see (\ref{eq:2cond})). Its modulus is always
smaller than $\frac{m_c^2}{m_W^2}$. So is the modulus of the first term in
the r.h.s. of (\ref{eq:ss1}). At the opposite, the modulus of  $T$ is,
as we mentioned,
 larger than $1.45$ for $\mu \geq 10\,MeV$. Accordingly, the coefficient of
$T$ in (\ref{eq:ss1}) should be very small, which writes
\begin{equation}
\big|(s_{13}^2 - s_{23}^2 - s_{13}^2 s_{23}^2) \cos 2\theta_{12}
-\sin 2\theta_{23}s_{13} \sin 2\theta_{12}\big| \approx \left|\frac{\frac12 \frac{m_d^2-m_s^2}{m_W^2} 
-\frac{m_c^2(1+s_{23}^2)-m_u^2(1+s_{13}^2)}{m_W^2} \cos 2\theta_{12}}{T(m_t, m_W, \mu)
}\right| \leq 2\,10^{-4} \ll 1.
\label{eq:ss2}
\end{equation}
There are two ways to consider the relation (\ref{eq:ss2}):\newline
*\ the first is to directly plug in the experimental values for $s_{23}$ and
$s_{13}$ and see whether they correspond to a suitable value of
 the Cabibbo angle $\theta_{12}$.
Experimentally, $\theta_{12} \approx 13^o$,
 $s_{13} \approx V_{ub} \approx 4.1\,10^{-3}$, $s_{23}\approx
V_{cb} \approx 42\, 10^{-3}$, such that
the l.h.s. of (\ref{eq:ss2}) is found approximately equal to $1.5\,
10^{-3}$ instead of a few $10^{-4}$.
The agreement is far from being good;\newline
*\ eqs.~(\ref{eq:bigt}) and (\ref{eq:ss}) show that the l.h.s. of
(\ref{eq:ss2})  scales, when $m_t$ gets larger and larger, like
$\lambda_1\displaystyle\frac{m_c^2}{m_t^2\big(1 + \lambda_2\,
\ln\frac{m_t^2}{\mu^2}\big)}$, and goes accordingly to $0$ when the
hierarchy $\frac{m_t}{m_c}$ increases. When $m_t$ gets very large $m_t \gg
m_W$, the CKM angles must therefore satisfy the condition 
\begin{equation}
\tan 2\theta_{12} \approx \frac{s_{13}^2 - s_{23}^2
-s_{13}^2s_{23}^2}{s_{13}\sin 2\theta_{23}}.
\label{eq:ss3}
\end{equation}
If one plugs in (\ref{eq:ss3}) the observed values of $\theta_{12}$ and
$\theta_{13}$, one finds that this corresponds to $\theta_{12}\approx
38^o$.
Reciprocally, plugging in a realistic value $|\tan
2\theta_{12}|\approx \frac12$ for the Cabibbo angle, one gets
 $s_{13}\approx \frac{\sqrt{5}-1}{2} \tan \theta_{23} \approx .618
\tan \theta_{23}$. Though the precise values  disagree with experiment, they
satisfy, as observed, $\theta_{13} < \theta_{23}$.

As we show now, a very heavy top quark tends to drag the value of the Cabibbo angle
down from quasi-maximal (which is the prediction for 2 generations)
to a smaller value. For that purpose, let us perform
the same study assuming now that $m_t \ll m_W$, only, for example, slightly
heavier than the bottom quark.  Instead of the system (\ref{eq:6const}),
eqs.~(\ref{eq:6cond}) now yield
\begin{subequations}
\begin{eqnarray}\label{subeq:6c1}
\alpha^u = \beta^u &:& \frac12 (m_d^2-m_s^2)
 = -c_{13}^2\cos 2\theta_{12}(m_u^2-m_t^2)\cr
&&\hskip -1cm -\left[\cos 2\theta_{12}(-c_{23}^2 + s_{13}^2
s_{23}^2) +\sin 2\theta_{12}\sin 2\theta_{23} s_{13}\cos\delta\right]
(m_c^2-m_t^2);
\end{eqnarray}
\begin{eqnarray}\label{subeq:6c2}
\beta^u = \gamma^u &:& 
\frac12 (m_s^2-m_b^2)
= -(s_{12}^2 c_{13}^2 - s_{13}^2)(m_u^2-m_t^2)\cr
&& \hskip -2cm-\left[ c_{12}^2 c_{23}^2 +s_{23}^2(-c_{13}^2 + s_{12}^2
s_{13}^2) -\frac12 \sin 2\theta_{12}\sin 2\theta_{23} s_{13}\cos\delta
\right](m_c^2-m_t^2);
\end{eqnarray}
\begin{eqnarray}\label{subeq:6c3}
\alpha^u = \gamma^u &:& 
\frac12 (m_d^2-m_b^2)
= -(c_{12}^2 c_{13}^2 - s_{13}^2)(m_u^2-m_t^2)\cr
&&\hskip -2cm\ -\left[ s_{12}^2 c_{23}^2 + s_{23}^2 ( c_{12}^2 s_{13}^2 -c_{13}^2)
+\frac12 \sin 2\theta_{12}\sin 2\theta_{23} s_{13}\cos\delta \right]
(m_c^2-m_t^2);
\end{eqnarray}

\vbox{
\begin{eqnarray}\label{subeq:6c4}
\alpha^d = \beta^d &:&   \frac12 (m_u^2-m_c^2) =
-\left[c_{12}^2(c_{13}^2 -s_{23}^2s_{13}^2) - s_{12}^2c_{23}^2
-\frac12 \sin2\theta_{12}\sin 2\theta_{23} s_{13}\cos\delta\right]
(m_d^2-m_s^2)\cr
&&  -\left[s_{12}^2(c_{13}^2 -s_{23}^2s_{13}^2) -
c_{12}^2c_{23}^2 +\frac12 \sin2\theta_{12}\sin 2\theta_{23}
s_{13}\cos\delta\right](m_s^2-m_b^2);
\end{eqnarray}
}

\vbox{
\begin{eqnarray}\label{subeq:6c5}
\beta^d = \gamma^d &:& \frac12 (m_c^2-m_t^2)
= -\left[\cos 2\theta_{23}(s_{12}^2 - c_{12}^2 s_{13}^2)
+  \sin2\theta_{12}\sin 2\theta_{23} s_{13}\cos\delta\right](m_d^2-m_b^2)\cr
&& \hskip 1cm - \left[\cos 2\theta_{23}(c_{12}^2 -s_{12}^2 s_{13}^2)
-  \sin2\theta_{12}\sin 2\theta_{23} s_{13}\cos\delta  \right](m_s^2-m_b^2);
\end{eqnarray}
}

\begin{eqnarray}\label{subeq:6c6}
\alpha^d = \gamma^d &:& \frac12 (m_u^2-m_t^2) = 
 - \left[ c_{12}^2 (c_{13}^2 -c_{23}^2 s_{13}^2)
-s_{12}^2 s_{23}^2 +\frac12 \sin2\theta_{12}\sin 2\theta_{23}
s_{13}\cos\delta \right](m_d^2-m_b^2)\cr
&& \hskip 1cm - \left[ s_{12}^2 (c_{13}^2 - c_{23}^2 s_{13}^2)
-c_{12}^2 s_{23}^2 -\frac12 \sin2\theta_{12}\sin 2\theta_{23}
s_{13}\cos\delta \right](m_s^2-m_b^2).
\end{eqnarray}
\label{eq:6c}
\end{subequations}
Neglecting $m_d \ll m_s$, $m_s \ll m_b$, $m_d \ll m_b$,
 $m_u \ll m_c$, $m_u \ll m_t$ and
supposing also that $m_c \ll m_t$, (\ref{subeq:6c1}) approximates to
\begin{equation}
\frac12 (m_d^2-m_s^2) -(m_c^2-m_u^2)\cos 2\theta_{12} \approx
m_t^2 \Big[ (-s_{13}^2 c_{23}^2 + s_{23}^2) \cos 2\theta_{12} + s_{13} \sin
2\theta_{23} \sin 2\theta_{12}\Big];
\label{eq:6co1}
\end{equation}
the sum of (\ref{subeq:6c2}) and (\ref{subeq:6c3}) yields
\begin{equation}
\frac{m_b^2}{m_t^2} \approx 1 - 3\,c_{13}^2 c_{23}^2;
\label{eq:6co2}
\end{equation}
Eq.~(\ref{subeq:6c5}) becomes
\begin{equation}
\frac{m_b^2}{m_t^2} \approx -\frac12\; \frac{1}{c_{13}^2\cos 2\theta_{23}},
\label{eq:6co3}
\end{equation}
and (\ref{subeq:6c6})
\begin{equation}
\frac{m_b^2}{m_t^2} \approx -\frac12\; \frac{1}{c_{13}^2 c_{23}^2 -s_{13}^2}
\equiv -\frac12\; \frac{1}{\cos 2\theta_{13} -s_{23}^2 c_{13}^2}.
\label{eq:6co4}
\end{equation}
Eqs.~(\ref{eq:6co3}) and (\ref{eq:6co4}) can only be simultaneously verified if
 $c_{13}^2\approx1$, such that $\theta_{13} \approx 0$.  Plugging this
result into (\ref{eq:6co2}) requires $c_{23}^2 \approx
\frac13\left(1-\frac{m_b^2}{m_t^2}\right)$. This entails $\theta_{23} \geq
\arccos \frac{1}{\sqrt{3}} \approx 54^o$. Then, (\ref{eq:6co1}) yields
$\frac12(m_d^2-m_s^2) = \cos 2\theta_{12}\left[(m_c^2-m_u^2) + m_t^2
\Big(\frac23 +\frac{m_b^2}{3m_t^2}\Big) \right]$. Because of the term
proportional to $m_t^2$, the corresponding modulus
of $\cos 2\theta_{12}$ gets accordingly  smaller than for 2 generations;
this corresponds to a larger Cabibbo angle, thus still closer to maximal.
This is the opposite of what happens when the top quark gets much heavier
than the $W$. So, as announced, by going across the electroweak scale
and getting more and more massive, the top quark shifts down the modulus 
of the 1-loop Cabibbo angle with respect to the 2-generation case.

\subsection{Solving the constraints for 3 generations of leptons}
\label{subsec:3nu}

The case that we just investigated, when all fermion masses for 3
generations stand below the $W$ scale corresponds {\em a priori} to the
leptonic sector. There, while one knows that $m_e \ll
m_\mu \ll m_\tau$, our knowledge about the  neutrino messes essentially
concerns the extreme smallness of their differences
\cite{BilenkyGiuntiGrimus}\cite{GiuntiKim}.

This is why all 3 equations (\ref{subeq:6c1}),
(\ref{subeq:6c2}) and (\ref{subeq:6c3}), in which the differences of
neutrino mass squared occurring in the r.h.s.'s are always much smaller than
the ones of charged leptons occurring in the l.h.s.'s, can never be
satisfied. This leaves only (b), (c), (d), (e) and (f) as possible
solutions of (\ref{eq:a2o})  for charged leptons. (b) corresponds to
general mass-flavor alignment; in (c) and (e), 1 flavor state is aligned
with the corresponding  mass state, while exact swapping, 2 by 2, occurs for
the remaining 4 states; for example,, for (c), $e_f=e_m, \mu_f=\tau_m,
\tau_f= -\mu_m$; in (d) and (f),  the 6 states are swapped 2 by 2, with no
alignment for any pair. This corroborates the common, but never
demonstrated statement, that charged leptons do not oscillate
\cite{Akhmedov}.

As for equations (\ref{subeq:6c4}),
(\ref{subeq:6c5}) and (\ref{subeq:6c6}), the extreme smallness of their
l.h.s.'s forces their r.h.s.'s to be practically vanishing.
(\ref{subeq:6c5}) and (\ref{subeq:6c6}) become respectively
\begin{equation}
m_\tau^2 c_{13}^2 \cos 2\theta_{23}\approx 0 
\label{eq:nu1}
\end{equation}
and
\begin{equation}
m_\tau^2 (\cos 2\theta_{13} -s_{23}^2 c_{13}^2) = 0.
\label{eq:nu2}
\end{equation}
Excluding $\theta_{13}= \pm\frac{\pi}{2}$, (\ref{eq:nu1}) yields
$\cos 2\theta_{23}=0 \Rightarrow \theta_{23}$ maximal $\Rightarrow c_{23}^2
= \frac12 = s_{23}^2$; when plugged into (\ref{eq:nu2}), this entails
$\tan^2\theta_{13} = c_{23}^2 = \frac12 \Rightarrow \theta_{13}\approx
\pm 35^o$. One has $s_{13}\approx \pm .577, c_{13} \approx .816$.
When the numerical values of $s_{23}^2$ and $c_{23}^2$ are plugged in
 (\ref{subeq:6c4}), it becomes
\begin{eqnarray}
&& \Big[c_{12}^2 \left( c_{13}^2 -\frac12 s_{13}^2\right) -\frac12 s_{12}^2
-\frac12 \sin 2\theta_{12}s_{13}\cos\delta \Big](m_e^2 - m_\mu^2)\cr
&& +\Big[s_{12}^2 \left( c_{13}^2 -\frac12 c_{13}^2\right) +\frac12 s_{12}^2
-\frac12 \sin 2\theta_{12}s_{13}\cos\delta \Big](m_\mu^2 - m_\tau^2)=0.
\label{eq:nu3}
\end{eqnarray}
Neglecting $m_e \ll m_\mu$, $m_\mu \ll m_\tau$, the approximate solution of
(\ref{eq:nu3}) writes $\tan \theta_{12} \approx -\frac{2s_{13}
\cos\delta}{3 c_{13}^2}\stackrel{\delta=0}{\approx} \mp .577 \Rightarrow
 |\theta_{12}| \approx 30^o$.

The values that we have found for $\theta_{12}$ and $\theta_{23}$ are very
close to the experimental values. We furthermore predict  $|\theta_{13}|
\approx 35^o$, which is still to be measured in future experiments.

Before concluding on the neutrino sector, and in relation with the
common prejudice that $\theta_{13}$ is small, let us check that no other
solution among (\ref{eq:a2o}) can accommodate for such a small angle. The
only one that could eventually fit is (o). Then, the equivalent of
(\ref{subeq:6c6}) writes (taking $\theta_{23} = \frac{\pi}{2},
\theta_{13}\approx 0$)
\begin{equation}
\frac12 (m_e^2 - m_\tau^2) \approx (c_{12}^2 -s_{12}^2) (m_{\nu_e}^2 -
m_{\nu_\mu}^2),
\label{eq:6c6nu}
\end{equation}
which, due to the strong hierarchy $(m_\tau^2 - m_e^2) \gg (m_{\nu_\mu}^2
-m_{\nu_e}^2)$, has no solution.

\section{Outlook}
\label{section:end}

We have paid in this study special attention to 
 1-loop transitions and to their role in fermionic mixing.
They spoil the diagonality of kinetic terms which
must be, first, cast back into their canonical form before the mass matrix
is  re-diagonalized and  orthogonal  mass eigenstates
suitably determined.

A first property that we encountered is that, for non-degenerate systems,
bare mass states and 1-loop mass states are non-unitarily related.

A second property is that the 1-loop mixing matrix ${\mathfrak C}(p^2)$
 occurring in
charged currents (Cabibbo, PMNS \ldots) stays unitary at ${\cal O}(g^2)$.

The third point concerns the 1-loop value of the CKM angles,
and their equivalent for leptons. The classical standard model does not
provide any hint that could help connecting masses and mixing angles.
Therefore, most investigations have concerned special structures or
textures of classical mass matrices that could eventually be explained
by subtle and broken symmetries, the origin of which being itself
lying presumably  ``beyond the standard model'' \cite{BSM}.
 To make it short, there
are more free parameters than masses and mixing angles in the classical
standard model, and one is looking for  constraints that reduce
their number, so as to, ultimately, put masses and mixing in one-to-one
correspondence.

The classical SM is  like a smooth polished sphere and
 it is extremely hard to find a defect or asperity to break
in and  put it in jeopardy. The diagonalization of classical mass matrix by
bi-unitary transformations is perfectly adequate and kinetic terms keep
 unchanged
since they are chosen from the beginning to be proportional to the unit
matrix. Through the covariant derivative, this form of the kinetic terms
dictates that of gauge currents, in particular neutral currents, for which
FCNC can only occur at 1-loop with the so-called ``Cabibbo suppression'',
``unfortunately'' very successful, too. The last cornerstone which bears
this elegant construction  is the unitarity of the Cabibbo (CKM) matrix, which
ensures, in bare mass space, the closure of the $SU(2)_L$ algebra,
when embedded in $SU(2n_f)$ ($n_f$ is the number of flavors),
on a diagonal $T^3$ generator, in which both
$n_f \times n_f$ sub-blocks are proportional to the unit matrix.
The grain of salt that may grip this beautiful machinery is, for example,
if kinetic terms are no longer diagonal. Through gauge invariance and the
covariant derivative, neutral gauge currents are then no longer diagonal
either: extra FCNC have been generated, which we know is extremely
dangerous because these are very constrained by experiments. Now, experiments
concern physical states, which are defined at the poles of the full
propagator. Since for them the standard  CKM phenomenology is perfectly
successful, we think rather unlikely that ``something goes wrong'' in this
space.  Getting, there, a suitable $SU(2)_L$ algebra which closes on ``good
old diagonal $T^3$'' is therefore a suitable goal to achieve. This goes, for
example, with a unitary renormalized CKM matrix. Then, where can things go
``wrong''? If not in physical mass space, maybe in bare mass or flavor
space, the two of them being unitary related. Classically, physical and
bare mass spaces are identical. But they are not at 1-loop. Extra FCNC can
be generated in bare mass space if they are no longer unitarily related
with physical states. Since physical states are constructed to be
orthogonal (one diagonalizes the renormalized quadratic Lagrangian), a
non-unitary relation with bare mass states can only occur if the latter are
non-orthogonal {\em i.e.} if there exists non-diagonal transitions among
them. This is  the point that we exploited in this work. Bare mass
or flavor states are no longer orthogonal at 1-loop, and they can never be,
because of mass splittings. We show that it is much better, for the
stability of corrections, to introduce counterterms ``\`a la Shabalin'',
but they cannot completely restore the orthogonality of bare mass states
on mass shell, because the different mass-shells do not coincide.
So, some trace of non-orthogonality always subsists in this space,
and thus, a slight non-unitarity in the connection between physical
states and bare mass (or flavor) states always remains, too.
 Therefore, in these last bases, some extra FCNC are
always generated at 1-loop with respect to the classical SM.
This means in particular that, in there, the gauge structure (generators,
closure on nice $T^3$ \ldots) is not perturbatively stable. It might be
possible to cope with this, but, in this work, we chose to be very
conservative and to  perturbatively preserve the structure of the
Lagrangian that was chosen at the classical level.
We therefore asked that these extra FCNC vanish or, at least, be strongly
damped. Since they depend on the classical CKM (or PMNS) angles,
on the fermion and $W$ masses (and on one renormalization scheme $\mu$),
the constraints that we obtained connect these parameters.

Shabalin's counterterms play a decisive role. They are very seldom
introduced, though they were already proved to be determinant in the
calculation of the electric dipole moment of the quarks \cite{Shabalin}.
We have shown that, in their absence, quantum corrections to mixing angle
go all the more out of control as fermions come closer to degeneracy.
One then faces technical problems such that results of perturbative
calculations cannot manifestly be trusted.
As we explicitly saw in the case of two generations, they furthermore allow
for non-trivial solutions to the suppression of extra FCNC.  In their
absence, while mass-flavor quasi-alignment  occurs for the fermion pair the
farthest from degeneracy, no special condition arises concerning
the Cabibbo angle. Instead, in their presence, in addition to the trivial,
aligned, solution, quasi-maximal mixing for the fermion pair the closest
to degeneracy, associated with mass-flavor quasi-alignment for the other
pair comes out as another suitable possibility. 
In the case of three generations, we systematically introduced them, which
had also the technical advantage to largely ease the calculations
because they ``nearly'' cancel non-diagonal kinetic terms.

The results that we obtained in the leptonic sector have the twofold
advantage to be quite encouraging (nice agreement for $\theta_{12}$ and
$\theta_{23}$) and also easily falsifiable in coming neutrinos experiments
since we also predict a large $\theta_{13} \approx 35^o$.
The quark sector looks
more problematic. We have been unable to get a small Cabibbo angle,
and the other two CKM angles also come out much too large. The only
encouraging point is the role of a heavy quark $m_t \ll m_W$ which
decreases the value of the 1-loop $\theta_{12}$ possibly down to $38^o$.
Unfortunately. this value is still much too large. So, what is happening
in the hadronic sector
\footnote{A solution has been proposed in
\cite{DorsnerSmirnov} in which, in the quark sector, $(d,s)$ and $(u,c)$
mixing angles largely cancel each other while, in the lepton sector, the
opposite occurs.}?
The role of leptons and quarks seem to have been interchanged
because, while, previously, the large values of the neutrino mixing angles
were problematic, it is now the small values of the ones of quarks that
are hard to accounted for.
One could be tempted to invoke the eventual existence of more super-heavy
fermions that could eventually drag down still more the renormalized mixing
angles. But the complexity of calculations in the presence of extra
generations of fermions rises so dramatically that it can only be the
object of a (long and tedious) forthcoming work. More
simply, the small measured values could just be thought of as second order
corrections to the trivial solution with general mass-flavor alignment
for all quark species.  Unfortunately, 2-loops calculations in the presence
of Shabalin's like counterterms stand at present also beyond our technical
abilities.

Should physics ``beyond the standard model'' be invoked? Suppose
that the leptonic $\theta_{13}$ is measured to be large $\approx 35^o$ as
we predict. The conservative conjecture of ours that
Shabalin's counterterms are enough to cancel  extra FCNC with
respect to the standard CKM phenomenology looks then reliable and
presumably carries some part of truth. Then, if BSM physics is needed,
it is to find a theoretical more sound basis to this statement.
The situation looks different for hadrons, but one should not be too much
in a hurry to invoke BSM physics before calculations of 2-loop corrections
have been achieved.

We end up this work  by pointing out at some differences with previous
approaches of the subject.
This study is based on the mandatory (re)-diagonalization of the sum of
kinetic and mass terms  to suitably determine an orthogonal
set of mass eigenstates. While this requirement is always and simply
taken care of at the classical
level by a bi-unitary diagonalization of the mass matrix, it is generally
overlooked as soon as radiative corrections are concerned
 \cite{DorsnerSmirnov} \cite{Casas-et-al} \cite{Balaji-et-al}
\cite{ChankowskiPokorski} \cite{KniehlSirlin}. 
In particular, only considering self-mass contributions to determine the
renormalized mass states from the renormalized mass matrix exposes to
the problem that they are not orthogonal since
there still exist kinetic-like transitions between them.  We show that
the  re-diagonalization of kinetic terms  can have important
effects.\newline
*\ First, and this is not a new result \cite{mixing} \cite{DuretMachet}
\cite{DuMaVy} \cite{Benes} but we confirm it,  bare mass (or flavor) states
are non-unitarily related to 1-loop mass eigenstates for non-degenerate
systems.
It turns out however, that, unlike individual mixing matrices, the 1-loop
Cabibbo matrix ${\mathfrak C}(p^2)$ occurring in charged currents stays
unitary (see however the caveat in appendix \ref{subsec:nonorphys}).
 It is a consequence of gauge invariance, which in particular
connects, through the covariant derivative of fermion fields,  kinetic
terms to gauge currents, both at the classical level and including radiative
corrections. The expression of the 1-loop Cabibbo matrix ${\mathfrak
C}(p^2)$ is thus directly dictated by that of the 1-loop
kinetic terms, which is one more reason to pay a special attention to
them;\newline
*\ then, by a cascade of mechanisms, mixing angles close to maximal
naturally appear if one wants to preserve the standard CKM phenomenology.

We hope to have convinced the reader that a reasonable argumentation exists 
that can account for large mixing angles by linking them with small mass
splittings without invoking BSM physics from the start.
If explaining both leptonic and hadronic sectors still
remains a challenge, at least 2 among the 3 neutrino mixing angles come out
with magnitudes which are close to their measured values.
Future lies accordingly in the hands and both experimentalists and
theorists, the first, in p[[articular, to
measure the leptonic $\theta_{13}$, and  the second to estimate
higher order corrections to mass-flavor quasi-alignment of quarks and see
whether their can account for the smallness of the CKM angles.

\bigskip
\begin{em}
\underline {Acknowledgments}:
It is a pleasure to thank M.I.~Vysotsky for comments and advice.
\end{em}

\appendix

\null

\section{The dependence on $\boldsymbol{p^2}$. 
Canceling  transitions between non-degenerate physical states}
\label{section:cterms}

\subsection{Non-orthogonality of non-degenerate physical states}
\label{subsec:nonorphys}

Eqs.~(\ref{eq:hij2}), (\ref{eq:hij3}), (\ref{eq:hij4}), (\ref{eq:hij5}),
which we obtained in the absence of Shabalin's counterterms,
 are only valid when $p^2
\ll m_W^2$, but it must kept in mind that all formu{\ae} 
depend on $p^2$, even though this dependence becomes very weak when
$p^2 \ll m_W^2$.

At the price, when no counterterms are introduced,
 of a high instability in the vicinity of degeneracy (see
subsection \ref{subsec:degen}) the Cabibbo procedure can be rescued and a
$p^2$-dependent, unitary renormalized Cabibbo matrix ${\mathfrak C}(p^2,
\ldots)$ be defined.
The 1-loop effective Lagrangian is made diagonal  (see section
\ref{section:fermions})  in the basis $d_{mL}(p^2,
\ldots), s_{mL}(p^2,\ldots)$, in which  $p_\mu$ stands for the common
4-momentum of $d$ and $s$ (see Fig.~2). This means that there exist no more
non-diagonal transitions between them, such that $d_{mL}(p^2,
\ldots)$ and $s_{mL}(p^2,\ldots)$  are, by definition, orthogonal at 1-loop.
However, as soon as  a mass splitting exists,
both cannot be simultaneously on mass-shell and the physical fermions
\begin{eqnarray}
&&\hskip -1cm d_{mL}^{phys} \equiv d_{mL}\big(p^2 = \mu_d^2(p^2)\big) =
[({\cal V}_d\,{\cal R}(\xi_d))^{-1}]_{11}\big(p^2=\mu_d^2(p^2)\big)\;
d^0_{mL}
+ [({\cal V}_d\,{\cal R}(\xi_d))^{-1}]_{12}\big(p^2=\mu_d^2(p^2)\big)\;
 s^0_{mL}, \cr
&&\hskip -1cm s_{mL}^{phys} \equiv s_{mL}\big(p^2
 = \mu_s^2(p^2)\big) = [({\cal V}_d\,{\cal
R}(\xi_d))^{-1}]_{21}\big(p^2=\mu_s^2(p^2)\big)\; d^0_{mL} + [({\cal
V}_d\,{\cal R}(\xi_d))^{-1}]_{22}\big(p^2=\mu_s^2(p^2)\big)\; s^0_{mL},\cr
&&
\label{eq:nonor}
\end{eqnarray}
 which belong to two different sets of orthogonal states,
are themselves expected to be non-orthogonal.
So, unless subtle cancellations take place,  non-diagonal transitions are 
expected to occur among them, which is akin to saying that
the 1-loop Lagrangian, despite it has been built by diagonalization,
is itself  not  diagonal when
re-expressed in terms physical non-degenerate eigenstates. 
At the same time, unlike  ${\mathfrak C}(p^2)$ in (\ref{eq:Cab2}),
 which is defined for an overall global $p^2$, the ``on mass-shell''
Cabibbo matrix is expected to exhibit some slight non-unitarity
\cite{DuretMachet} \cite{DuMaVy} \cite{Benes}.

More specifically, the 1-loop quadratic effective
Lagrangian (kinetic and mass terms) can be generically
rewritten in the basis of physical eigenstates 

\vbox{
\begin{eqnarray}
{\cal L}^{1-loop}&=&\left(\begin{array}{cc} \overline{d^{phys}_{mL}} & \overline{s^{phys}_{mL}} \end{array}\right)
\psl \left(\begin{array}{cc} g_1(p^2) & g_2(p^2) \cr g_3(p^2) & g_4(p^2)
\end{array}\right)
\left(\begin{array}{c} d^{phys}_{mL} \cr s^{phys}_{mL} \end{array}\right)
+  \left(\begin{array}{cc} \overline{d^{phys}_{mR}} & \overline{s^{phys}_{mR}} \end{array}\right)
 \psl\; {\mathbb I}\left(\begin{array}{c} d^{phys}_{mR} \cr s^{phys}_{mR} \end{array}\right)\cr
&& - \left(\begin{array}{cc} \overline{d^{phys}_{mL}} & \overline{s^{phys}_{mL}} \end{array}\right)
\left(\begin{array}{cc} \rho_1(p^2) & \rho_2(p^2) \cr \rho_3(p^2) &
\rho_4(p^2)
\end{array}\right) \left(\begin{array}{c} d^{phys}_{mR} \cr s^{phys}_{mR}
\end{array}\right)\cr
 &&- \left(\begin{array}{cc} \overline{d^{phys}_{mR}} & \overline{s^{phys}_{mR}} \end{array}\right)
\left(\begin{array}{cc} \sigma_1(p^2) & \sigma_2(p^2) \cr \sigma_3(p^2) &
\sigma_4(p^2)
\end{array}\right) \left(\begin{array}{c} d^{phys}_{mL} \cr s^{phys}_{mL}
\end{array}\right) + \ldots 
\label{eq:Lphys}
\end{eqnarray}
}

Indeed, combined with (\ref{eq:trans}) which relates bare mass
states to 1-loop mass eigenstates, (\ref{eq:nonor}) entails
that the coefficients of the linear relation between the latter and
physical states are functions of $(p^2, \ldots)$.
Hermiticity requires the (supposedly real and presumably ${\cal O}(g^2)$)
 quantities $g_2,g_3, \sigma_2, \sigma_3, \rho_2,\rho_3$ to satisfy the
relations $g_3=g_2, \rho_2=\sigma_3, \rho_3=\sigma_2$.
Furthermore, since right-handed fermions are not concerned by 1-loop
transitions, $(1+\gamma^5) d_m^{phys} = (1+\gamma^5) d^0_m$ and
$(1+\gamma^5) s_m^{phys} = (1+\gamma^5) s^0_m$.

\subsection{Recovering orthogonality on mass-shell}
\label{subsec:ortho}

Whether Shabalin's counterterms are included or not, the same
technique  of diagonalizing the effective, $p^2$-dependent,
quadratic Lagrangian  yields by definition orthogonal 1-loop mass eigenstates
$d_m(p^2), s_m(p^2)$, which are however not the physical states.
Therefore, an argumentation similar to the one used, in the absence of
counterterms, in subsection \ref{subsec:nonorphys}, 
can be invoked in their presence: 
non-diagonal transitions between physical mass eigenstates at
1-loop  are expected to occur, and, when expressed in terms of them, the
effective Lagrangian at 1-loop is  expected to also be of the form
(\ref{eq:Lphys}).

When classical physical states (which are nothing more than bare
mass states) and  1-loop physical states do not drastically differ (for
example would they differ by perturbative amounts), one expects 
the non-diagonal ``scalar products'' not to be drastically different either
within the two sets.
This cannot be guaranteed in the absence of Shabalin's
counterterms because of the non-perturbative nature of the link that
occurs, then, between the two sets. In their presence, instead,
they only differ by ``small amounts'' and the above property
is expected to be true: since non-diagonal transitions between bare
mass states are, then, canceled at ${\cal O}(g^2)$, this is certainly
also true  among 1-loop physical states. 

Higher order non-diagonal transitions that still exist, in the presence of
Shabalin's counterterms, between  on mass-shell 1-loop
$s_{mL}(p^2)$ and $d_{mL}(p^2)$ can always be canceled
by another set of counterterms. This is shown in subsection \ref{subsec:2lct}
below.
However,  being presumably of order higher than $g^2$, they should only
be introduced in the framework of a 2-loop calculation, which is out of
the scope of the present work.

\subsection{Expression of the additional counterterms in the basis of
 physical states}
\label{subsec:2lct}

From any Lagrangian of the form (\ref{eq:Lphys}), on-diagonal,
$p^2$-dependent transitions 
between on mass-shell fermions, like $\mu \leftrightarrow e$ are expected
This can be embarrassing  since  defining on mass-shell muon and
electron as asymptotic states seems then problematic.
They can however be themselves canceled by introducing counterterms,
as follows.
But for the fact that we are now working in the space of physical states,
the procedure is formally similar to the one used in 
\cite{Shabalin} to determine Shabalin's counterterms, which we recalled
in section \ref{subsec:shabal} (see also \cite{DuMaVy}, appendix A). 
Canceling, for example, (on mass-shell $s$) $\to$ (on mass-shell $d)$
transitions can be done by
adding to (\ref{eq:Lphys}) four kinetic and mass-like counterterms,
 concerning both chiralities of fermions:
\begin{equation}
-{\cal A}_d\, \overline{d^{phys}_{m}} \psl (1-\gamma^5) s^{phys}_{m}
- {\cal B}_d\, \overline{d^{phys}_{m}} (1-\gamma^5) s^{phys}_m
- {\cal E}_d\, \overline{d^{phys}_{m}} \psl (1+\gamma^5) s^{phys}_{m}
- {\cal D}_d\, \overline{d^{phys}_{m}} (1+\gamma^5) s^{phys}_m.
\label{eq:cterms2}
\end{equation}
Since $s_m^{phys}$ is on mass-shell, one gets the  condition
(we call respectively $\mu_s$ and $\mu_d$ the 1-loop physical masses of $s$
and $d$, that is, the square roots of the values of $p^2$ solutions of
$p^2 = \mu_s^2(p^2)$ and $p^2 = \mu_d^2(p^2)$
(see subsection \ref{subsub:renmass}))

\vbox{
\begin{eqnarray}
&& g_2(\mu_s^2)\,  \overline{d_m^{phys}} (1+\gamma^5) \mu_s\, s_m^{phys}
-\rho_2(\mu_s^2)\,\overline{d_m^{phys}}(1+\gamma^5) s_m^{phys} -
\sigma_2(\mu_s^2)\, \overline{d_m^{phys}} (1-\gamma^5) s_m^{phys} \cr
&& \hskip -.5cm = {\cal A}_d\, \overline{d_m^{phys}}(1+\gamma^5) \mu_s s_m^{phys}
+{\cal B}_d\, \overline{d^{phys}_m} (1-\gamma^5)s_m^{phys}
+ {\cal E}_d\,\overline{d_m^{phys}}(1-\gamma^5) \mu_s s_m^{phys}
+ {\cal D}_d\, \overline{d_m^{phys}} (1+\gamma^5) s^{phys}_m,\cr
&&
\label{eq:sshell}
\end{eqnarray}
}

and since $d_m^{phys}$ is also on mass-shell, 
\begin{eqnarray}
&& g_2(\mu_d^2)\,  \overline{d_m^{phys}} (1-\gamma^5) \mu_d\, s_m^{phys}
-\rho_2(\mu_d^2)\,\overline{d_m^{phys}}(1+\gamma^5) s_m^{phys} -
\sigma_2(\mu_d^2)\, \overline{d_m^{phys}} (1-\gamma^5) s_m^{phys} \cr
&&\hskip -.5cm = {\cal A}_d\, \overline{d_m^{phys}}(1-\gamma^5) \mu_d s_m^{phys}
+{\cal B}_d\, \overline{d^{phys}_m} (1-\gamma^5)s_m^{phys}
+ {\cal E}_d\,\overline{d_m^{phys}}(1+\gamma^5) \mu_d s_m^{phys}
+ {\cal D}_d\, \overline{d_m^{phys}} (1+\gamma^5) s^{phys}_m.\cr
&&
\label{eq:dshell}
\end{eqnarray}
Equating the terms with identical chiralities in (\ref{eq:sshell}) and
(\ref{eq:dshell}) yields the  four equations

\vbox{
\begin{eqnarray}
\mu_s\, g_2(\mu_s^2)  -\rho_2(\mu_s^2) &=&
\mu_s {\cal A}_d + {\cal D}_d,\cr
-\sigma_2(\mu_s^2) &=& \mu_s {\cal E}_d + {\cal B}_d,\cr
\mu_d\, g_2(\mu_d^2) - \sigma_2(\mu_d^2) &=&
\mu_d {\cal A}_d + {\cal B}_d,\cr
-\rho_2(\mu_d^2) &=& \mu_d {\cal E}_d + {\cal D}_d,
\end{eqnarray}
}

which have the ${\cal O}(g^2)$ solutions

\vbox{
\begin{eqnarray}
{\cal A}_d &=& \frac{ \mu_s^2\, g_2(\mu_s^2) -\mu_d^2\, g_2(\mu_d^2)
+ \mu_s
\left(\rho_2(\mu_d^2)-\rho_2(\mu_s^2)\right)-\mu_d\left(\sigma_2(\mu_s^2) -
\sigma_2(\mu_d^2)\right)}{\mu_s^2 - \mu_d^2},\cr
{\cal E}_d &=& \frac{\mu_d\mu_s\left(g_2(\mu_s^2) -g_2(\mu_d^2)\right)
+ \mu_d         
\left(\rho_2(\mu_d^2)-\rho_2(\mu_s^2)\right)-\mu_s\left(\sigma_2(\mu_s^2) -
\sigma_2(\mu_d^2)\right)}{\mu_s^2 - \mu_d^2},\cr
{\cal B}_d &=& -\sigma_2(\mu_s^2) -\mu_s {\cal E}_d,\cr
{\cal D}_d &=& -\rho_2(\mu_d^2) - \mu_d {\cal E}_d.
\end{eqnarray}
}

Likewise, four counterterms
$\tilde {\cal A}_d,\tilde E_d, \tilde {\cal B}_d, \tilde D_d$ can get rid of the
on mass-shell $d_m^{phys} \to s_m^{phys}$ transitions. Hermiticity (see
above) constrains them to satisfy $\tilde {\cal A}_d = {\cal A}_d, \tilde
{\cal E}_d = {\cal E}_d,
\tilde {\cal B}_d= {\cal D}_d, \tilde {\cal D}_d = {\cal B}_d$.
 Similar additions can be done in the $(u,c)$ sector.

As emphasized at the end of subsection \ref{subsec:ortho},
 when Shabalin's counterterms are already present,
the additional counterterms invoked here are presumably of higher order
in $g$.

\newpage
\begin{em}

\end{em}

\end{document}